\documentclass[floatfix,twocolumn,showpacs,aps,tightenlines]{revtex4-1}
\usepackage{graphicx}
\usepackage{color}
\usepackage{psfrag}
\usepackage{dcolumn}
\usepackage{bm}

\usepackage{amssymb}
\usepackage{amsmath}
\usepackage{amsfonts}
\usepackage{longtable}
\usepackage{xspace}
\usepackage{mathrsfs}
\usepackage[export]{adjustbox}

\newcommand{\beq}{\begin{equation}}
\newcommand{\eeq}{\end{equation}}

\newcommand{\be}{\begin{equation}}
\newcommand{\ee}{\end{equation}}
\newcommand{\bea}{\begin{eqnarray}}
\newcommand{\eea}{\end{eqnarray}}
\newcommand{\bes}{\begin{subequations}}
\newcommand{\ees}{\end{subequations}}

\usepackage{bm}

\def\prx{{\it Phys. Rev.} X~}

\def\apjl{{\it Astrophys. J.} Lett~}

\begin{document}

\title{Application of the third RIT binary black hole simulations catalog
to parameter estimation of gravitational waves 
signals from the LIGO-Virgo O1/O2 observational runs}
\author{James Healy}
\author{Carlos O. Lousto}
\author{Jacob Lange}
\author{Richard O'Shaughnessy}
\affiliation{Center for Computational Relativity and Gravitation,
School of Mathematical Sciences,
Rochester Institute of Technology, 85 Lomb Memorial Drive, Rochester,
New York 14623}

\date{\today}

\begin{abstract}
Using exclusively the 777 full numerical waveforms of the third 
Binary Black Holes RIT catalog, we  reanalyze the ten black hole 
merger signals reported in LIGO/Virgo's O1/O2 observation runs.  
We obtain  binary parameters, extrinsic parameters, 
and the remnant properties of these gravitational waves events which are 
consistent with, but not identical to previously presented results. 
We have also analyzed  three additional events 
(GW170121, GW170304, GW170727) reported in \cite{Venumadhav:2019lyq} 
and found closely matching parameters.
We finally assess the accuracy of our waveforms with convergence studies
applied to O1/O2 events and found them adequate for current estimation
of parameters.
\end{abstract}

 \maketitle

\section{Introduction}\label{sec:Intro}

The Advanced LIGO  \cite{2015CQGra..32g4001L} and Virgo \cite{TheVirgo:2014hva}
ground-based gravitational wave (GW) detectors have
identified several coalescing compact binaries
\cite{DiscoveryPaper,LIGO-O1-BBH,2017PhRvL.118v1101A,LIGO-GW170814,LIGO-GW170608,LIGO-GW170817-bns}
and characterized their properties with Bayesian inference
\cite{DiscoveryPaper,LIGO-O1-BBH,2017PhRvL.118v1101A,LIGO-GW170814,LIGO-GW170608,LIGO-GW170817-bns,LIGO-O2-Catalog,gwastro-PE-AlternativeArchitectures,gwastro-PENR-RIFT,gw-astro-PE-lalinference-v1}.
As observatories' sensitivities increase, many more observations are expected \cite{2016LRR....19....1A}, and some will
be even better resolved.  With more and some very informative events, these GW observations pose a challenge
to source parameter inference: barring substantial improvements, systematic uncertainty in our models 
will increasingly limit
our ability to draw the sharpest possible conclusions from each observation. 

  
Our understanding of the gravitational waves from merging binary black holes follows from numerical solutions to
Einstein's equations. 
Numerical relativity breakthroughs~\cite{Pretorius:2005gq,Campanelli:2005dd,Baker:2005vv}
led to detailed predictions of the gravitational waves from the late inspiral,
plunge, merger, and ringdown of black-hole-binary systems (BHB).
These predictions helped to accurately identify the first direct
detection \cite{TheLIGOScientific:2016wfe} of gravitational waves with
such binary black hole systems \cite{Abbott:2016blz,Abbott:2016nmj,TheLIGOScientific:2016pea,Abbott:2016wiq} and match them to 
targeted supercomputer simulations \cite{Abbott:2016apu,TheLIGOScientific:2016uux,Lovelace:2016uwp}.
There have been several significant efforts to coordinate numerical
relativity simulations to support gravitational wave observations.
These include the numerical injection analysis (NINJA) project
\cite{Aylott:2009ya, Aylott:2009tn, Ajith:2012az, Aasi:2014tra}, the
numerical relativity and analytical relativity (NRAR) collaboration
\cite{Hinder:2013oqa}, and the waveform catalogs released by the
SXS collaboration~\cite{Mroue:2013xna,Blackman:2015pia, Chu:2015kft,Boyle:2019kee},
Georgia Tech,~\cite{Jani:2016wkt}, and RIT~\cite{Healy:2017psd,Healy:2019jyf,Healy:2020vre}.
Numerical relativity simulations have been directly compared to GW observations to draw inferences about binary
parameters, starting with  GW150914 \cite{Abbott:2016apu,Lovelace:2016uwp,Healy:2019jyf} and continuing through GW170104 \cite{Abbott:2017vtc,Heal:2017abq},
GW170608 \cite{Abbott:2017gyy}, the analysis in
GWTC-1 \cite{LIGO-O2-Catalog}, and GW190521 \cite{LIGO-O3-GW190521-discovery,LIGO-O3-GW190521-implications}.    Further discussion of these methods can be found in 
\cite{Lange:2017wki,2017PhRvD..96j4041L,gwastro-mergers-nr-LangeMastersThesis}.
Previous  comparisons of GW observations to banks of NR simulations have used  heterogeneous sets of NR simulations, with differences in accuracy
standards and choices for initial starting separation.  
Only the analysis of GW190521 presented posteriors for all intrinsic parameters of a generic quasicircular
binary black hole, allowing for precessing spins.

In this work, we analyze all proposed candidate BBH observations reported before the latest observing run (O3) with a single, consistent set of numerical relativity
simulations: the simulations in the third release of the RIT public
catalog~\cite{Healy:2020vre}.   These simulations adopt  consistent 
resolutions and initial conditions.  We demonstrate that
direct comparisons to numerical relativity simulations can recover all astrophysically interesting properties of merging
binary black holes, including the effect of misaligned spin.

This paper is organized as follows.  
In Sec.~\ref{sec:Application} we review the methods we use to infer the intrinsic and extrinsic parameters of compact
binary sources, via direct comparison to our specific set of numerical relativity simulations. 
Specifically, following \cite{Lange:2017wki}, on the grid of simulations we evaluate the Bayesian likelihood maximized over extrinsic
parameters, using RIFT \cite{Lange:2018pyp}.  We generate posterior distributions by interpolating the resulting (marginal) likelihood distribution. 
In Sec.~\ref{sec:O1O2} we use the
waveform catalog to estimate the binary black hole parameters that
best match the ten  BBH signals reported in the first and second  LIGO-Virgo
observing runs   \cite{LIGOScientific:2018mvr}.   We find our method can produce posteriors quite consistent with
previously reported results.  Our headline posterior inferences differ principally because we adopt different prior
distributions for the binary mass, mass ratio, and spin
\cite{Vitale:2017cfs}.
 We conclude in
Sec.~\ref{sec:Discussion} with a discussion of the future use of this
catalog for parameter inference of new gravitational waves events and
the extensions of this work to more generic precessing binaries.


\section{Application of the waveforms catalog to parameter estimation of Binary Black holes}\label{sec:Application}

\subsection{Simulations}

The third release of the RIT public
catalog~\cite{Healy:2020vre} of numerical relativity black-hole-binary
waveforms \url{http://ccrg.rit.edu/~RITCatalog}
consists of 777 accurate simulations that include 300 precessing
and 477 nonprecessing binary systems with mass ratios
$q=m_1/m_2$  
in the
range $1/15\leq q\leq1$ and individual spins up to $S_i/m_i^2=0.95$.
The catalog also provides
initial parameters of the binary, 
trajectory information, peak radiation, and final remnant black hole
properties. 
The catalog includes all waveform modes $\ell\leq4 $ of $\psi_4$ and the strain $h$
(both extrapolated to null-infinity) and is updated to correct for
the center of mass displacement during inspiral and after merger
\cite{Healy:2020vre}.

The third RIT public catalog has two families of simulations salient to our comparison.  First, the RIT catalog has many
nonprecessing simulations,  displayed in Fig.~5 of Ref.~\cite{Healy:2020vre}.  
The RIT catalog also has many precessing simulations, performed with similar settings and to a consistent standard.  
To simplify the large precessing parameter space, we focus on
systems where one black hole is nonspinning, and vary the spin orientation 
of the other.  Currently this set of simulations consists
of nine different mass ratio families as displayed in Fig.~8 of 
Ref.~\cite{Healy:2020vre} with up to 40 different spin orientations per
family.  We supplement the new simulations in this catalog release 
with those reported in Ref.~ \cite{Lousto:2012gt,Zlochower:2015wga}


\subsection{Direct comparison of NR to GW observations}
We can directly compare any of our simulations to real or synthetic gravitational wave observations by scaling that
simulation and its predictions to a specific total redshifted mass $M_z$ and then marginalizing the likelihood for the gravitational
wave data over all extrinsic parameters
\cite{2015PhRvD..92b3002P,Abbott:2016apu,2017PhRvD..96j4041L,2017CQGra..34n4002O,2018arXiv180510457L}: 
 the seven coordinates characterizing the spacetime coordinates and orientation of the binary relative to the earth.  
Specifically the likelihood of the data given Gaussian noise has the form  (up to normalization)
\begin{equation}
\label{eq:lnL}
\ln {\cal L}(\bm{\lambda} ;\theta )=-\frac{1}{2}\sum\limits_{k}\langle h_{k}(\bm{\lambda} ,\theta )-d_{k} |h_{k}(\bm{\lambda} ,\theta )-d_{k}\rangle _{k}-\langle d_{k}|d_{k}\rangle _{k},
\end{equation}
where $h_{k}$ are the predicted response of the k$^{th}$ detector due to a source with parameters ($\bm{\lambda}$, $\theta$) and
$d_{k}$ are the detector data in each instrument k; $\bm{\lambda}$ denotes the combination of redshifted mass $M_{z}$ and the
remaining intrinsic parameters (mass ratio and spins; with eccentricity~$\approx0$)
needed to uniquely specify the binary's dynamics; $\theta$ represents the
seven extrinsic parameters (4 spacetime coordinates for the coalescence event and 3 Euler angles for the binary's
orientation relative to the Earth); and $\langle a|b\rangle_{k}\equiv
\int_{-\infty}^{\infty}2df\tilde{a}(f)^{*}\tilde{b}(f)/S_{h,k}(|f|)$ is an inner product implied by the k$^{th}$ detector's
noise power spectrum $S_{h,k}(f)$. 
In practice we adopt a low-frequency cutoff f$_{\rm min}$ so all inner products are modified to
\begin{equation}
\label{eq:overlap}
\langle a|b\rangle_{k}\equiv 2 \int_{|f|>f_{\rm min}}df\frac{[\tilde{a}(f)]^{*}\tilde{b}(f)}{S_{h,k}(|f|)}.
\end{equation}
For our analysis of GW150914, we adopt the same noise power spectrum employed in previous work 
\cite{Abbott:2016apu,2018arXiv180510457L}.
For each simulation and each detector-frame mass $M_z=(1+z)M$, we then compute the marginalized likelihood
\begin{equation}
\label{eq:Lmarg}
{\cal L}_{\rm marg}(\lambda) = \int d\theta p(\theta) {\cal L}(\bm{\lambda}, \theta)
\end{equation}
where $\lambda$ denotes the simulation parameters and the redshifted mass $M_z$ and where $p(\theta)$ is a conventional
prior on the extrinsic parameters.

For each simulation,  the marginalized likelihood in Eq. (\ref{eq:Lmarg}) is a one-dimensional function of $M_z$.  In
practice, we explore a small range of redshifted masses for each simulation, to be sure we cover the region near the
peak values well; see \cite{Abbott:2016apu,2017PhRvD..96j4041L}.





\subsection{Intrinsic coordinate systems and priors for binaries}
\label{sub:priors}
We characterize the intrinsic parameters of BH binaries with the (redshifted) components masses $m_{1,z},m_{2,z}$ and dimensionless spins
$\bm{\chi}_i$.
However, we will also use several other coordinates to characterize binary properties when performing parameter inference.  We use the familiar total mass $M_z=m_{1,z}+m_{2,z}$ and mass
ratio $q=m_{1,z}/m_{2,z}$, where we require $m_{2,z}>m_{1,z}$.
For binary spins, we principally characterize the effects of aligned spin in the strong field with  $S_{hu}$, defined by
\beq
M^2\,S_{hu}=\left((1+\frac{1}{2q})\,\vec{S}_1+(1+\frac{1}{2}q)\,\vec{S}_2\right)\cdot\hat{L},
\eeq
to describe the leading effect of hangup on
the full numerical waveforms \cite{Healy:2018swt}. 
Motivated by work on post-Newtonian inspiral, we also use the variable \cite{Ajith:2009bn} 
$$M^2\,\chi_{eff}=\left((1+\frac1q)\,\vec{S}_1+(1+q)\,\vec{S}_2\right)\cdot\hat{L}$$ to characterize the effects of
aligned spins. 

We adopt four distinct joint prior distributions over these intrinsic parameters.   For the first family, appropriate
to generic quasicircular binaries, we adopt the
generic quasicircular priors used in previous work \cite{LIGO-O2-Catalog,Lange:2018pyp}: jointly uniform in $m_{1,z},m_{2,z}$; a uniformly isotropic spin orientation distribution
for both spins; and magnitudes $|\bm{\chi}_1|$, $|\bm{\chi}_2|$ both uniform from 0 to 1.   
For the second family, appropriate to nonprecessing binaries, we adopt a familiar nonprecessing prior
\cite{LIGO-O2-Catalog,Lange:2018pyp}: jointly uniform in $m_{1,z},m_{2,z}$; both spin angular momenta aligned with
$\hat{L}$ (also denoted the $z$ axis); and $\chi_{i,z}$ drawn from the ``z prior'' between $[-1,1]$
\cite{Lange:2018pyp}.   
[The ``z prior'' is equal to the marginal distribution of $\chi_{i,z}$ if $\bm{\chi}_i$ are isotropic and have
  magnitudes uniform from 0 to 1.  To the extent transverse spins have no impact on the likelihood, an aligned result
  with the z prior will agree with generic quasicircular inference using the isotropic/uniform-magnitude prior.]
For the third family, also appropriate to nonprecessing binaries, we adopt a different prior:
$M_z,q,\chi_{1,z},\chi_{2,z}$ all jointly uniform over their allowed range, with $q\le 1$ and $|\bm{\chi}_i|<1$.     
For the fourth family, appropriate to a single precessing spin, we allow the polar spin angles to be uniform in
$\theta,\phi$; $|\bm{\chi}_i|$ to be uniform from [0,1]; the parameter $q\in[0.2,2]$, where $q<1$ means the more massive
BH is spinning and $q>1$ means the less massive BH is spinning; and $M_z$ is uniform. 
Unless otherwise noted, all posterior distributions and credible intervals are generated using the last two sets of
prior assumptions.

The precessing simulations used here explore only one of the two precessing degrees of freedom.    For the short binary
black hole GW signals studied here, however, observations only weakly constrain the subdominant effect of the smaller
objects' spin.   Targeted studies of precessing, two spin configurations have 
been performed for GW170104~\cite{Heal:2017abq} and GW190521~\cite{LIGO-O3-GW190521-discovery}.

\subsection{Likelihood interpolation and posterior generation }

Following previous work
\cite{Lange:2018pyp,Lange:2017wki}, we interpolate the marginal likelihood between simulation parameters.   Because of
the two distinct groupings and limited parameter space coverage, we perform two independent interpolations over the two
distinct sets of simulations: nonprecessing and precessing. 
In both cases, we use Gaussian process (GPR) regression to interpolate between and extrapolate outside the parameter space
covered by our simulations.   Posterior distributions are generated by sampling from our prior distributions, weighting
by the likelihood.   When employing the first two conventional priors, we use 
the ``construct intrinsic posteriors'' (CIP) program, an interpolation and posterior-generating
code provided by RIFT
(Rapid Inference via iterative fitting), which uses a  squared-exponential plus white noise kernel \cite{Lange:2018pyp}.  
When employing the uniform in $M,q,|\bm{\chi}|$ prior, we use an independent  implementation, also using a GPR with
a squared exponential and white noise kernel.

Specifically, for one family of results, we only use nonprecessing simulations to compare to GW observations.
For the other family of results, we use only precessing simulations.  For the final black hole parameters, we use the nonprecessing simulations.




For nonprecessing simulations, we can also perform a 3-dimensional GPR fit of $\ln {\cal L}(q,\chi_1,\chi_2)$ by maximizing
over the total mass for each simulation.  We perform both analyses and compare results
to check consistency and robustness of the algorithm.  Finally, for the purposes of illustration, we can also perform a
2-dimensional GPR fit of $\ln{\cal L}(q,S_{hu})$ or ${\cal L}(q,\chi_{eff})$ by maximizing over total mass and assuming
the remaining spin degrees of freedom do not impact the marginal likelihood.

\subsection{Estimation of extrinsic parameters}

To roughly estimate the extrinsic parameters of the events, we look at up to 100
of the top (precessing and non-precessing)
simulations per event and output the samples from the RIFT analysis before marginalization.
We then apply the technique described in \cite{Lange:2018pyp} to infer the extrinsic parameters, assuming the intrinsic
parameters of NR simulations cover  a representative region of the posterior.   Unlike the interpolation-based methods
used for intrinsic parameters, we do not presently correct for the finite, discrete simulation coverage over the
intrinsic parameter space when inferring extrinsic binary parameters.


\subsection{Simulated versus signal waveform comparison}

We use standard techniques \cite{Abbott:2016apu,Lange:2017wki} to directly compare GW150914 (See Fig. 10 of \cite{Healy:2019jyf}) and other O1/O2 BBH signals to our simulations.
For each simulation, direct comparison of our waveforms to the
data selects a fiducial total mass which best fits the
observations, as measured by the marginalized likelihood.
We can for each simulation select the binary extrinsic parameters,
like event time and sky location which maximize the likelihood of the data,
given our simulation and mass. Then, using these extrinsic parameters,
we evaluate the expected detector response in the LIGO Hanford (H1)
and Livingston (L1) instruments.
For each of the ten O1/O2 signals we will display
these reconstructions for the highest log-likelihood NR waveform of the
nonprecessing and precessing simulations in our catalog.
They directly compare to the signals as observed by LIGO H1 and L1 
(and Virgo, when available)
and with each other. The lower panels show the residuals of the
signals with respect to the RIT simulations.
A similar analysis was performed in Ref.~\cite{Healy:2017abq}, Figures 4-6,
for the GW170104 event.



\section{Parameter estimation of the BBH signals in O1/O2 LIGO runs}\label{sec:O1O2}

In this section, we analyze the ten events reported by LIGO and three additional binary black hole candidates, using
 direct comparison to numerical relativity.   
We provide two sets of our own estimates of binary intrinsic parameters, derived first assuming strict spin-orbit
alignment (Table~\ref{tab:aligned-PE}) and then allowing for a single precessing spin (Table~\ref{tab:precessing-PE}),
both calculated using our preferred priors (i.e., uniform in $M_z,q$).  
We also provide our estimates for binary extrinsic parameters (Table~\ref{tab:extrinsic-PE}), using a standard flat 
$\Lambda$CDM 
cosmology with Hubble parameter $H_0=67.9$km/s/Mpc 
and matter density parameter $\Omega_m=0.306$ \cite{Ade:2015xua}.  
We provide quantitative comparison between our results and previous analyses of these events using the Jensen-Shannon divergence
\cite{Mateos:2017gvm} (JSD) (Table
\ref{tab:jsd-table}), both when using our preferred priors and when using priors consistent with previous work.
To simplify our presentation, we have selected four exemplary events to illustrate our inferences in greater detail:
GW159014, GW170104, GW170729, and GW170814.


\begin{table*}
  \caption{Parameter estimation of the mass ratio $q$, the individual spins $a_1$ and $a_2$, 
    the total mass of the system in the detector frame, $M_{total}=m_1+m_2$ and 
    the effective spin variables $S_{hu}$ and $\chi_{eff}$, at the 5,50,95 percentiles. 
    The last column gives the Bayes Factor between uniform aligned spins and nonspinning systems.
\label{tab:aligned-PE}}
\begin{ruledtabular}
\renewcommand{\arraystretch}{1.5}
\begin{tabular}{lccrrrrrrr}
  Event & $f_{min}$ & Max($\ln{\cal L}$) & $q=m_1/m_2$ & $a_1$ & $a_2$ & $M_{total}/M_{\odot}$ & $S_{hu}$ & $\chi_{eff}$ & B.F.\\
\hline
GW150914 & 30 & 296.6 & $0.9436^{+0.0520}_{-0.2216}$ & $-0.4434^{+1.1750}_{-0.4468}$ & $0.3388^{+0.3602}_{-1.1620}$  & $71.7^{+4.}_{-4.1}$  & $-0.0342^{+0.1122}_{-0.1092}$  & $-0.0418^{+0.1166}_{-0.1048}$ & 0.295 \\
GW151012 & 50 & 23.7  & $0.7111^{+0.2621}_{-0.4510}$ & $0.0768^{+0.7832}_{-0.9114}$  & $0.3218^{+0.5694}_{-0.8580}$  & $47.5^{+20.4}_{-8.4}$  & $0.1924^{+0.4518}_{-0.4018}$  & $0.1826^{+0.4464}_{-0.3888}$ & 0.865\\
GW151226 & 80 & 27.4  & $0.6782^{+0.2741}_{-0.3301}$ & $0.2056^{+0.6858}_{-1.0484}$  & $0.2524^{+0.6656}_{-0.8224}$  & $23.3^{+4.5}_{-3.9}$  & $0.2034^{+0.3908}_{-0.5448}$  & $0.1962^{+0.3930}_{-0.5240}$ & - \\
GW170104 & 30 & 75.7  & $0.9167^{+0.0732}_{-0.3412}$ & $-0.1328^{+1.0370}_{-0.7962}$ & $-0.0490^{+0.9612}_{-0.8476}$ & $61.0^{+5.4}_{-6.1}$  & $-0.0212^{+0.1896}_{-0.2530}$  & $-0.0216^{+0.1884}_{-0.2520}$ & 0.404 \\
GW170608 & 80 & 54.2  & $0.6952^{+0.2585}_{-0.3289}$ & $0.3476^{+0.5460}_{-0.8888}$  & $0.2302^{+0.6312}_{-0.5390}$  & $22.0^{+3.1}_{-2.9}$  & $0.2878^{+0.3600}_{-0.4578}$  & $0.2948^{+0.3368}_{-0.4630}$ & - \\
GW170729 & 20 & 40.5  & $0.6302^{+0.3262}_{-0.2194}$ & $-0.1216^{+0.9132}_{-0.7676}$ & $0.6184^{+0.3286}_{-0.6140}$  & $125.8^{+15.9}_{-17.1}$  & $0.3568^{+0.2100}_{-0.2632}$  & $0.3236^{+0.2368}_{-0.2582}$ & 3.145 \\
GW170809 & 30 & 56.0  & $0.8653^{+0.1247}_{-0.3600}$ & $0.1476^{+0.7020}_{-1.0518}$  & $0.0334^{+0.8758}_{-0.5578}$  & $72.2^{+4.7}_{-6.9}$  & $0.1160^{+0.1554}_{-0.2418}$  & $0.1112^{+0.1566}_{-0.2286}$ & 0.392 \\
GW170814 & 30 & 118.6 & $0.7949^{+0.1828}_{-0.1438}$ & $-0.2334^{+0.7426}_{-0.6790}$ & $-0.0392^{+0.9190}_{-0.4422}$ & $58.1^{+4.3}_{-3.1}$  & $-0.0942^{+0.1624}_{-0.1310}$  & $-0.0942^{+0.1544}_{-0.1298}$ & 0.254 \\
GW170818 & 30 & 48.0  & $0.8758^{+0.1107}_{-0.2936}$ & $-0.2590^{+1.0278}_{-0.6498}$ & $0.0984^{+0.7848}_{-0.8246}$  & $76.5^{+8.3}_{-7.4}$  & $-0.0304^{+0.2372}_{-0.2562}$  & $-0.0348^{+0.2310}_{-0.2504}$ & 0.237\\
GW170823 & 30 & 53.0  & $0.8367^{+0.1508}_{-0.3031}$ & $-0.0836^{+0.9032}_{-0.7876}$ & $0.1642^{+0.6982}_{-0.8892}$  & $90.2^{+12.7}_{-10.9}$  & $0.0528^{+0.2344}_{-0.2608}$  & $0.0468^{+0.2332}_{-0.2502}$  & 0.295\\
\hline
GW170121 & 30 & 31.5  & $0.8519^{+0.1327}_{-0.3086}$ & $-0.2910^{+0.9282}_{-0.5978}$ & $-0.2568^{+0.7024}_{-0.6102}$ & $70.9^{+10.9}_{-8.3}$  & $-0.2520^{+0.2942}_{-0.3036}$  & $-0.2498^{+0.2872}_{-0.3010}$ & 0.933\\
GW170304 & 20 & 24.3  & $0.7948^{+0.1867}_{-0.3228}$ & $0.0606^{+0.7756}_{-0.8774}$  & $0.3262^{+0.5890}_{-0.7760}$  & $106.1^{+17.5}_{-15.1}$  & $0.2066^{+0.2602}_{-0.3074}$  & $0.1966^{+0.2654}_{-0.2974}$ & 0.822 \\
GW170727 & 20 & 19.6  & $0.8261^{+0.1569}_{-0.3062}$ & $-0.1136^{+0.9198}_{-0.7688}$ & $0.0200^{+0.7954}_{-0.8106}$  & $103.0^{+17.5}_{-15.3}$  & $-0.0108^{+0.3004}_{-0.3632}$  & $-0.0136^{+0.2926}_{-0.3558}$ & 0.414 \\
\end{tabular}
\end{ruledtabular}
\end{table*}


\begin{table*}
  \caption{Doing a GPR fit to find the highest $\ln{\cal L}$ from the 300 precessing simulations. Parameter estimation of
    the mass ratio $q$, the initial spin angle $\theta$ and $\varphi$, and the total mass of the system in the detector frame, $M_{total}/M_{\odot}$, at the mean of $\ln{\cal L}$ and its $90\%$ confidence ranges.
\label{tab:precessing-PE}}
\begin{ruledtabular}
\renewcommand{\arraystretch}{1.5}
\begin{tabular}{lcrrrrr}
  Event & $f_{min}$ & Max($\ln{\cal L}$) & $q$ & $\theta$ & $\varphi$ & $M_{total}/M_{\odot}$\\
\hline
GW150914 & 30 & 296.6  & $0.9853^{+0.1928}_{-0.1664}$  & $1.6346^{+0.2727}_{-0.2454}$  & $4.1796^{+1.3440}_{-3.5406}$  & $72.2^{+5.2}_{-7.7}$ \\
GW151012 & 50 & 23.7  & $0.9898^{+0.3447}_{-0.4484}$  & $1.4338^{+0.9705}_{-1.2302}$  & $4.0376^{+1.5588}_{-3.3854}$  & $45.2^{+7.9}_{-6.8}$ \\
GW151226 & 80 & 27.4  & $0.6004^{+0.2396}_{-0.0309}$  & $2.9566^{+0.1454}_{-0.1772}$  & $3.6298^{+2.3424}_{-3.0473}$  & $14.6^{+1.6}_{-0.7}$ \\
GW170104 & 30 & 75.7  & $0.6110^{+0.3656}_{-0.0867}$  & $2.3201^{+0.6119}_{-0.7597}$  & $3.6505^{+2.3776}_{-3.3137}$  & $54.9^{+8.1}_{-3.6}$ \\
GW170608 & 80 & 54.2  & $0.6010^{+0.0183}_{-0.0178}$  & $3.0609^{+0.0728}_{-0.1631}$  & $4.4296^{+1.6349}_{-3.9571}$  & $11.7^{+0.8}_{-0.6}$ \\
GW170729 & 20 & 40.5  & $0.7130^{+0.6074}_{-0.2810}$  & $0.4907^{+0.7031}_{-0.4392}$  & $3.3200^{+2.5843}_{-3.0065}$  & $126.9^{+11.5}_{-12.2}$ \\
GW170809 & 30 & 56.0  & $0.8661^{+0.4389}_{-0.3400}$  & $1.6142^{+0.9650}_{-0.9469}$  & $4.0074^{+2.0326}_{-3.5186}$  & $68.6^{+8.4}_{-9.1}$ \\
GW170814 & 30 & 118.6  & $1.0890^{+0.2306}_{-0.3629}$  & $1.6160^{+0.5646}_{-0.4112}$  & $3.7617^{+2.1948}_{-3.3847}$  & $60.1^{+3.6}_{-4.2}$ \\
GW170818 & 30 & 48.0  & $0.8929^{+0.2240}_{-0.3186}$  & $1.7876^{+0.7850}_{-0.4870}$  & $3.6656^{+2.1489}_{-3.0857}$  & $76.4^{+6.1}_{-7.7}$ \\
GW170823 & 30 & 53.0  & $1.0036^{+0.3539}_{-0.4174}$  & $1.4502^{+0.8174}_{-0.9947}$  & $3.2547^{+2.7583}_{-2.9123}$  & $88.9^{+14.2}_{-20.3}$ \\
\hline
GW170121 & 30 & 31.5  & $1.1050^{+0.2736}_{-0.5313}$  & $2.7502^{+0.3380}_{-0.9014}$  & $3.1290^{+2.8670}_{-2.8381}$  & $70.4^{+5.0}_{-4.7}$ \\
GW170304 & 20 & 24.3  & $0.9935^{+0.3574}_{-0.5184}$  & $0.8781^{+0.9566}_{-0.7713}$  & $3.1937^{+2.7716}_{-2.8827}$  & $107.6^{+6.1}_{-8.9}$ \\
GW170727 & 20 & 19.6  & $1.0757^{+0.2890}_{-0.4808}$  & $1.7577^{+1.1376}_{-1.1253}$  & $3.1165^{+2.8582}_{-2.7948}$  & $95.8^{+21.5}_{-13.5}$ \\
\end{tabular}
\end{ruledtabular}
\end{table*}



\begin{table*}
  \caption{Using samples from the top $\ln{\cal L}$ simulations to estimate the extrinsic parameters. Shown are the
    the luminosity distance $D$, sky location r.a. and declination, and the euler angles, $\phi_{orb}$, $\iota$, and $\psi$
    at the 5,50,95 percentiles.  Note that the priors in this analysis is a discrete set of simulations, so the ranges 
    are not comprehensive. 
\label{tab:extrinsic-PE}}
\begin{ruledtabular}
\renewcommand{\arraystretch}{1.5}
\begin{tabular}{lrrrrrr}
  Event & $D$ & r.a. & declination & $\phi_{orb}$ & $\iota$ & $\psi$\\
\hline
GW150914   & $541.7794^{+130.1726}_{-229.9624}$  & $2.4741^{+0.1738}_{-1.5149}$  & $-1.1023^{+0.1662}_{-0.1442}$  & $3.1280^{+2.8179}_{-2.8181}$  & $2.6698^{+0.3474}_{-1.3776}$  & $3.1578^{+2.7897}_{-2.8731}$ \\
GW151012   & $1131.1394^{+571.6106}_{-504.8474}$  & $-0.6554^{+2.5200}_{-1.6497}$  & $-0.0541^{+1.1431}_{-0.9482}$  & $3.0769^{+2.8817}_{-2.7397}$  & $1.6932^{+1.2072}_{-1.4446}$  & $3.1209^{+2.8288}_{-2.7922}$ \\
GW151226   & $408.8335^{+254.4875}_{-192.2955}$  & $-0.7096^{+2.6668}_{-2.1212}$  & $-0.0609^{+0.9958}_{-1.1217}$  & $3.1968^{+2.7599}_{-2.8648}$  & $1.8386^{+1.0505}_{-1.5661}$  & $3.1457^{+2.8230}_{-2.8430}$ \\
GW170104   & $1211.1260^{+395.9440}_{-516.9810}$  & $2.2381^{+0.2963}_{-2.4172}$  & $0.7532^{+0.4371}_{-0.9003}$  & $2.9597^{+3.0383}_{-2.6647}$  & $0.8017^{+2.1117}_{-0.6215}$  & $3.1029^{+2.7929}_{-2.7636}$ \\
GW170608   & $362.5212^{+118.6758}_{-150.2092}$  & $2.1704^{+0.0739}_{-0.2628}$  & $0.8439^{+0.3593}_{-0.4402}$  & $3.1150^{+2.8649}_{-2.8441}$  & $1.6875^{+1.1957}_{-1.4282}$  & $3.1349^{+2.8241}_{-2.8781}$ \\
GW170729   & $2980.4779^{+1566.0921}_{-1446.0979}$  & $-1.1476^{+3.2056}_{-0.3336}$  & $-0.6694^{+0.9918}_{-0.4752}$  & $3.0461^{+2.8902}_{-2.6350}$  & $2.0288^{+0.8297}_{-1.6937}$  & $3.1207^{+2.7493}_{-2.5717}$ \\
GW170809   & $1128.9550^{+358.0350}_{-416.3770}$  & $0.2851^{+0.1743}_{-0.0935}$  & $-0.4489^{+0.2638}_{-0.2179}$  & $3.2394^{+2.7236}_{-2.8990}$  & $2.6348^{+0.3662}_{-0.5185}$  & $3.1487^{+2.8320}_{-2.8376}$ \\
GW170814   & $533.8521^{+210.9939}_{-223.3711}$  & $0.7956^{+0.0674}_{-0.1274}$  & $-0.7954^{+0.4434}_{-0.0903}$  & $2.9504^{+2.9985}_{-2.6109}$  & $0.8458^{+1.9090}_{-0.6125}$  & $3.1625^{+2.8355}_{-2.7779}$ \\
GW170818   & $1299.2956^{+450.8644}_{-517.1766}$  & $-0.3259^{+0.0258}_{-0.0260}$  & $0.3716^{+0.0855}_{-0.0922}$  & $2.8997^{+3.0485}_{-2.5737}$  & $2.6126^{+0.3846}_{-0.5074}$  & $3.1506^{+2.8441}_{-2.8513}$ \\
GW170823   & $2110.2011^{+847.9689}_{-1008.9611}$  & $-1.8321^{+3.0728}_{-0.4451}$  & $-0.2937^{+1.2662}_{-0.5735}$  & $3.0280^{+2.9715}_{-2.7297}$  & $1.7803^{+1.1610}_{-1.5703}$  & $3.1497^{+2.8297}_{-2.8413}$ \\
\hline
GW170121   & $1368.2817^{+894.8583}_{-782.9907}$  & $-0.2286^{+3.0993}_{-2.4309}$  & $-0.0644^{+1.0534}_{-0.9907}$  & $2.9501^{+3.0039}_{-2.6214}$  & $1.1122^{+1.7782}_{-0.8860}$  & $3.1555^{+2.8553}_{-2.8948}$ \\
GW170304   & $2849.7316^{+1298.5984}_{-1367.1816}$  & $-0.4891^{+2.0592}_{-0.5098}$  & $0.3703^{+0.3234}_{-1.0038}$  & $3.1210^{+2.8499}_{-2.8125}$  & $1.5930^{+1.3327}_{-1.3730}$  & $3.1409^{+2.8467}_{-2.8372}$ \\
GW170727   & $2851.2071^{+1442.8729}_{-1351.0571}$  & $0.9501^{+1.9558}_{-3.9880}$  & $-0.2456^{+1.5040}_{-0.4526}$  & $3.2438^{+2.7327}_{-2.9254}$  & $1.8114^{+1.1084}_{-1.5859}$  & $3.1391^{+2.8362}_{-2.8382}$ \\
\end{tabular}
\end{ruledtabular}
\end{table*}

\begin{table*}
  \caption{Jensen-Shannon divergence for the mass-ratio, spin magnitudes, and total mass in the detector frame between our preferred analysis with uniform priors and GWTC-1 LIGO posteriors (first number in each column) and between the CIP analysis and GWTC-1 using the same priors (second number in each column). 
\label{tab:jsd-table}}
\begin{ruledtabular}
\renewcommand{\arraystretch}{1.5}
\begin{tabular}{lcccc}
  Event & $q$ & $a_1$ & $a_2$ & $M_{total}/M_{\odot}$\\
\hline
GW150914   & 0.1501, 0.0201 & 0.4061, 0.0203 & 0.4453, 0.0113 & 0.0135, 0.0147\\
GW151012   & 0.0263, 0.0167 & 0.1616, 0.0029 & 0.1999, 0.0271 & 0.0840, 0.0461\\
GW151226   & 0.0551, 0.0287 & 0.1478, 0.0038 & 0.1279, 0.1193 & 0.2004, 0.3512\\
GW170104   & 0.2758, 0.0186 & 0.2954, 0.0033 & 0.2615, 0.0012 & 0.0280, 0.0188\\
GW170608   & 0.0261, 0.0106 & 0.2249, 0.0387 & 0.2465, 0.0804 & 0.3281, 0.1553\\
GW170729   & 0.0118, 0.0478 & 0.0562, 0.0070 & 0.0545, 0.0224 & 0.0041, 0.0699\\
GW170809   & 0.1145, 0.0056 & 0.2846, 0.0060 & 0.2221, 0.0004 & 0.0352, 0.0234\\
GW170814   & 0.2945, 0.0508 & 0.2228, 0.0042 & 0.5244, 0.0023 & 0.6033, 0.3592\\
GW170818   & 0.0704, 0.0104 & 0.1671, 0.0003 & 0.1278, 0.0054 & 0.0439, 0.0229\\
GW170823   & 0.0506, 0.0137 & 0.1644, 0.0006 & 0.1450, 0.0038 & 0.0291, 0.0366\\
\end{tabular}
\end{ruledtabular}
\end{table*}

\subsection{Discussion: Mass and spin estimates}

\begin{figure*}[h!]
\includegraphics[width=1.9\columnwidth]{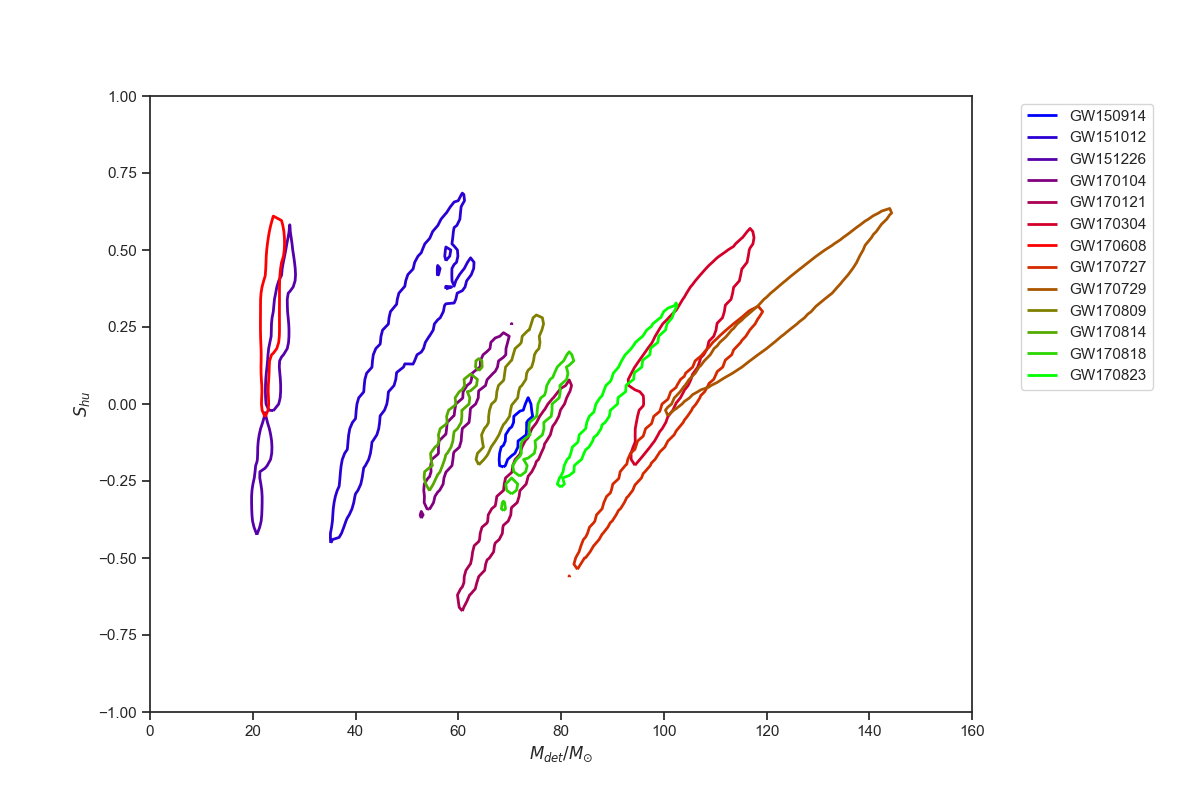}\\
\includegraphics[width=2\columnwidth]{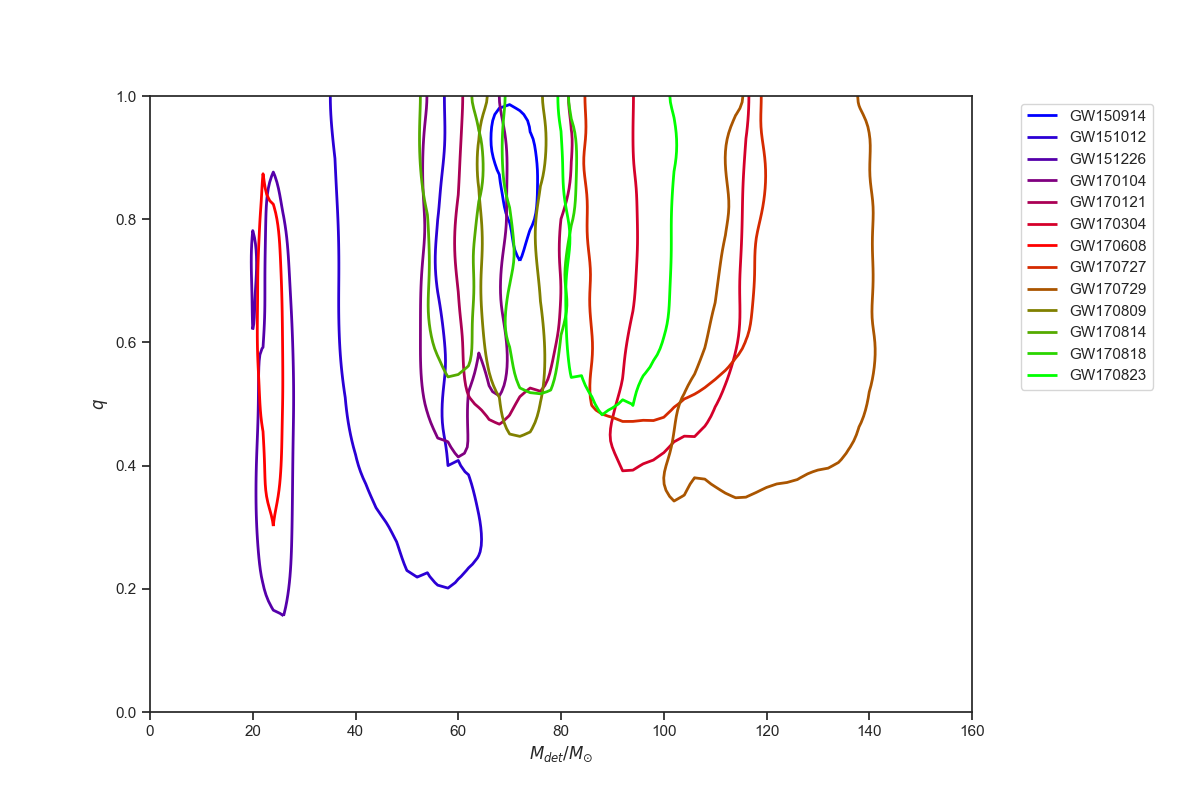}
\caption{\label{fig:all:proto}
Binary black hole observed 90\% credible intervals for selected two-dimensional marginal
  distributions.  All calculations above are derived assuming component spins are aligned relative to the
  orbital angular momentum, and adopting priors which are uniform in $M_z,q$; see Sec.~\ref{sub:priors} for details. }
\end{figure*}

\begin{figure*}
\includegraphics[width=1.9\columnwidth]{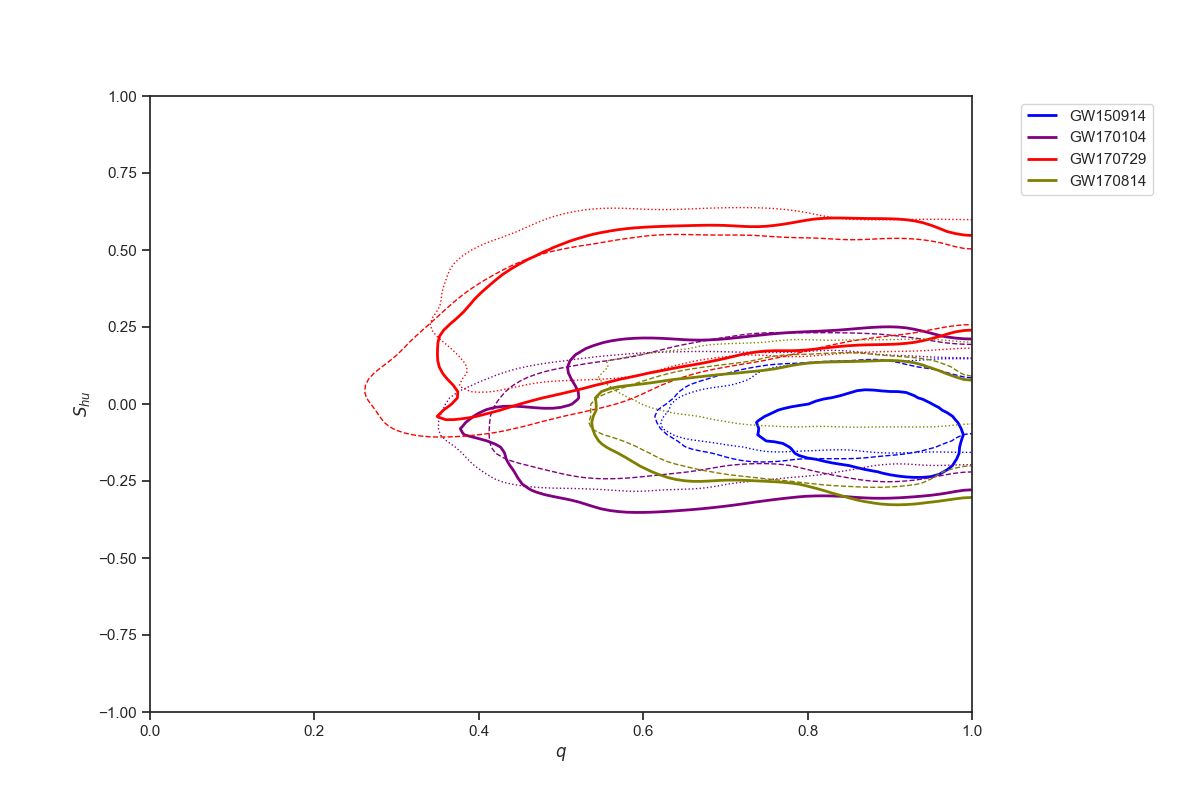}\\
\includegraphics[width=1.9\columnwidth]{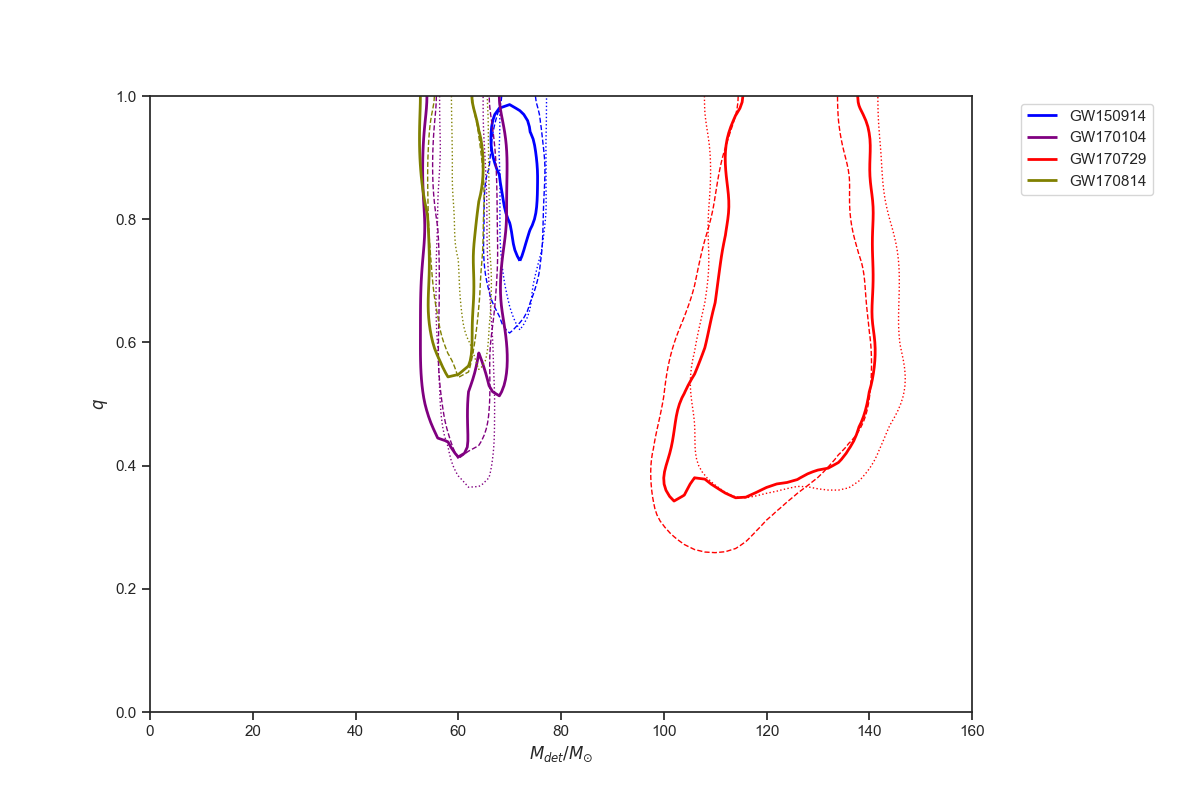}
\caption{\label{fig:exemplary:proto} Comparison of inferred 90\% credible intervals for selected binary black holes,
  using three different inference methods.  Solid lines show our preferred results, as shown in Figure
  \ref{fig:all:proto}. Dotted lines show the corresponding bound obtained from the published, public
  fiducial GWTC-1 analyses of these events, which notably adopt a different prior than our fiducial results and which
  employ a mixture of approximations to general relativity.  Dashed
  lines provide the corresponding bound from our analysis, after reanalyzing and adopting the same prior as used in GWTC-1.  See Table
  \ref{tab:jsd-table} for a quantitative comparison between the marginal (one-dimensional) posteriors.
}
\end{figure*}

Figure \ref{fig:all:proto} illustrates our inferences about binary masses and spin, using our preferred prior assumptions.
Figure \ref{fig:exemplary:proto} shows the recovered mass and spin
distributions for the four exemplary events in greater detail, for different prior choices, as well as previously-published
fiducial LIGO results.

\begin{figure*}[h]
  \includegraphics[angle=0,width=0.9\columnwidth]{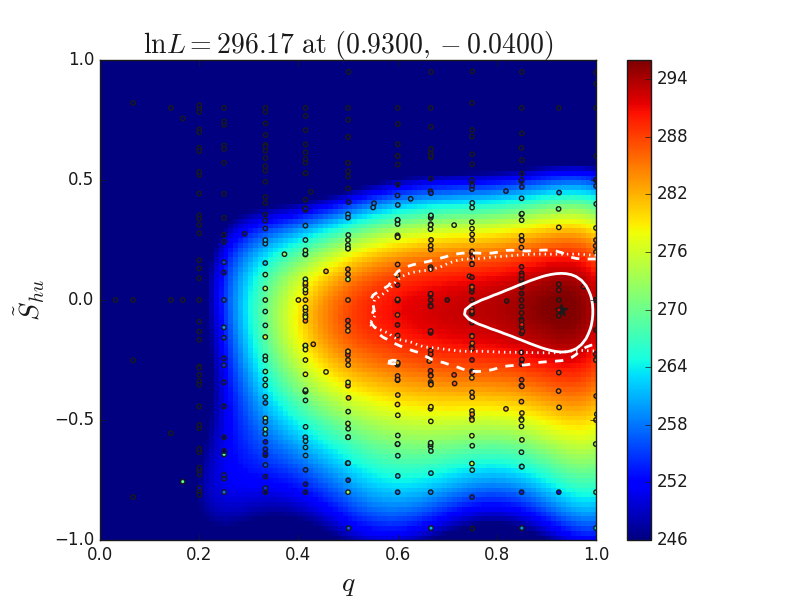} \hfill
\includegraphics[angle=0,width=1.1\columnwidth]{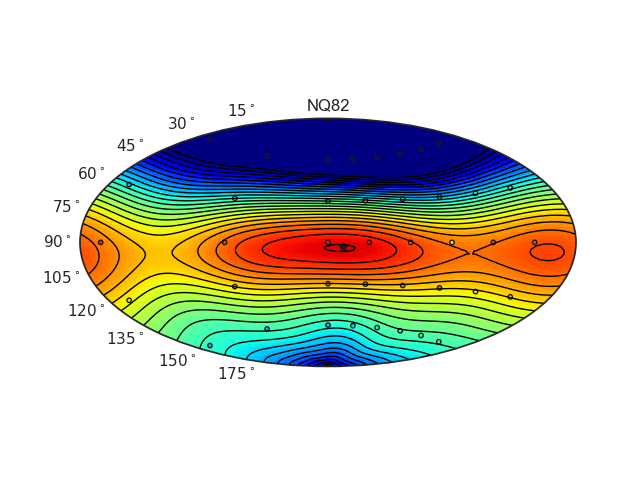}\\
    \includegraphics[angle=0,width=0.9\columnwidth]{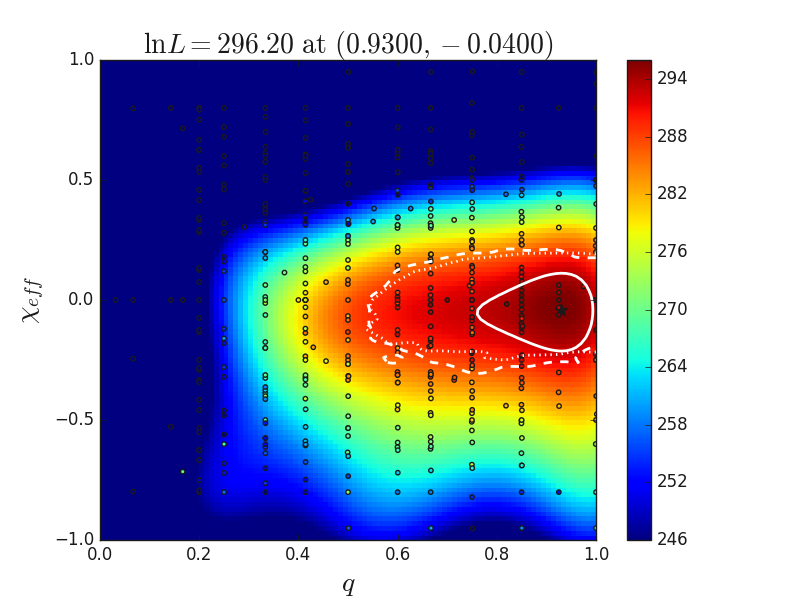} 
\includegraphics[angle=0,width=1.1\columnwidth]{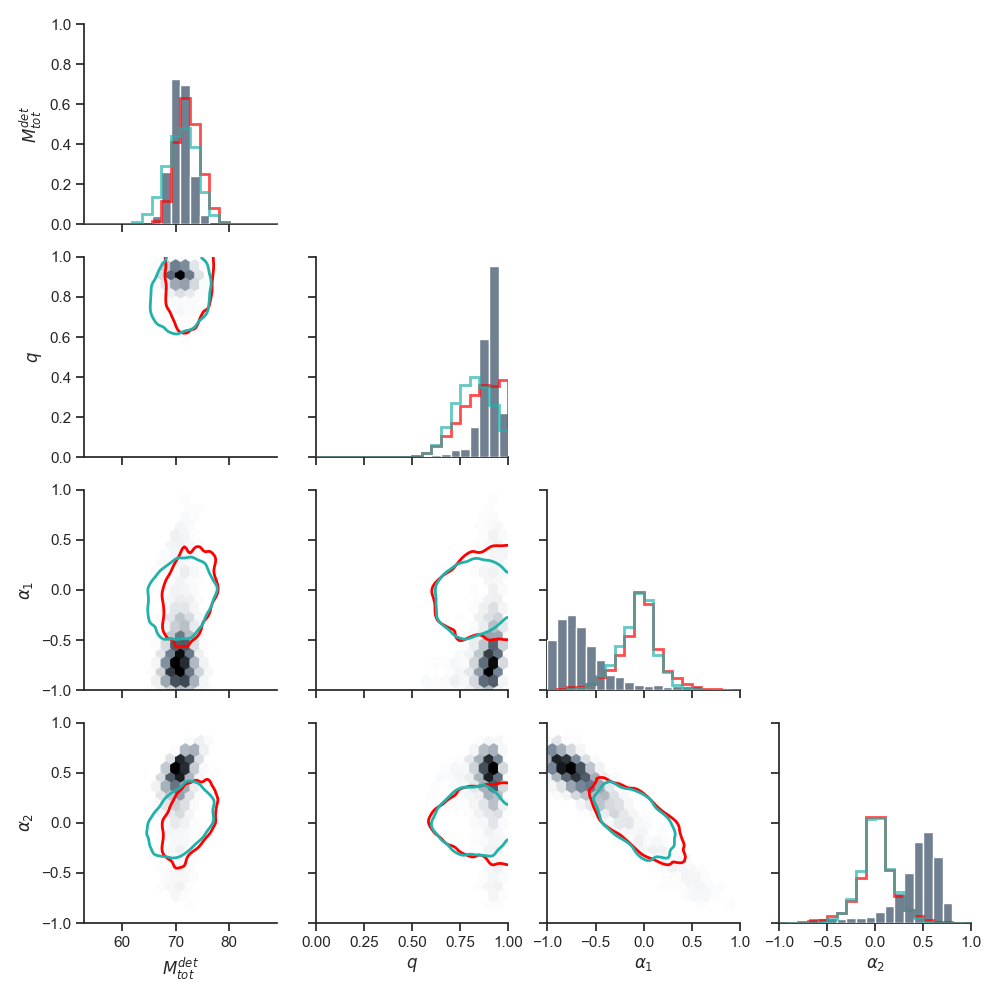}
  \caption{\emph{Left panels}: Comparative analysis of the $S_{hu}$ and $\chi_{eff}$ spins versus $q$ for GW150914 using
    the 477 nonprecessing binaries.   The points show the parameters of these nonprecessing simulations.  As described
    in the text, the color scale is based on an interpolation between each simulation's maximum ${\cal L}$ (i.e.,  $\max_M{\cal
      L}(M,q,\chi_{1,z},\chi_{2,z})$) over (only) the two parameter dimensions shown in this plot.   
  As in Figure \ref{fig:exemplary:proto}, the contours are 90\% credible intervals of a posterior based on our
full 4-dimensional  interpolated likelihood (solid);  a reanalysis of the same likelihood, using conventional priors
consistent with GWTC-1 (dashed); and the LIGO GWTC-1 analysis itself
  (dotted).
      \label{fig:GW150914ShuChi}
\emph{Top right panel}: Top likelihood panel for the binary spin orientation for GW150914 using the 300 precessing simulations. The star labels the most likely orientation of the spin and NQ82 label means a mass ratio $q=0.82$.
\emph{Bottom right panel}: 
Estimation of the (aligned) binary parameters $(M_{Total},q,\chi_1,\chi_2)$ for GW150914 using the 477 nonprecessing
simulations.  The weighted histogram is the posterior associated with the uniform priors. For
      comparison, we also superimpose the LIGO GWTC-1 posteriors in red, and the CIP analysis
      of the same dataset assuming the same priors as the LIGO GWTC-1 published results in green. 
      The fiducial pseudo-precessing spin prior adopted by GWTC-1 strongly
      disfavors extreme positive or negative spins, relative to our fiducial uniform prior.
}
\end{figure*}

To illustrate the process of likelihood interpolation, we follow previous work
\cite{Abbott:2016apu,Lovelace:2016uwp,Healy:2019jyf} and show the likelihood versus event
parameters, superimposed with a (lower-dimensional, simplified) interpolated likelihood. 
We use GW150914 to illustrate our nonprecessing and precessing analyses' likelihoods over the aligned
and precessing binary spin parameters, relative to our grid of simulated binary black holes.  The left panels of Figure  \ref{fig:GW150914ShuChi} shows our interpolated marginal likelihood
versus nonprecessing binary parameters.  We emphasize the interpolation shown for illustration is performed only in two
dimensions: for each simulation, we find the single largest value of $max_M {\cal L}(M,q,\chi_{1,z},\chi_{2,z})$, then
use GPR to interpolate in $q$ and one spin degree of freedom, treating the other as a nuisance variable.   
This figure shows  first that our simulation grid is quite dense relative to the support of the likelihood.  The
reconstructed likelihood varies smoothly over our parameter range.   Second, this figure shows that, for our fiducial
prior, marginal likelihood contours are in good agreement with our inferred posterior distribution.
Third, the two left hand figures compare two different parameterizations for the dominant spin parameter: $\chi_{\rm
  eff}$ and $S_{hu}$.  In this case, both produce consistent answers, showing the data prefers a narrow range of net aligned spins.
Finally, the bottom right hand panel of Figure  \ref{fig:GW150914ShuChi} shows a posterior corner plot.  Each
hexagon's weight reflects the likelihood of parameters associated with that hexagon. 

We also use GW190514 to illustrate our precessing analysis, using the 300 precessing single-spin simulations of the RIT catalog,
again using a slightly simplified version of the analysis used for our tabulated results.
The right panels of Figure \ref{fig:GW150914ShuChi} show slices through our parameter space corresponding to specific
mass ratios and spin magnitudes.  (This simulation catalog has only binaries where one spin has magnitude
$|\bm{\chi}|=0.8$.)  As above, for each simulation we select the largest value of $\ln {\cal L}$ for a specific
simulation, maximizing over mass, and then interpolate this function over each two-dimensional slice.  As expected from
the purely aligned analysis, we find that the data disallows spins with significant components aligned or antialigned
with the orbital angular momentum.  Further, the best mass ratio slice in this simplified analysis corresponds to
$q=0.82$, consistent with the purely aligned analysis.
Despite exploring only one spin degree of freedom, we draw consistent 
conclusions about the precessing spin, which we can usefully compare 
to targeted studies with two precessing spin binaries.

Figure \ref{fig:exemplary:proto} shows the 90\% credible intervals of our inferred posterior distribution for GW150914,
using our fiducial assumptions (solid) and using the assumptions of GWTC-1 (dotted).  For comparison, the dashed
curves in this figure also show the GWTC-1 results.   When adopting consistent priors, we obtain posterior distributions
which are exceptionally consistent with previous work.  With our fiducial prior, however, the inferred masses and spins
differ.   These differences are due to our preferred choice of priors, best illustrated in the bottom right hand panel of Figure \ref{fig:GW150914ShuChi}.  

As seen in    Figure \ref{fig:exemplary:proto}, inferences about the other  exemplary events' masses and spins are less
prior-dependent.  For GW170814 and to a lesser extent GW170104, the posterior distributions' 90\% credible intervals are
qualitatively and quantitatively quite similar.   For GW170729, modest differences exist between our analysis and
GWTC-1, which may reflect model systematics; see discussion in GWTC-1 \cite{LIGO-O2-Catalog} and
elsewhere \cite{gwastro-170729HM-Katerina,gwastro-170729HM-TechDoc}.

\subsection{Discussion: Extrinsic parameters}
\begin{figure}[h!]
  \includegraphics[angle=0,width=0.95\columnwidth]{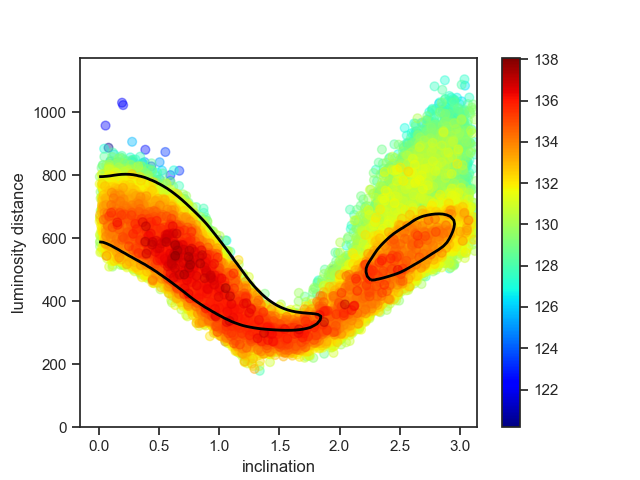}
  \includegraphics[angle=0,width=0.95\columnwidth]{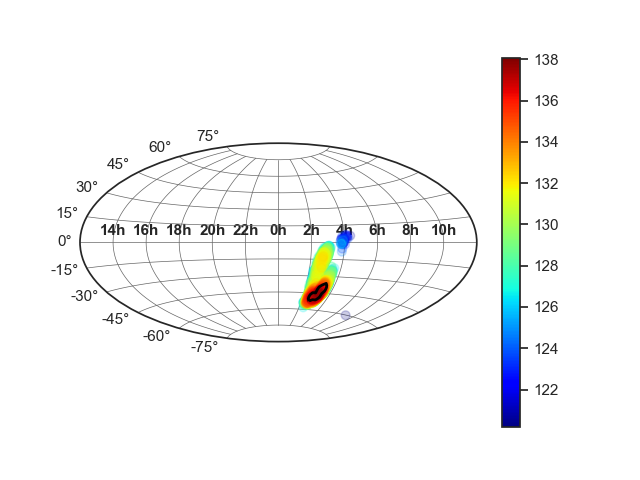}
  \caption{Sky localization on top and  luminosity distance versus inclination angle of the binary on the bottom for GW170814.  RIFT samples are colored by $\ln{L}$ and LIGO GWTC-1 posteriors are shown as solid black lines.
      \label{fig:GW170814_extrinsic}}
\end{figure}

GW170814's sky location was exceptionally tightly isolated via triple-coincident data.   To
illustrate our ability to also reconstruct binary extrinsic parameters, in Figure  \ref{fig:GW170814_extrinsic} we show
the inferred 2d marginal distributions for distance-inclination  (top panel) and sky location (bottom panel).  The color
scale shows our marginal distributions; the solid black line shows previously published
results. 
We find good agreement, despite only using a very sparse set of simulations.

\subsection{Discussion: Remnant properties}
GW170729 was a massive binary black hole whose inferred net-aligned-spins ($\chi_{\rm eff}$ or $S_{hu}$ were likely
significantly positive. As a result, its remnant properties are most different from the other merging black
holes, with the largest remnant BH spin.  
Table~\ref{tab:aligned-merger} gives the final merged black hole properties
and the peak frequency, luminosity and amplitude of the merger waveform.
For those estimates we have used the nonprecessing 477 simulations and
those values can be confronted with those given in Table III of the 
Ligo-Virgo Catalog \cite{LIGOScientific:2018mvr} (given in the source
frame). We observe again 
a large superposition of the 90\% confidence intervals in all events.
The results can also be confronted with the prediction from the remnant
formulas given in \cite{Healy:2014yta,Healy:2016lce,Healy:2018swt}
where they use as an input the binary parameters given in 
Table~\ref{tab:aligned-PE}.

\begin{table*}
  \caption{Parameter estimation of
    the final black hole mass, $m_f$, spin, $a_f$, and its recoil velocity, $v_f$,
    and the peak luminosity, $p_L$, waveform frequency $p_O$ at the maximum amplitude $p_A$ of the strain, at the mean of $\ln{\cal L}$ and its $90\%$ confidence ranges from the nonprecessing simulations.
\label{tab:aligned-merger}}
\begin{ruledtabular}
\renewcommand{\arraystretch}{1.5}
\begin{tabular}{lrrrrrrr}
  Event & $m_f$ & $a_f$ & $v_f$ & $10^{3}p_L$ & $p_O$ & $p_A$\\
\hline
GW150914   & $0.9526^{+0.0030}_{-0.0035}$  & $0.6788^{+0.0362}_{-0.0434}$  & $215.5^{+181.1}_{-158.2}$  & $0.974^{+0.083}_{-0.045}$  & $0.3562^{+0.0087}_{-0.0089}$  & $0.3928^{+0.0027}_{-0.0100}$ \\
GW151012   & $0.9508^{+0.0192}_{-0.0185}$  & $0.7292^{+0.1376}_{-0.1696}$  & $119.7^{+213.2}_{-101.6}$  & $1.005^{+0.301}_{-0.480}$  & $0.3681^{+0.0439}_{-0.0348}$  & $0.3785^{+0.0250}_{-0.1273}$ \\
GW151226   & $0.9502^{+0.0201}_{-0.0202}$  & $0.7264^{+0.1364}_{-0.2384}$  & $135.6^{+212.0}_{-109.9}$  & $0.959^{+0.343}_{-0.380}$  & $0.3673^{+0.0423}_{-0.0452}$  & $0.3763^{+0.0234}_{-0.0800}$ \\
GW170104   & $0.9525^{+0.0080}_{-0.0061}$  & $0.6746^{+0.0644}_{-0.0996}$  & $221.2^{+229.7}_{-201.4}$  & $0.991^{+0.114}_{-0.173}$  & $0.3543^{+0.0172}_{-0.0180}$  & $0.3925^{+0.0060}_{-0.0305}$ \\
GW170608   & $0.9476^{+0.0162}_{-0.0193}$  & $0.7468^{+0.1306}_{-0.1418}$  & $104.0^{+151.0}_{-88.1}$  & $1.106^{+0.254}_{-0.392}$  & $0.3742^{+0.0444}_{-0.0333}$  & $0.3790^{+0.0215}_{-0.0757}$ \\
GW170729   & $0.9430^{+0.0176}_{-0.0136}$  & $0.8018^{+0.0610}_{-0.1284}$  & $112.2^{+124.4}_{-86.2}$  & $1.125^{+0.229}_{-0.373}$  & $0.3838^{+0.0254}_{-0.0298}$  & $0.3714^{+0.0252}_{-0.0504}$ \\
GW170809   & $0.9489^{+0.0111}_{-0.0058}$  & $0.7172^{+0.0612}_{-0.1110}$  & $197.3^{+188.0}_{-167.2}$  & $1.061^{+0.097}_{-0.260}$  & $0.3660^{+0.0139}_{-0.0227}$  & $0.3901^{+0.0085}_{-0.0436}$ \\
GW170814   & $0.9553^{+0.0040}_{-0.0067}$  & $0.6468^{+0.0700}_{-0.0492}$  & $113.5^{+285.5}_{-87.4}$  & $0.944^{+0.112}_{-0.056}$  & $0.3502^{+0.0121}_{-0.0101}$  & $0.3879^{+0.0083}_{-0.0130}$ \\
GW170818   & $0.9531^{+0.0082}_{-0.0080}$  & $0.6750^{+0.0828}_{-0.1084}$  & $149.0^{+263.7}_{-131.2}$  & $0.987^{+0.129}_{-0.185}$  & $0.3553^{+0.0188}_{-0.0206}$  & $0.3910^{+0.0066}_{-0.0276}$ \\
GW170823   & $0.9512^{+0.0096}_{-0.0082}$  & $0.6984^{+0.0792}_{-0.1172}$  & $175.8^{+210.7}_{-144.4}$  & $1.017^{+0.143}_{-0.212}$  & $0.3610^{+0.0200}_{-0.0231}$  & $0.3896^{+0.0075}_{-0.0364}$ \\
\hline
GW170121   & $0.9586^{+0.0078}_{-0.0075}$  & $0.5950^{+0.1056}_{-0.1306}$  & $155.1^{+231.4}_{-127.9}$  & $0.885^{+0.136}_{-0.178}$  & $0.3390^{+0.0218}_{-0.0220}$  & $0.3897^{+0.0071}_{-0.0350}$ \\
GW170304   & $0.9470^{+0.0132}_{-0.0113}$  & $0.7452^{+0.0844}_{-0.1286}$  & $135.6^{+188.6}_{-109.4}$  & $1.084^{+0.191}_{-0.287}$  & $0.3724^{+0.0249}_{-0.0282}$  & $0.3873^{+0.0103}_{-0.0490}$ \\
GW170727   & $0.9532^{+0.0108}_{-0.0101}$  & $0.6748^{+0.1052}_{-0.1516}$  & $172.5^{+214.9}_{-139.4}$  & $0.976^{+0.179}_{-0.232}$  & $0.3558^{+0.0253}_{-0.0289}$  & $0.3890^{+0.0080}_{-0.0392}$ \\
\end{tabular}
\end{ruledtabular}
\end{table*}

\subsection{Contrasts with previously reported results}
\label{sub:jsd}
The figures and results emphasized above and derived from our fiducial priors  are qualitatively similar but noticeably quantitatively different from
previously-published results.  However, as noted above, we have also performed all our calculations with the default
priors adopted in previous work: uniform in $m_{1,z},m_{2,z}$ and compatible with spins which are isotropic and uniform
in $|\bm{\chi}|$.   Table \ref{tab:jsd-table} provides a quantitative comparison between our two sets of results (for
nonprecessing binaries), and between our results and the analyses published in GWTC-1.
Critically, this table shows that when we adopt similar priors to previous work, we find quite similar results, modulo a
few exceptions.  Though waveform systematics still plays some role, the principal difference between our inferences and
previous work is our choice of prior assumptions.

As previously noted \cite{2016PhRvL.116x1103A,LIGO-GW170608}, the long-duration signals GW170608 and GW151226 are more challenging
to analyze: few of our simulations have comparable duration,.  Because we  had to adopt a larger starting frequency
$f_{\rm min}$,  a pure NR analysis of these events is necessarily less tightly constraining than an analysis that could
incorporate lower frequency information.   As a result, for these two events we expect and observe more substantial
differences between our inferences and inferences performed using longer-duration waveform models.

\subsection{Events reported by external groups}


In addition to the ten BBH GW events reported in the first LIGO and Virgo
Gravitational-Wave Transient Catalog \cite{LIGOScientific:2018mvr}, there
have been studies of other potentially astrophysical $(p_{astro}>0.98)$ events,
such as  (GW170121, GW170304, GW170727) reported in 
\cite{Venumadhav:2019lyq}. Here we will apply our technique 
to have an
independent parameter estimation of those events with highest
claimed significance ($p_{astro}$). 
All inferred event parameters are described in the previously discussed Table \ref{tab:aligned-PE} (aligned inferences),
Table \ref{tab:precessing-PE} (precessing inferences), and Table~\ref{tab:extrinsic} (extrinsic parameters); all are
consistent with the published results in Ref.~\cite{Venumadhav:2019lyq},
for instance for the estimated mass ratios $q$ and $\chi_{eff}$ and their
90\% confidence intervals in all three signals.
All analyses are performed using on-source PSDs.
As a concrete example, we will discuss one event in greater detail: GW170304.

\begin{figure}
  \includegraphics[angle=0,width=1.0\columnwidth]{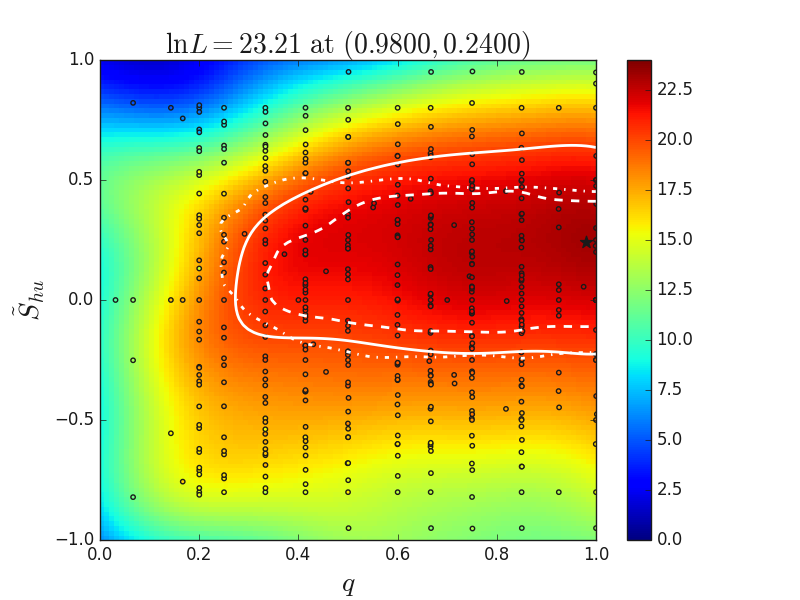}
  \caption{90\% Confidence level of the binary parameters for GW170304
in a two dimensional $(S_{hu},q)$, search (solid), CIP reanalysis (dashed), and published Princeton posterior (dash-dotted).  
      \label{fig:GW170304qShuN}}
\end{figure}
\begin{figure}
  \includegraphics[angle=0,width=1.0\columnwidth]{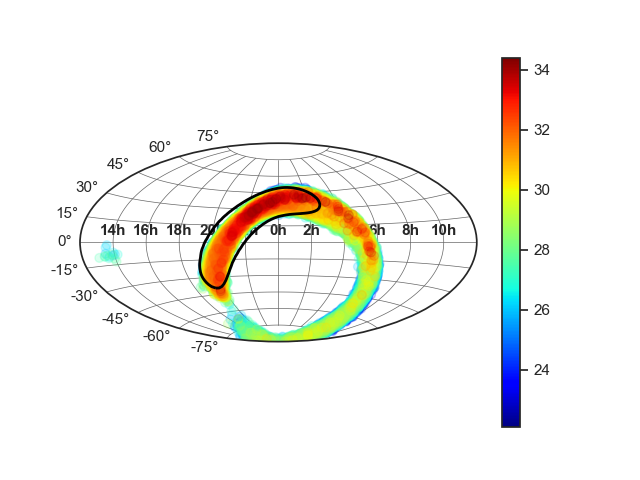}
  \includegraphics[angle=0,width=1.0\columnwidth]{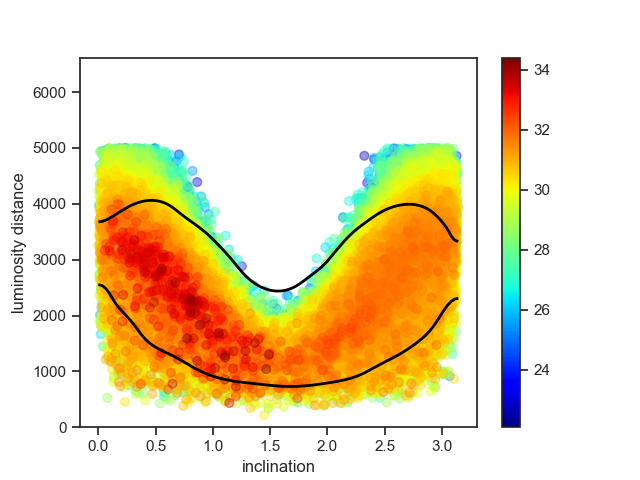}
  \caption{Sky localization on top and  luminosity distance versus inclination angle of the binary on the bottom for GW170304. 
RIFT samples are colored by $\ln{L}$ and LIGO GWTC-1 posteriors are shown as solid black lines.
      \label{fig:GW170304extr}}
\end{figure}

For these low-significance events, a large fraction of our simulation space is consistent with the observations for some
mass scale, suggesting that systematics from simulation placement will be particularly small.
Figure ~\ref{fig:GW170304qShuN} illustrates the (peak) marginal likelihood for each simulation, versus $q$ and $S_{hu}$,
as previously interpolated by a Gaussian process; compare to the corresponding Figure \ref{fig:GW150914ShuChi} for GW150914.
For comparison, the heavy dashed line shows the inferred two-dimensional posterior distribution, using our fiducial
prior as expected, its credible intervals are in good agreement with the
marginal likelihood contours.
Similarly, Figure ~\ref{fig:GW170304extr} shows the recovered sky location and joint distance-inclination posterior for this 
event, 
using all the the simulations within the 90\% confidence limit; for comparison, the black contour shows the published
reconstructed sky location.

\subsection{Waveform reconstructions}

\begin{figure*}
\includegraphics[width=0.62\columnwidth]{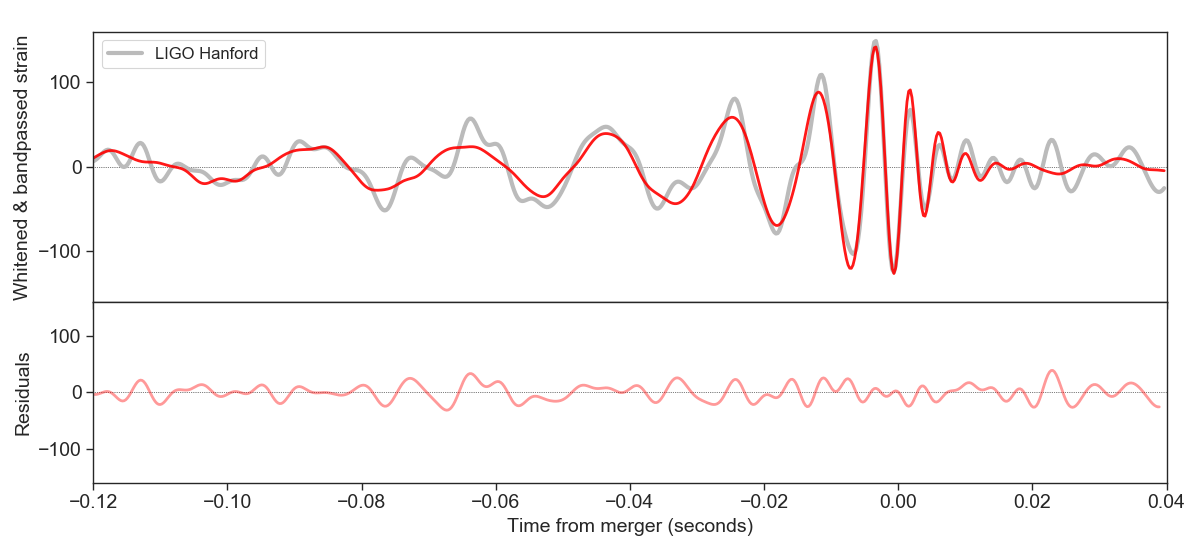}
\includegraphics[width=0.62\columnwidth]{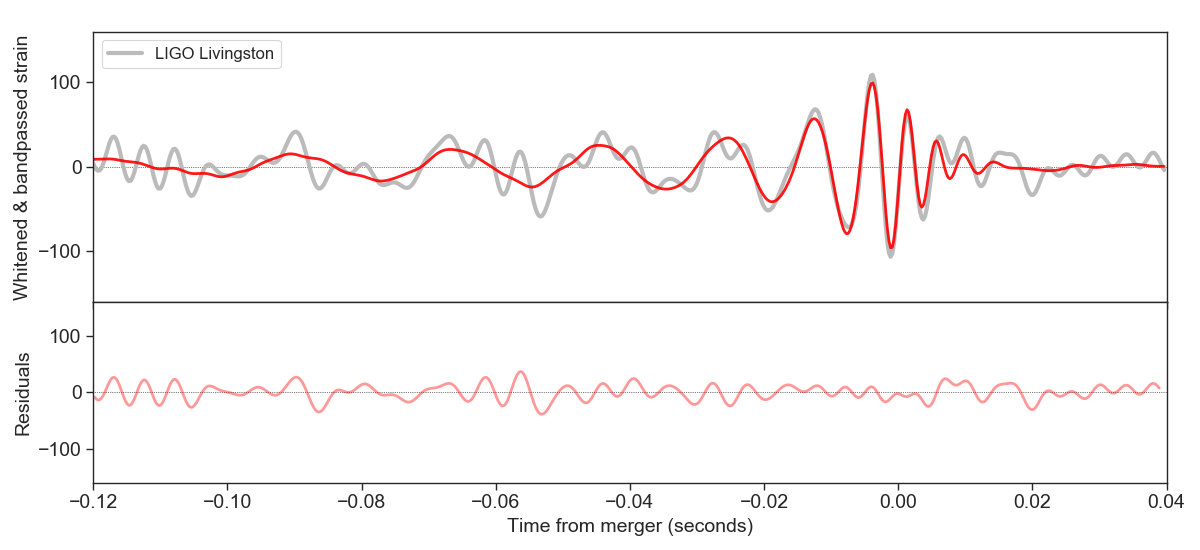} \\
\includegraphics[width=0.62\columnwidth]{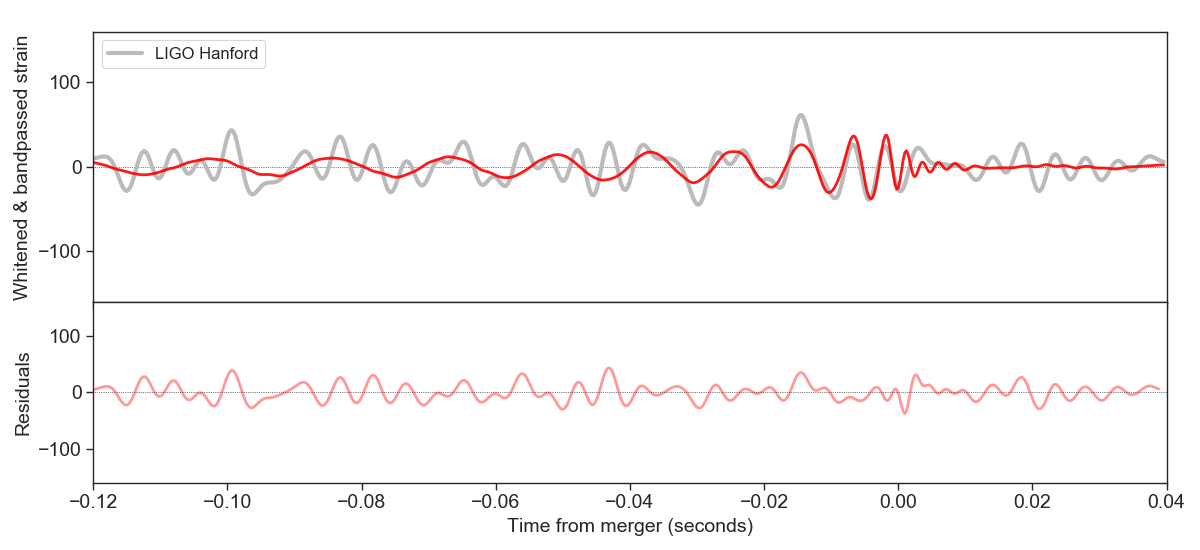}
\includegraphics[width=0.62\columnwidth]{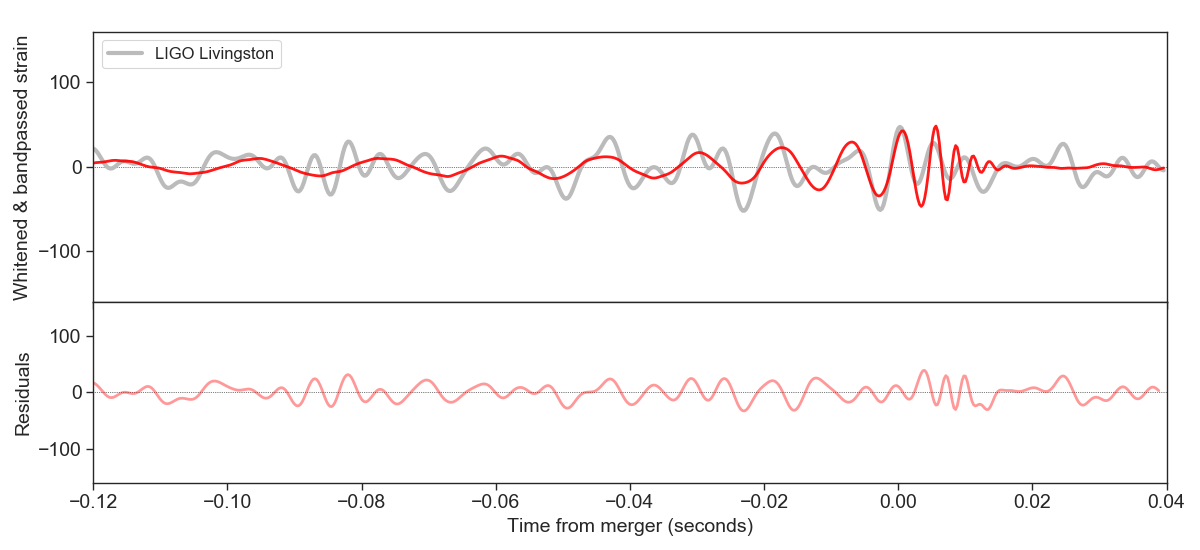}\\
\includegraphics[width=0.62\columnwidth]{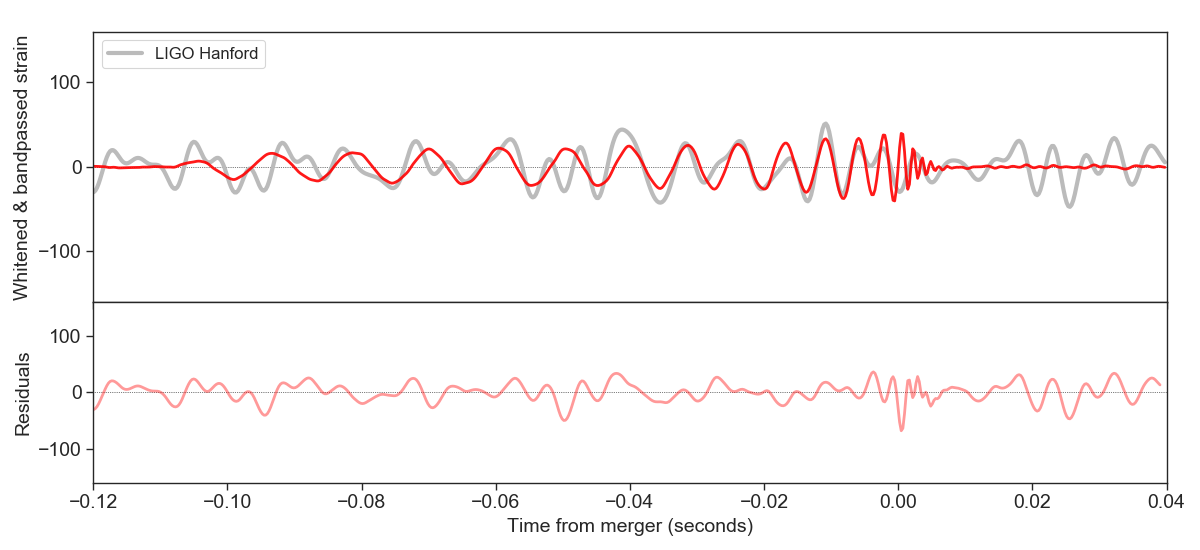}
\includegraphics[width=0.62\columnwidth]{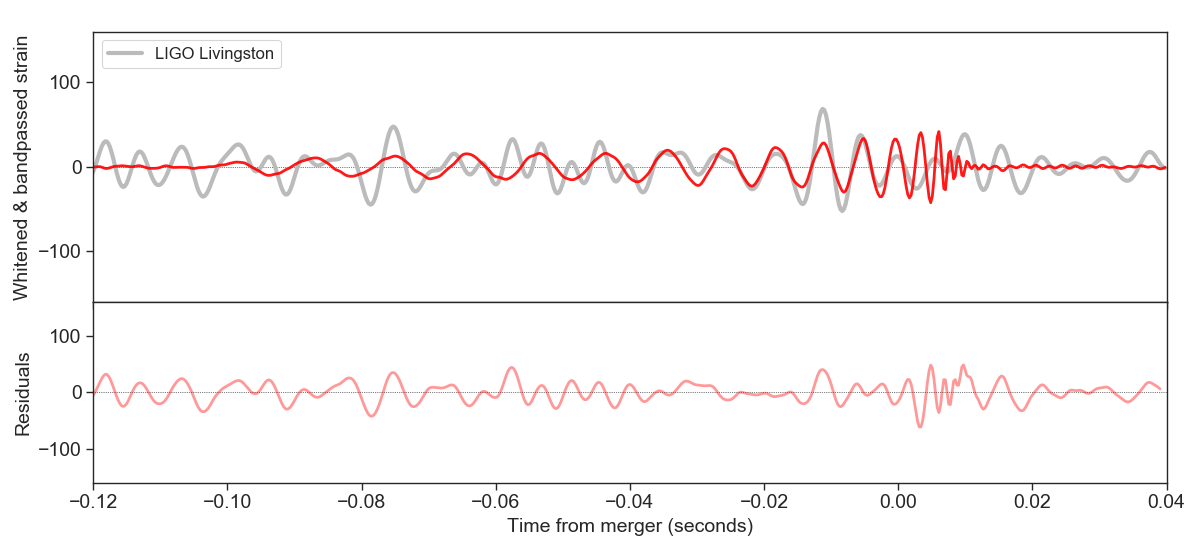}\\
\includegraphics[width=0.62\columnwidth]{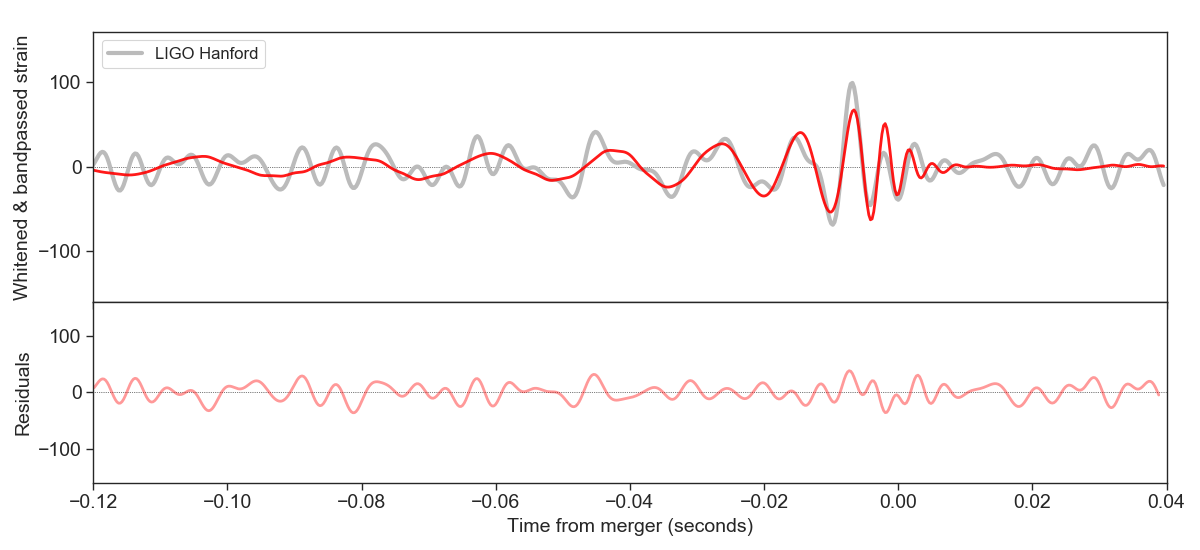}
\includegraphics[width=0.62\columnwidth]{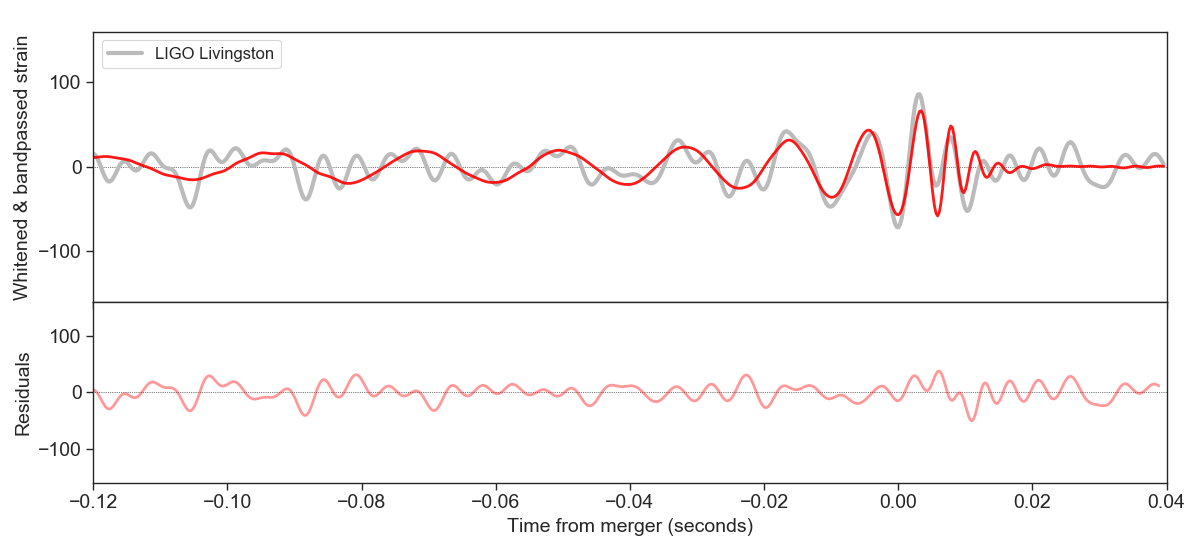}\\
\includegraphics[width=0.62\columnwidth]{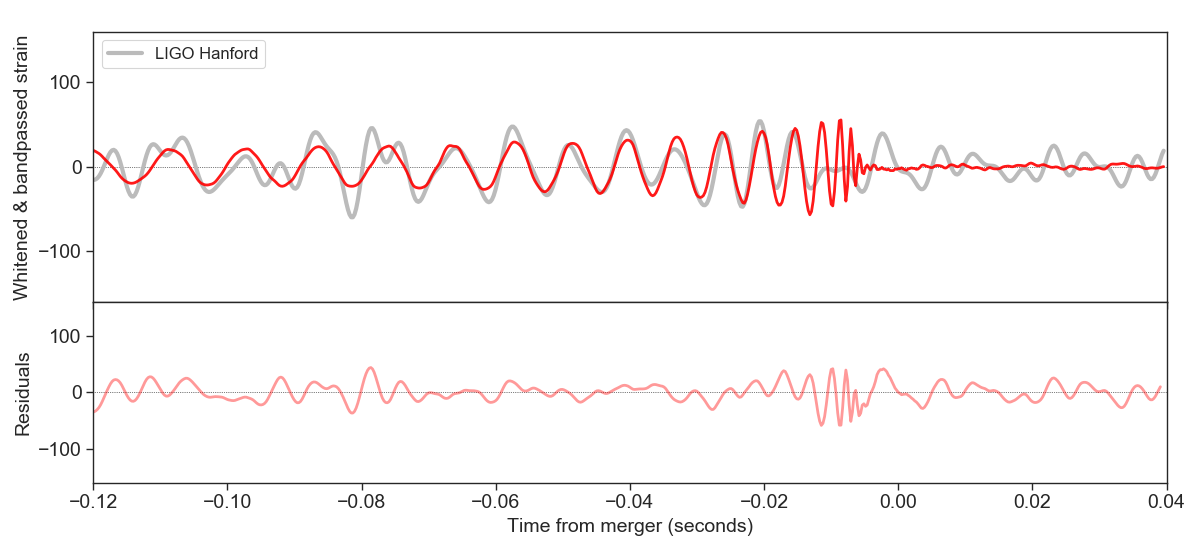}
\includegraphics[width=0.62\columnwidth]{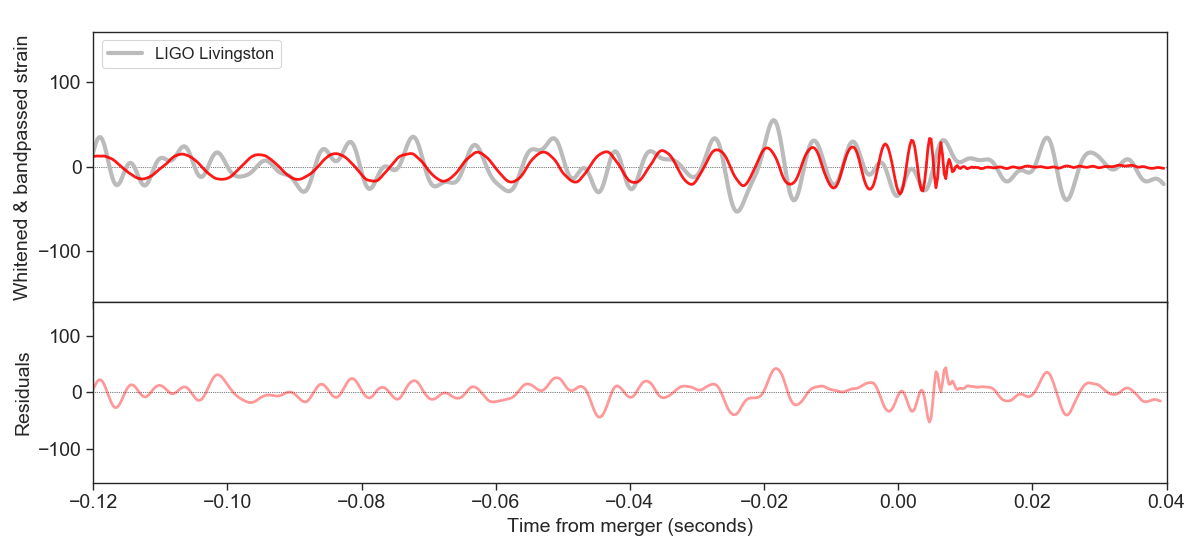}\\
\includegraphics[width=0.62\columnwidth]{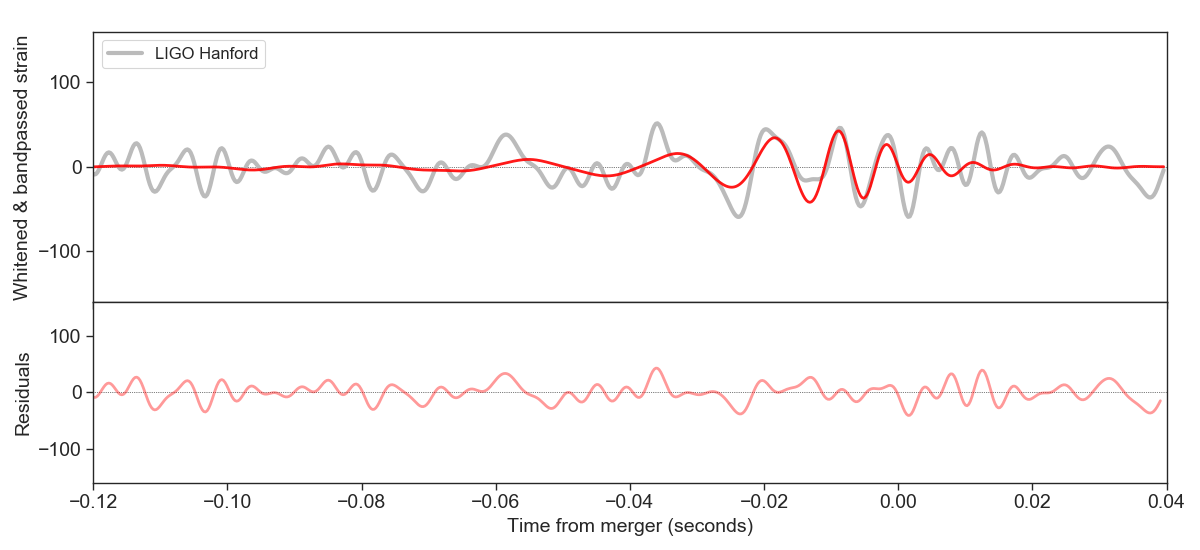}
\includegraphics[width=0.62\columnwidth]{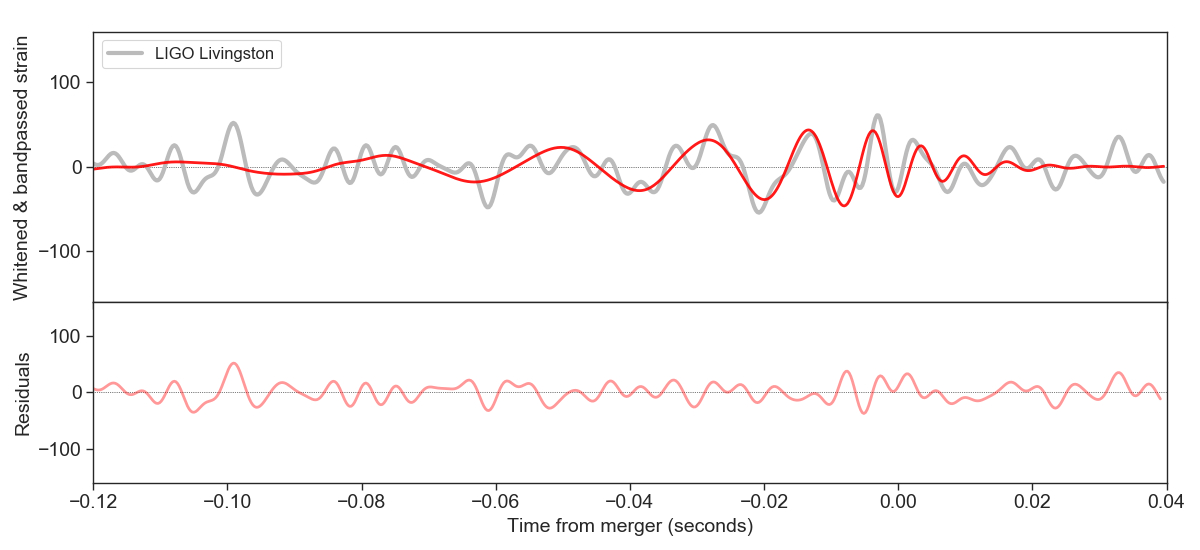}\\
\includegraphics[width=0.62\columnwidth]{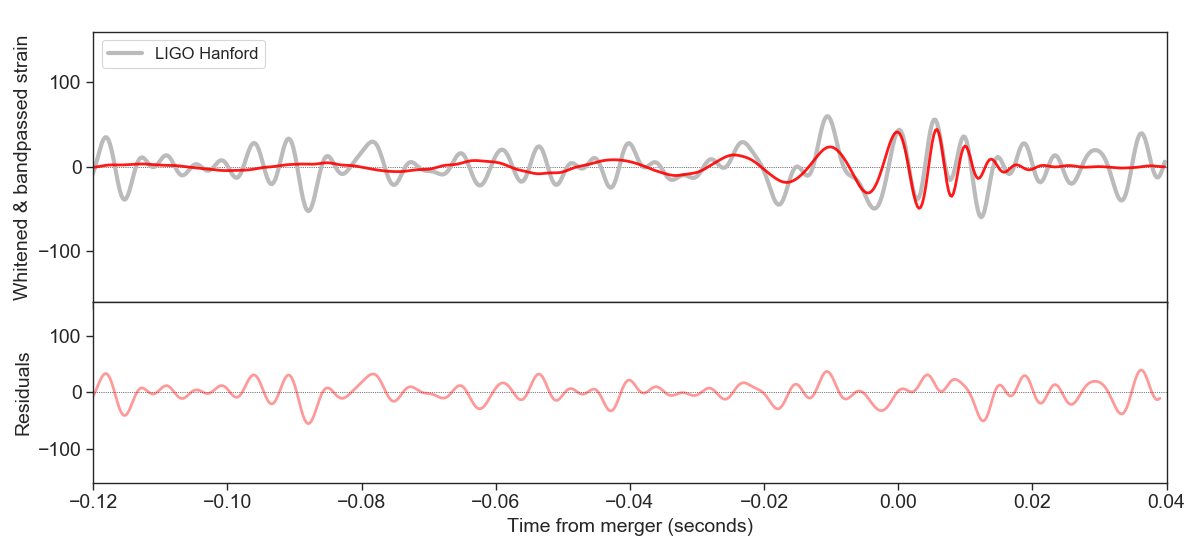}
\includegraphics[width=0.62\columnwidth]{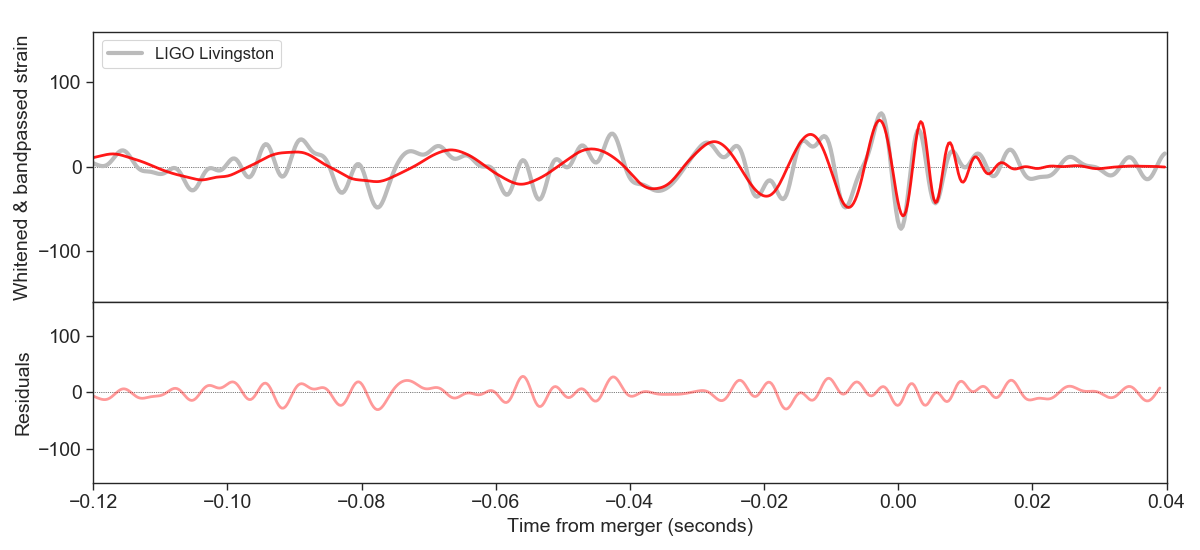}
\includegraphics[width=0.62\columnwidth]{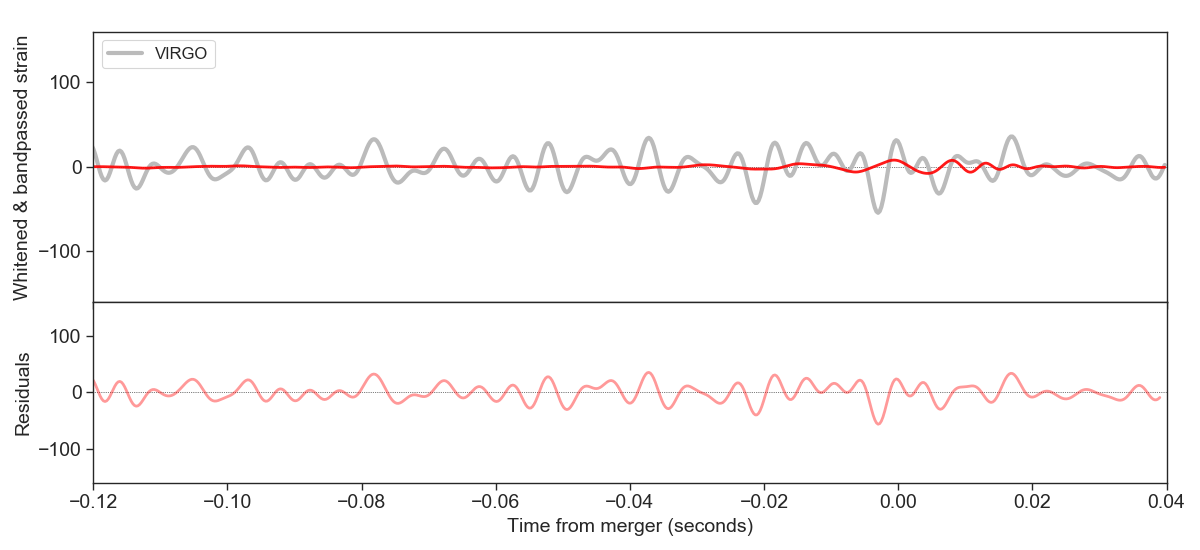}\\
\includegraphics[width=0.62\columnwidth]{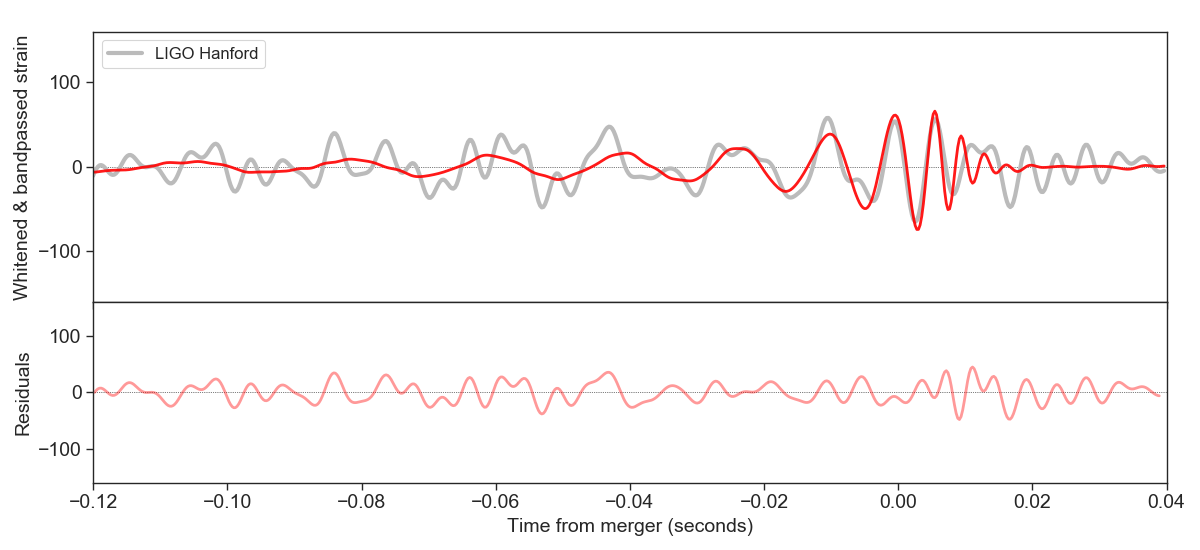}
\includegraphics[width=0.62\columnwidth]{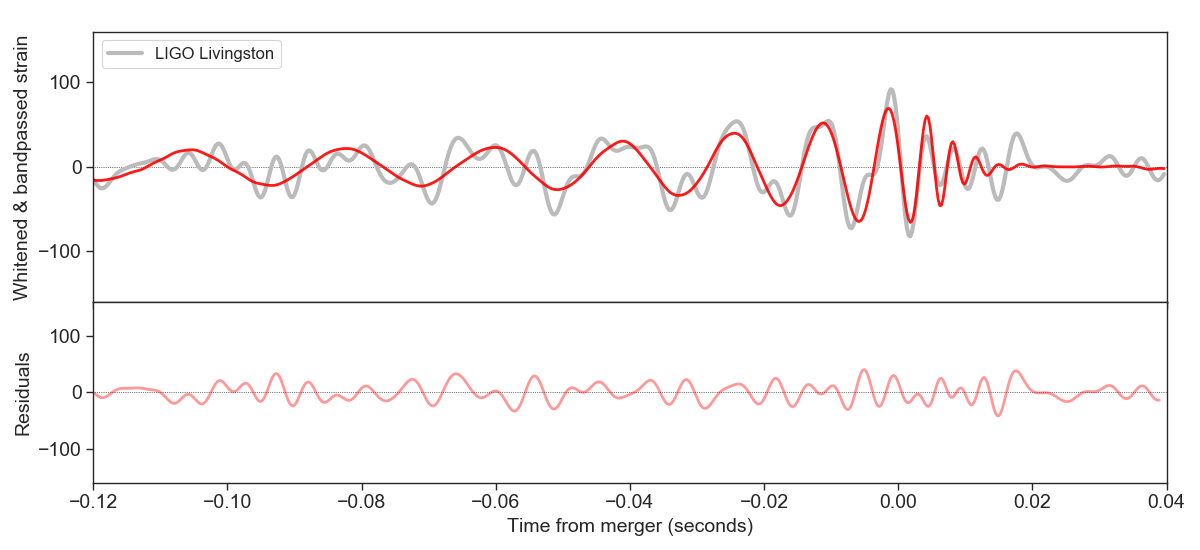}
\includegraphics[width=0.62\columnwidth]{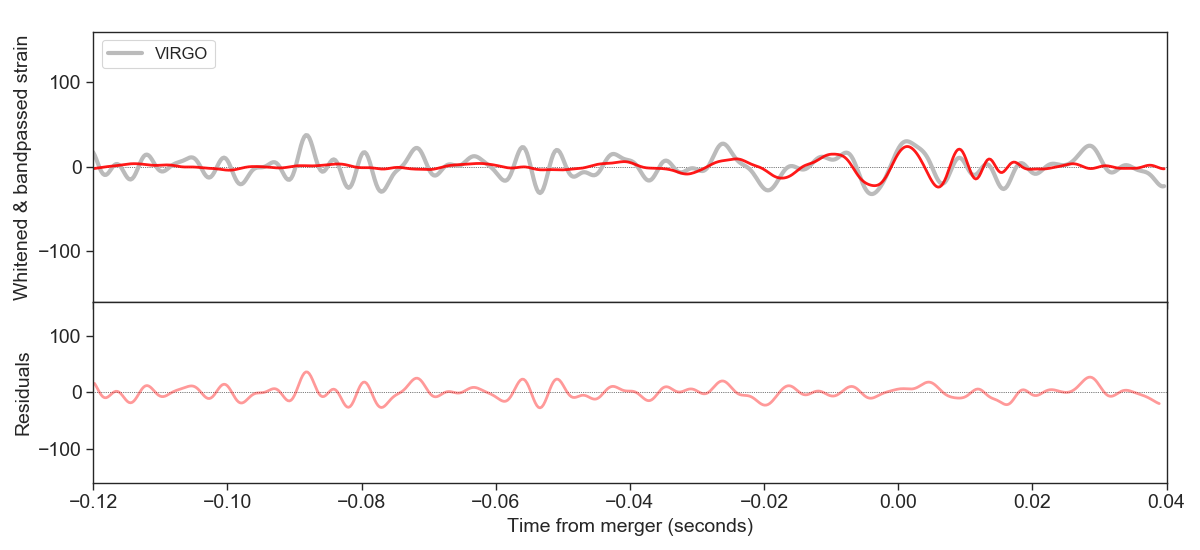}\\
\includegraphics[width=0.62\columnwidth]{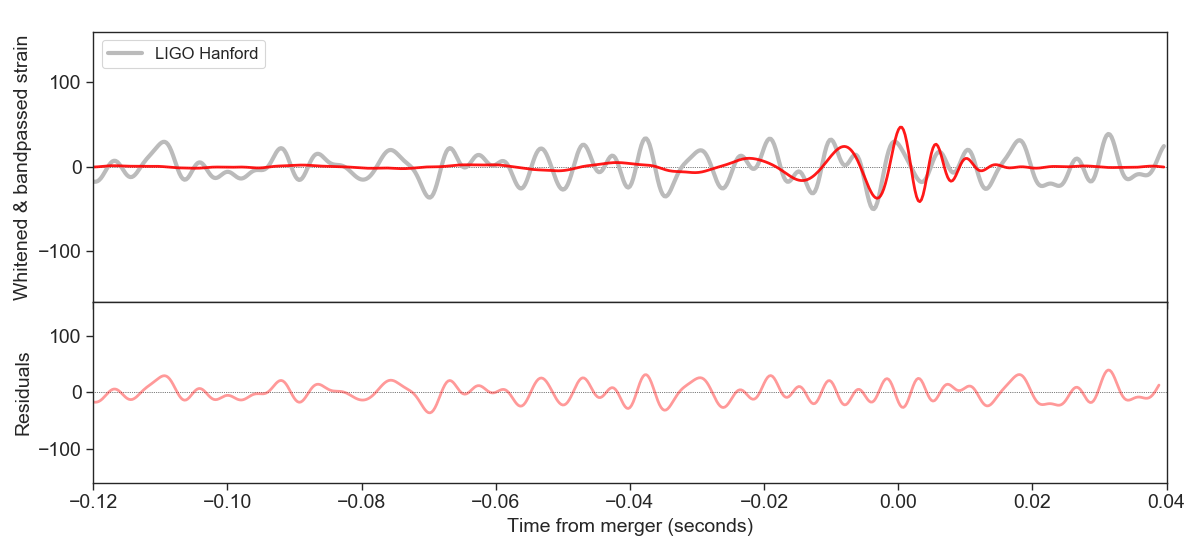}
\includegraphics[width=0.62\columnwidth]{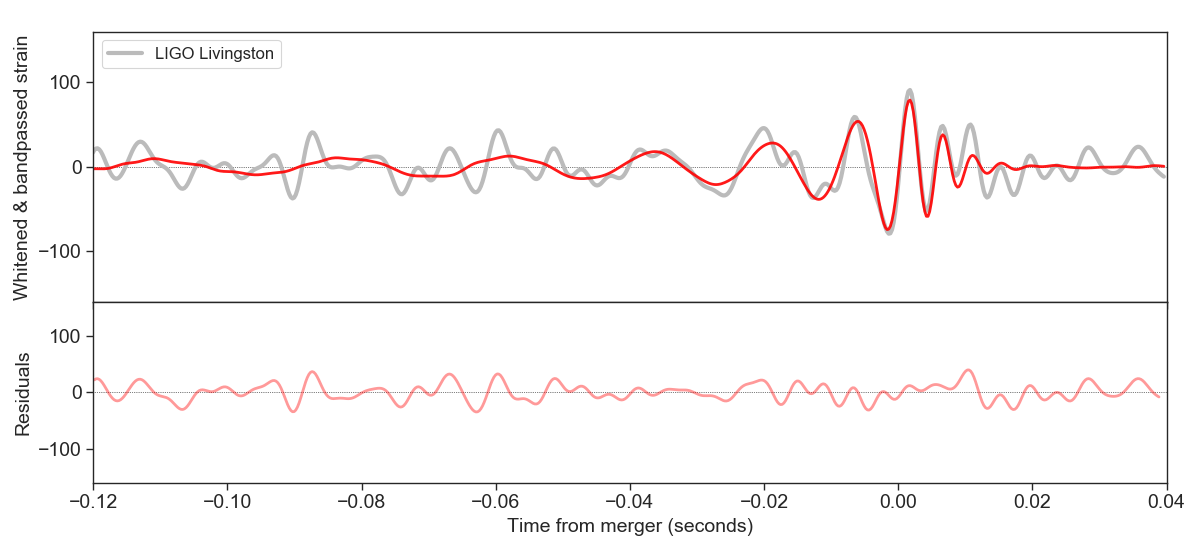}
\includegraphics[width=0.62\columnwidth]{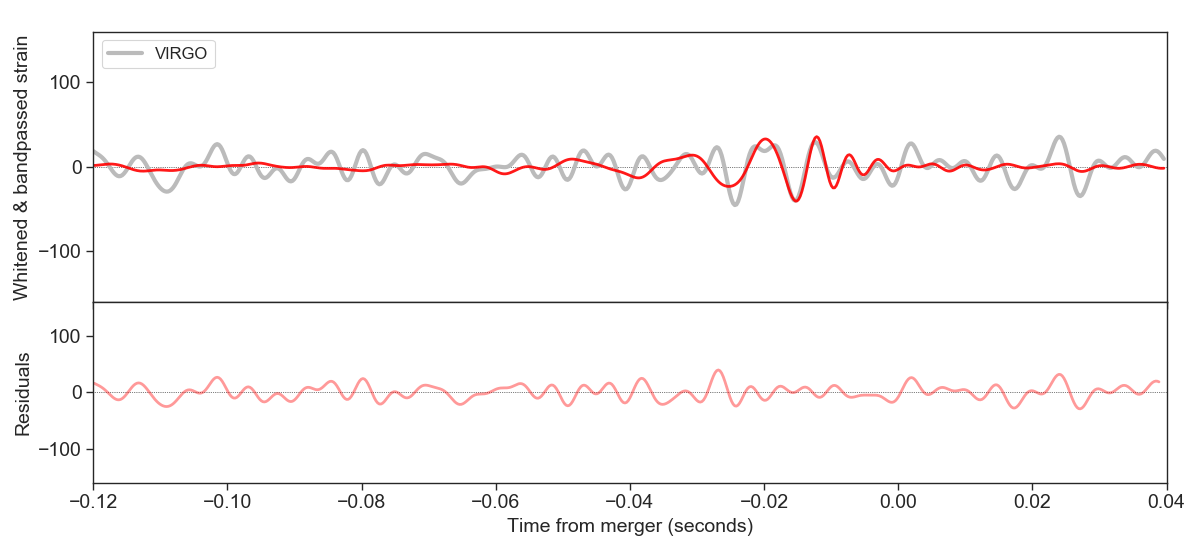}\\
\includegraphics[width=0.62\columnwidth]{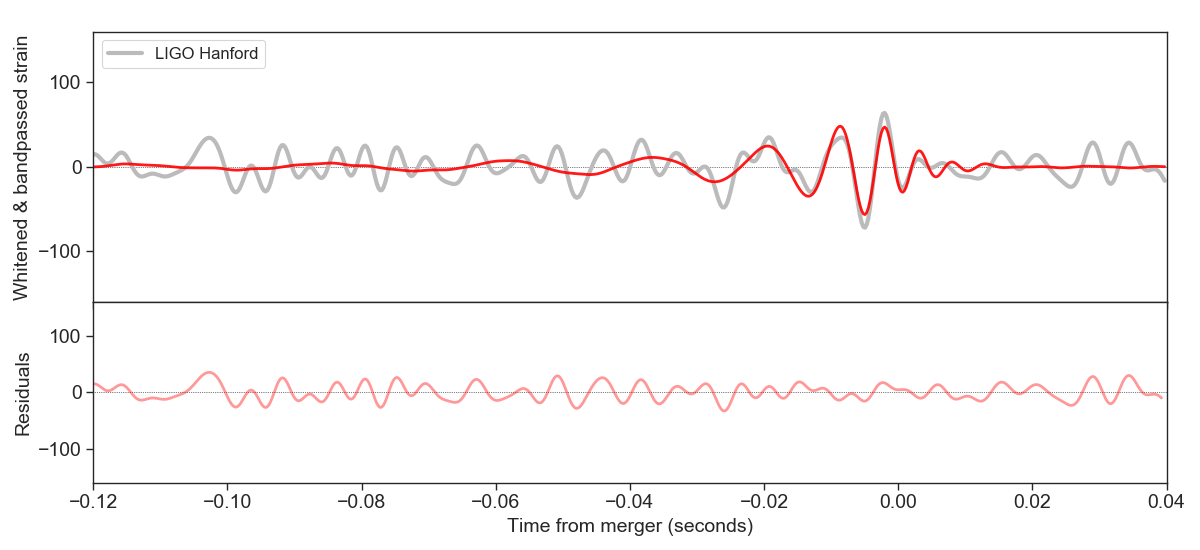}
\includegraphics[width=0.62\columnwidth]{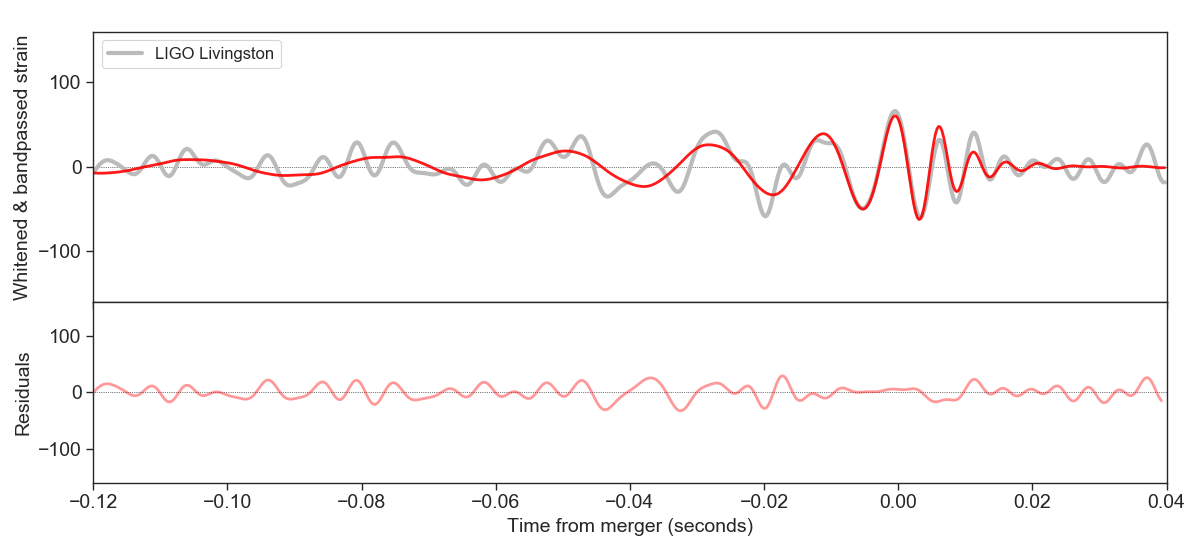}
\caption{\label{fig:reconstruct}For each observed GW signal (rows) and each interferometer (columns), our best point estimate for the GW strain
  (solid red line
  in top panel); the whitened GW data (gray line); and the residual difference between the two (bottom panel). 
Table \ref{tab:extrinsic} provides the specific simulations and extrinsic parameters used to estimate the strain for
each event.
}
\end{figure*}

For each simulation in our catalog, we identify the optimal mass and extrinsic parameters, then generate a point
estimate for the likely
response of each detector.  Table \ref{tab:extrinsic} provides the specific simulations (from the RIT catalog \url{http://ccrg.rit.edu/~RITCatalog})
and extrinsic parameters used to estimate the strain for
each event.
 In Figure \ref{fig:reconstruct}, we compare these point estimates to the whitened GW data in
each interferometer, for all the events in GWTC-1; rows correspond to events, presented in chronological order.
Examining these plots, this subtraction doesn't seem to leave behind a notably significant or correlated residual for
most events.   As expected, for the two low-mass events, where our analysis is suboptimal due to limited simulation
duration, some correlated high-frequency residual does remain near the merger epoch.

\begin{table*}
\caption{Extrinsic parameters used to reconstruct the best fitting numerical relativity
waveform in Fig.~\ref{fig:reconstruct}.
\label{tab:extrinsic}}
\begin{ruledtabular}
\begin{tabular}{llccccccc}
Event    & Sim.      & r.a.   & decl    & $\phi$ & $\theta$ & $\psi$ & $D_L$   & $z$   \\
\hline
GW150914 & RIT:BBH:0160 & 1.9298 & -1.2710 & 4.4278 & 2.9828   & 1.8366 & 571.64  & 0.1188\\
GW151012 & RIT:BBH:0040 & 3.9358 & -0.0134 & 4.6488 & 1.3702   & 2.0121 & 516.54  & 0.1081\\
GW151226 & RIT:BBH:0573 & 4.5194 & -1.1913 & 3.5168 & 1.1708   & 1.8474 & 306.91  & 0.0661\\
GW170104 & RIT:BBH:0162 & 2.1092 &  0.3889 & 3.0760 & 1.1386   & 1.3337 & 586.65  & 0.1217\\
GW170608 & RIT:BBH:0555 & 2.1540 &  1.0300 & 5.7880 & 1.6134   & 0.7532 & 125.37  & 0.0277\\
GW170729 & RIT:BBH:0015 & 5.2211 & -0.8403 & 5.5634 & 0.8868   & 2.3039 & 2468.73 & 0.4345\\
GW170809 & RIT:BBH:0204 & 0.2594 & -0.5293 & 5.5287 & 2.6277   & 2.2101 & 1038.26 & 0.2047\\
GW170814 & RIT:BBH:0661 & 0.8068 & -0.8197 & 1.1063 & 0.2078   & 1.7285 & 635.37  & 0.1310\\
GW170818 & RIT:BBH:0664 & 5.9577 &  0.3449 & 0.7399 & 2.0777   & 1.3192 & 562.29  & 0.1170\\
GW170823 & RIT:BBH:0017 & 4.0601 & -0.1352 & 1.1855 & 2.4861   & 3.0995 & 1653.22 & 0.3084\\
\end{tabular}
\end{ruledtabular}
\end{table*}

\subsection{Numerical Waveforms accuracy}\label{sec:convergence}

In addition to the numerical convergence studies performed for our code,
we have evaluated the impact our standard numerical resolutions have
on the accuracy of our simulations for the first gravitational waves
event GW150914 in \cite{Lovelace:2016uwp},
where we performed comparative convergence studies of RIT waveforms
with the completely independent numerical approach to solve the binary
black hole problem by the SXS group. 
We have also performed those comparative studies for the first O2 event,
GW170104 in \cite{Healy:2017abq}.
Both studies display convergence to each other's approach with increasing
numerical resolution and display that the lower resolutions used performs
an excellent match to the signals.

\begin{table*}
  \caption{Variation of the maximum of $\ln{\cal L}$ with the numerical
resolution of selected events. In the case of GW170823 an extrapolation
of the result to infinite resolution (n$\to\infty$) and order of convergence
of $\ln{\cal L}$.
\label{tab:convergence}}
\begin{ruledtabular}
\begin{tabular}{lcccccc}
  Event & Simulation & low & medium & high & n$\to\infty$ & order \\
\hline
GW170729 & RIT:BBH:0166& 36.78& 36.58& 36.59 & --& -- \\
GW170809 & RIT:BBH:0198& 58.46& 58.47& 58.44 & --& -- \\
GW170814 & RIT:BBH:0062&148.18&148.22&148.23 & --& --\\
GW170823 & RIT:BBH:0113& 57.32& 57.77& 58.03 & 58.66& 2.25\\
\end{tabular}
\end{ruledtabular}
\end{table*}

Here we extend those analysis to several additional O2 events:
GW170729, GW170809, GW170814, GW170823. We evaluate the maximum
$\ln{\cal L}$ for a set of three numerical waveforms 
(see details of the simulations in \cite{Healy:2017psd,Healy:2019jyf})
with increasing
resolution (typically those labeled with n100, n120, n140;
see \url{http://ccrg.rit.edu/~RITCatalog}), and compare
the results in Table~\ref{tab:convergence}. Those show that
there is very little differences between low, medium, and high resolution
runs regarding the evaluation of the likelihood and that even in
the case of GW170823, where we find enough differences to extrapolate
to infinite resolution, the extrapolated value lies within 1-sigma
from the lowest resolution.

\subsection{Null test}
\begin{figure}
  \includegraphics[angle=0,width=1.0\columnwidth]{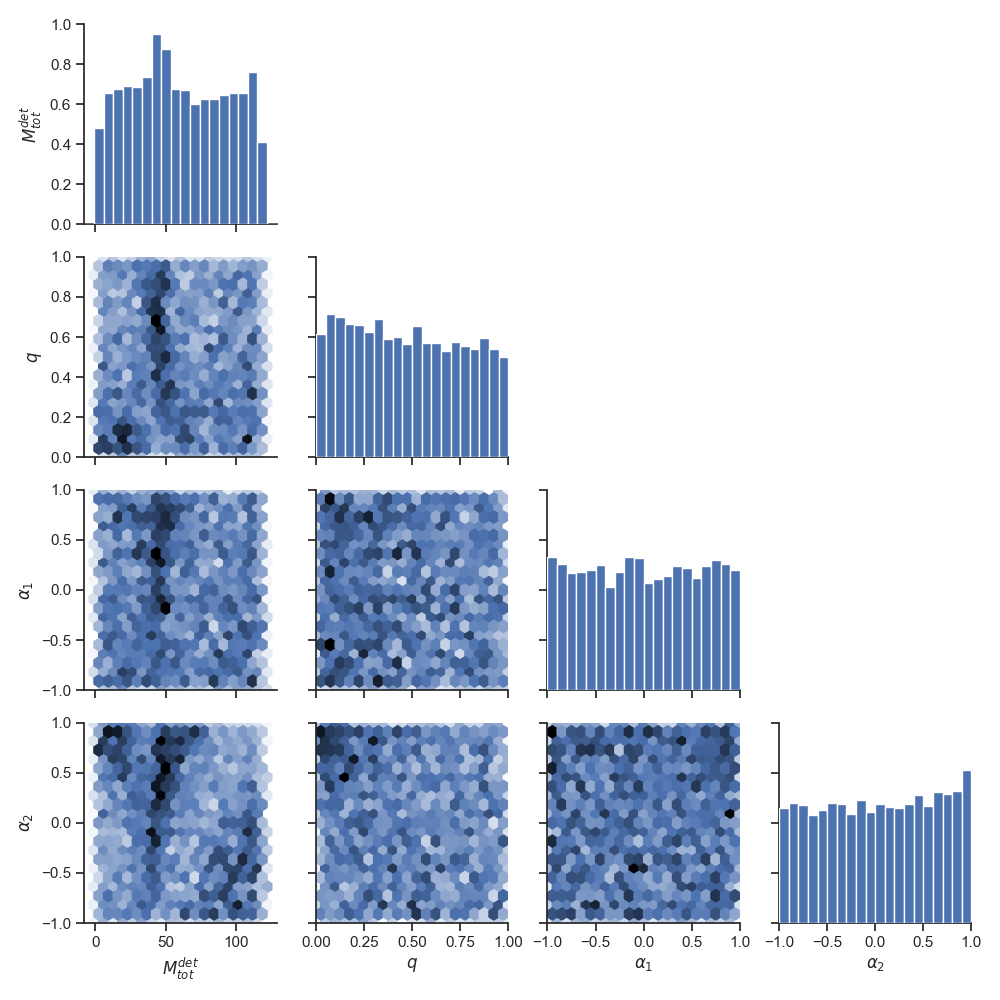}
  \caption{Estimation of the (aligned) binary parameters $(M_{Total},q,\chi_1,\chi_2)$ for Null Test using the 477 nonprecessing simulations.
      \label{fig:NullTestaligned_4d_intrinsic}}
\end{figure}

The specific finite set of simulation parameters in principle impacts the posterior distributions we recover.  To
assess this effect, we 
  study a pure-noise signal as a control case.
Figure \ref{fig:NullTestaligned_4d_intrinsic} shows the parameters estimated
for the BBH in the detector frame  using 477 aligned spins simulations. 
The recovered posterior is consistent with our adopted prior, modulo small modulations principally in total binary mass.



\section{Conclusions and Discussion}\label{sec:Discussion}

The breakthroughs~\cite{Pretorius:2005gq,Campanelli:2005dd,Baker:2005vv}
in numerical relativity were instrumental in identifying the first
detection of gravitational waves \cite{TheLIGOScientific:2016wfe} with
the merger of two black holes.  The comparison of
different approaches to solve the binary black hole problem has
produced an excellent agreement for the GW150914
\cite{Lovelace:2016uwp} and GW170104~\cite{Healy:2017abq}, including
higher (up to $\ell=5$) modes.  We have shown in this paper that the
use of numerical relativity waveform catalogs 
(See also Refs.~\cite{Abbott:2016apu,Lange:2017wki,Kumar:2018hml})
allows the application of a consistent
method for parameter estimation (of merging binary black holes) 
of the observed gravitational waves in the observation runs O1/O2. 
This method of direct comparison of the gravitational wave
signals with numerical waveforms does not rely {\it at all} on any 
information from phenomenological models 
\cite{Babak:2016tgq,Hannam:2013oca} (either Phenom or SEOBNR).

It also shows that with the current aligned spin coverage one can
successfully carry out parameter estimations with results,
at least as good as with contemporary phenomenological models
\cite{TheLIGOScientific:2016wfe}, particularly modulo uncertainty in priors. In particular, we included coverage of spins above 0.95 in
magnitude up to mass ratios 2:1. 
Note that we have used a different set of priors, as discussed in Sec.~\ref{sub:jsd}
but those differences display a measure of the uncertainties expected
in the parameter estimations.

New forthcoming simulations  (for instance targeted to
followup any new detection or catalog expansions) will contribute to
improve the binary parameter coverage, thus reducing the interpolation
error. The next step will be to reduce the extrapolation error
at very high spins by adding more simulations with spin magnitudes 
above 0.95. In addition, new simulations will extend the family displayed
in Fig.~8 of Ref.~\cite{Healy:2020vre} with single spinning binaries
to smaller mass ratios, i.e. $q<1/5$. 
Coverage for
low total binary masses (below $20M_\odot$), in turn, would require longer
full numerical simulations or hybridization of the current NR waveforms
with post-Newtonian waveforms \cite{Sadiq:2020hti}.

The next area of development for the numerical relativity waveform
catalogs is the coverage of precessing binaries. Those require
expansion of the parameter space to seven dimensions (assuming
negligible eccentricity), and is being carried out in a hierarchical
approach by first neglecting the effects of the spin of the secondary black
holes, which is a good assumption for small mass ratios. This approach
has proven also successful when applied to all O1/O2 events.  
It required an homogeneous set of
simulations since the differences in $\ln {\cal L}$ are subtle.
In a second stage, a follow up of the first determined spin orientations
can be performed with a two spin search.
Another line of extension of the use of NR waveforms is its use in
searches of GW (in addition to that of parameter estimation). A first
implementation of the nonspinning waveforms 
(using for instance the simulations reported in \cite{Healy:2017mvh})
would produce a prototype of this search analysis.
The success of the current systematic study can be carried out to the next
LIGO-Virgo observational run O3ab, and in particular to focus
on studies of interesting gravitational waves events and perform
targeted simulations to extract independent parameter estimations.


\begin{acknowledgments}

The authors also gratefully acknowledge the National Science Foundation
(NSF) for financial support from Grants No.\ PHY-1607520,
No.\ PHY-1707946, No.\ ACI-1550436, No.\ AST-1516150,
No.\ ACI-1516125, No.\ PHY-1726215, No.\ PHY-1707965, and No.\ PHY-2012057.  This work used the Extreme
Science and Engineering Discovery Environment (XSEDE) [allocation
  TG-PHY060027N], which is supported by NSF grant No. ACI-1548562.
Computational resources were also provided by the NewHorizons, BlueSky
Clusters, and Green Prairies at the Rochester Institute of Technology,
which were supported by NSF grants No.\ PHY-0722703, No.\ DMS-0820923,
No.\ AST-1028087, No.\ PHY-1229173, and No.\ PHY-1726215. 
The authors are grateful for computational resources provided by the Leonard E Parker Center for Gravitation, Cosmology and Astrophysics at the University of Wisconsin-Milwaukee and LIGO Laboratories at CIT and LLO supported by National Science Foundation Grants PHY-1626190, PHY-1700765, PHY-0757058 and PHY-0823459.
\end{acknowledgments}


\bibliographystyle{apsrev4-1}
\bibliography{../../Bibtex/references,paperexport}

\begin{thebibliography}{69}%
\makeatletter
\providecommand \@ifxundefined [1]{%
 \@ifx{#1\undefined}
}%
\providecommand \@ifnum [1]{%
 \ifnum #1\expandafter \@firstoftwo
 \else \expandafter \@secondoftwo
 \fi
}%
\providecommand \@ifx [1]{%
 \ifx #1\expandafter \@firstoftwo
 \else \expandafter \@secondoftwo
 \fi
}%
\providecommand \natexlab [1]{#1}%
\providecommand \enquote  [1]{``#1''}%
\providecommand \bibnamefont  [1]{#1}%
\providecommand \bibfnamefont [1]{#1}%
\providecommand \citenamefont [1]{#1}%
\providecommand \href@noop [0]{\@secondoftwo}%
\providecommand \href [0]{\begingroup \@sanitize@url \@href}%
\providecommand \@href[1]{\@@startlink{#1}\@@href}%
\providecommand \@@href[1]{\endgroup#1\@@endlink}%
\providecommand \@sanitize@url [0]{\catcode `\\12\catcode `\$12\catcode
  `\&12\catcode `\#12\catcode `\^12\catcode `\_12\catcode `\%12\relax}%
\providecommand \@@startlink[1]{}%
\providecommand \@@endlink[0]{}%
\providecommand \url  [0]{\begingroup\@sanitize@url \@url }%
\providecommand \@url [1]{\endgroup\@href {#1}{\urlprefix }}%
\providecommand \urlprefix  [0]{URL }%
\providecommand \Eprint [0]{\href }%
\providecommand \doibase [0]{http://dx.doi.org/}%
\providecommand \selectlanguage [0]{\@gobble}%
\providecommand \bibinfo  [0]{\@secondoftwo}%
\providecommand \bibfield  [0]{\@secondoftwo}%
\providecommand \translation [1]{[#1]}%
\providecommand \BibitemOpen [0]{}%
\providecommand \bibitemStop [0]{}%
\providecommand \bibitemNoStop [0]{.\EOS\space}%
\providecommand \EOS [0]{\spacefactor3000\relax}%
\providecommand \BibitemShut  [1]{\csname bibitem#1\endcsname}%
\let\auto@bib@innerbib\@empty
\bibitem [{\citenamefont {Venumadhav}\ \emph {et~al.}(2020)\citenamefont
  {Venumadhav}, \citenamefont {Zackay}, \citenamefont {Roulet}, \citenamefont
  {Dai},\ and\ \citenamefont {Zaldarriaga}}]{Venumadhav:2019lyq}%
  \BibitemOpen
  \bibfield  {author} {\bibinfo {author} {\bibfnamefont {T.}~\bibnamefont
  {Venumadhav}}, \bibinfo {author} {\bibfnamefont {B.}~\bibnamefont {Zackay}},
  \bibinfo {author} {\bibfnamefont {J.}~\bibnamefont {Roulet}}, \bibinfo
  {author} {\bibfnamefont {L.}~\bibnamefont {Dai}}, \ and\ \bibinfo {author}
  {\bibfnamefont {M.}~\bibnamefont {Zaldarriaga}},\ }\href {\doibase
  10.1103/PhysRevD.101.083030} {\bibfield  {journal} {\bibinfo  {journal}
  {Phys. Rev. D}\ }\textbf {\bibinfo {volume} {101}},\ \bibinfo {pages}
  {083030} (\bibinfo {year} {2020})},\ \Eprint
  {http://arxiv.org/abs/1904.07214} {arXiv:1904.07214 [astro-ph.HE]}
  \BibitemShut {NoStop}%
\bibitem [{\citenamefont {{LIGO Scientific Collaboration}}\ \emph
  {et~al.}(2015)\citenamefont {{LIGO Scientific Collaboration}}, \citenamefont
  {{Aasi}}, \citenamefont {{Abbott}}, \citenamefont {{Abbott}}, \citenamefont
  {{Abbott}}, \citenamefont {{Abernathy}}, \citenamefont {{Ackley}},
  \citenamefont {{Adams}}, \citenamefont {{Adams}}, \citenamefont {{Addesso}},\
  and\ \citenamefont {et~al.}}]{2015CQGra..32g4001L}%
  \BibitemOpen
  \bibfield  {author} {\bibinfo {author} {\bibnamefont {{LIGO Scientific
  Collaboration}}}, \bibinfo {author} {\bibfnamefont {J.}~\bibnamefont
  {{Aasi}}}, \bibinfo {author} {\bibfnamefont {B.~P.}\ \bibnamefont
  {{Abbott}}}, \bibinfo {author} {\bibfnamefont {R.}~\bibnamefont {{Abbott}}},
  \bibinfo {author} {\bibfnamefont {T.}~\bibnamefont {{Abbott}}}, \bibinfo
  {author} {\bibfnamefont {M.~R.}\ \bibnamefont {{Abernathy}}}, \bibinfo
  {author} {\bibfnamefont {K.}~\bibnamefont {{Ackley}}}, \bibinfo {author}
  {\bibfnamefont {C.}~\bibnamefont {{Adams}}}, \bibinfo {author} {\bibfnamefont
  {T.}~\bibnamefont {{Adams}}}, \bibinfo {author} {\bibfnamefont
  {P.}~\bibnamefont {{Addesso}}}, \ and\ \bibinfo {author} {\bibnamefont
  {et~al.}},\ }\href {\doibase 10.1088/0264-9381/32/7/074001} {\bibfield
  {journal} {\bibinfo  {journal} {Classical and Quantum Gravity}\ }\textbf
  {\bibinfo {volume} {32}},\ \bibinfo {eid} {074001} (\bibinfo {year}
  {2015})},\ \Eprint {http://arxiv.org/abs/1411.4547} {arXiv:1411.4547 [gr-qc]}
  \BibitemShut {NoStop}%
\bibitem [{\citenamefont {Acernese}\ \emph {et~al.}(2015)\citenamefont
  {Acernese} \emph {et~al.}}]{TheVirgo:2014hva}%
  \BibitemOpen
  \bibfield  {author} {\bibinfo {author} {\bibfnamefont {F.}~\bibnamefont
  {Acernese}} \emph {et~al.} (\bibinfo {collaboration} {VIRGO}),\ }\href
  {\doibase 10.1088/0264-9381/32/2/024001} {\bibfield  {journal} {\bibinfo
  {journal} {Class. Quant. Grav.}\ }\textbf {\bibinfo {volume} {32}},\ \bibinfo
  {pages} {024001} (\bibinfo {year} {2015})},\ \Eprint
  {http://arxiv.org/abs/1408.3978} {arXiv:1408.3978 [gr-qc]} \BibitemShut
  {NoStop}%
\bibitem [{\citenamefont {{The LIGO Scientific Collaboration and the Virgo
  Collaboration}}(2016)}]{DiscoveryPaper}%
  \BibitemOpen
  \bibfield  {author} {\bibinfo {author} {\bibnamefont {{The LIGO Scientific
  Collaboration and the Virgo Collaboration}}},\ }\href {\doibase
  http://link.aps.org/doi/10.1103/PhysRevLett.116.061102} {\bibfield  {journal}
  {\bibinfo  {journal} {\prl}\ }\textbf {\bibinfo {volume} {16}},\ \bibinfo
  {pages} {061102} (\bibinfo {year} {2016})}\BibitemShut {NoStop}%
\bibitem [{\citenamefont {{Abbott et al.\ (The LIGO Scientific Collaboration
  and the Virgo Collaboration)}}(2016)}]{LIGO-O1-BBH}%
  \BibitemOpen
  \bibfield  {author} {\bibinfo {author} {\bibfnamefont {B.}~\bibnamefont
  {{Abbott et al.\ (The LIGO Scientific Collaboration and the Virgo
  Collaboration)}}},\ }\href {\doibase 10.1103/PhysRevX.6.041015} {\bibfield
  {journal} {\bibinfo  {journal} {\prx}\ }\textbf {\bibinfo {volume} {6}},\
  \bibinfo {pages} {041015} (\bibinfo {year} {2016})},\ \Eprint
  {http://arxiv.org/abs/1606.04856} {arXiv:1606.04856 [gr-qc]} \BibitemShut
  {NoStop}%
\bibitem [{\citenamefont {{Abbott}}\ \emph {et~al.}(2017)\citenamefont
  {{Abbott}}, \citenamefont {{Abbott}}, \citenamefont {{Abbott}}, \citenamefont
  {{Acernese}}, \citenamefont {{Ackley}}, \citenamefont {{Adams}},
  \citenamefont {{Adams}}, \citenamefont {{Addesso}}, \citenamefont
  {{Adhikari}}, \citenamefont {{Adya}},\ and\ \citenamefont
  {et~al.}}]{2017PhRvL.118v1101A}%
  \BibitemOpen
  \bibfield  {author} {\bibinfo {author} {\bibfnamefont {B.~P.}\ \bibnamefont
  {{Abbott}}}, \bibinfo {author} {\bibfnamefont {R.}~\bibnamefont {{Abbott}}},
  \bibinfo {author} {\bibfnamefont {T.~D.}\ \bibnamefont {{Abbott}}}, \bibinfo
  {author} {\bibfnamefont {F.}~\bibnamefont {{Acernese}}}, \bibinfo {author}
  {\bibfnamefont {K.}~\bibnamefont {{Ackley}}}, \bibinfo {author}
  {\bibfnamefont {C.}~\bibnamefont {{Adams}}}, \bibinfo {author} {\bibfnamefont
  {T.}~\bibnamefont {{Adams}}}, \bibinfo {author} {\bibfnamefont
  {P.}~\bibnamefont {{Addesso}}}, \bibinfo {author} {\bibfnamefont {R.~X.}\
  \bibnamefont {{Adhikari}}}, \bibinfo {author} {\bibfnamefont {V.~B.}\
  \bibnamefont {{Adya}}}, \ and\ \bibinfo {author} {\bibnamefont {et~al.}},\
  }\href {\doibase 10.1103/PhysRevLett.118.221101} {\bibfield  {journal}
  {\bibinfo  {journal} {Physical Review Letters}\ }\textbf {\bibinfo {volume}
  {118}},\ \bibinfo {eid} {221101} (\bibinfo {year} {2017})},\ \Eprint
  {http://arxiv.org/abs/1706.01812} {arXiv:1706.01812 [gr-qc]} \BibitemShut
  {NoStop}%
\bibitem [{\citenamefont {{The LIGO Scientific Collaboration}}\ \emph
  {et~al.}(2017{\natexlab{a}})\citenamefont {{The LIGO Scientific
  Collaboration}}, \citenamefont {{the Virgo Collaboration}}, \citenamefont
  {{Abbott}}, \citenamefont {{Abbott}}, \citenamefont {{Abbott}}, \citenamefont
  {{Acernese}}, \citenamefont {{Ackley}}, \citenamefont {{Adams}},
  \citenamefont {{Adams}}, \citenamefont {{Addesso}}, \citenamefont
  {{Adhikari}}, \citenamefont {{Adya}},\ and\ \citenamefont
  {et~al.}}]{LIGO-GW170814}%
  \BibitemOpen
  \bibfield  {author} {\bibinfo {author} {\bibnamefont {{The LIGO Scientific
  Collaboration}}}, \bibinfo {author} {\bibnamefont {{the Virgo
  Collaboration}}}, \bibinfo {author} {\bibfnamefont {B.~P.}\ \bibnamefont
  {{Abbott}}}, \bibinfo {author} {\bibfnamefont {R.}~\bibnamefont {{Abbott}}},
  \bibinfo {author} {\bibfnamefont {T.~D.}\ \bibnamefont {{Abbott}}}, \bibinfo
  {author} {\bibfnamefont {F.}~\bibnamefont {{Acernese}}}, \bibinfo {author}
  {\bibfnamefont {K.}~\bibnamefont {{Ackley}}}, \bibinfo {author}
  {\bibfnamefont {C.}~\bibnamefont {{Adams}}}, \bibinfo {author} {\bibfnamefont
  {T.}~\bibnamefont {{Adams}}}, \bibinfo {author} {\bibfnamefont
  {P.}~\bibnamefont {{Addesso}}}, \bibinfo {author} {\bibfnamefont {R.~X.}\
  \bibnamefont {{Adhikari}}}, \bibinfo {author} {\bibfnamefont {V.~B.}\
  \bibnamefont {{Adya}}}, \ and\ \bibinfo {author} {\bibnamefont {et~al.}},\
  }\href {\doibase 10.1103/PhysRevLett.119.141101} {\bibfield  {journal}
  {\bibinfo  {journal} {\prl}\ }\textbf {\bibinfo {volume} {119}},\ \bibinfo
  {eid} {141101} (\bibinfo {year} {2017}{\natexlab{a}})},\ \Eprint
  {http://arxiv.org/abs/1709.09660} {arXiv:1709.09660 [gr-qc]} \BibitemShut
  {NoStop}%
\bibitem [{\citenamefont {{The LIGO Scientific Collaboration}}\ \emph
  {et~al.}(2017{\natexlab{b}})\citenamefont {{The LIGO Scientific
  Collaboration}}, \citenamefont {{the Virgo Collaboration}}, \citenamefont
  {{Abbott}}, \citenamefont {{Abbott}}, \citenamefont {{Abbott}}, \citenamefont
  {{Acernese}}, \citenamefont {{Ackley}}, \citenamefont {{Adams}},
  \citenamefont {{Adams}}, \citenamefont {{Addesso}}, \citenamefont
  {{Adhikari}}, \citenamefont {{Adya}},\ and\ \citenamefont
  {et~al.}}]{LIGO-GW170608}%
  \BibitemOpen
  \bibfield  {author} {\bibinfo {author} {\bibnamefont {{The LIGO Scientific
  Collaboration}}}, \bibinfo {author} {\bibnamefont {{the Virgo
  Collaboration}}}, \bibinfo {author} {\bibfnamefont {B.~P.}\ \bibnamefont
  {{Abbott}}}, \bibinfo {author} {\bibfnamefont {R.}~\bibnamefont {{Abbott}}},
  \bibinfo {author} {\bibfnamefont {T.~D.}\ \bibnamefont {{Abbott}}}, \bibinfo
  {author} {\bibfnamefont {F.}~\bibnamefont {{Acernese}}}, \bibinfo {author}
  {\bibfnamefont {K.}~\bibnamefont {{Ackley}}}, \bibinfo {author}
  {\bibfnamefont {C.}~\bibnamefont {{Adams}}}, \bibinfo {author} {\bibfnamefont
  {T.}~\bibnamefont {{Adams}}}, \bibinfo {author} {\bibfnamefont
  {P.}~\bibnamefont {{Addesso}}}, \bibinfo {author} {\bibfnamefont {R.~X.}\
  \bibnamefont {{Adhikari}}}, \bibinfo {author} {\bibfnamefont {V.~B.}\
  \bibnamefont {{Adya}}}, \ and\ \bibinfo {author} {\bibnamefont {et~al.}},\
  }\href {\doibase 10.3847/2041-8213/aa9f0c} {\bibfield  {journal} {\bibinfo
  {journal} {\apjl}\ }\textbf {\bibinfo {volume} {851}},\ \bibinfo {eid} {L35}
  (\bibinfo {year} {2017}{\natexlab{b}})}\BibitemShut {NoStop}%
\bibitem [{\citenamefont {{The LIGO Scientific Collaboration}}\ \emph
  {et~al.}(2017{\natexlab{c}})\citenamefont {{The LIGO Scientific
  Collaboration}}, \citenamefont {{the Virgo Collaboration}}, \citenamefont
  {{Abbott}}, \citenamefont {{Abbott}}, \citenamefont {{Abbott}}, \citenamefont
  {{Acernese}}, \citenamefont {{Ackley}}, \citenamefont {{Adams}},
  \citenamefont {{Adams}}, \citenamefont {{Addesso}},\ and\ \citenamefont
  {et~al.}}]{LIGO-GW170817-bns}%
  \BibitemOpen
  \bibfield  {author} {\bibinfo {author} {\bibnamefont {{The LIGO Scientific
  Collaboration}}}, \bibinfo {author} {\bibnamefont {{the Virgo
  Collaboration}}}, \bibinfo {author} {\bibfnamefont {B.~P.}\ \bibnamefont
  {{Abbott}}}, \bibinfo {author} {\bibfnamefont {R.}~\bibnamefont {{Abbott}}},
  \bibinfo {author} {\bibfnamefont {T.~D.}\ \bibnamefont {{Abbott}}}, \bibinfo
  {author} {\bibfnamefont {F.}~\bibnamefont {{Acernese}}}, \bibinfo {author}
  {\bibfnamefont {K.}~\bibnamefont {{Ackley}}}, \bibinfo {author}
  {\bibfnamefont {C.}~\bibnamefont {{Adams}}}, \bibinfo {author} {\bibfnamefont
  {T.}~\bibnamefont {{Adams}}}, \bibinfo {author} {\bibfnamefont
  {P.}~\bibnamefont {{Addesso}}}, \ and\ \bibinfo {author} {\bibnamefont
  {et~al.}},\ }\href {\doibase 10.1103/PhysRevLett.119.161101} {\bibfield
  {journal} {\bibinfo  {journal} {\prl}\ }\textbf {\bibinfo {volume} {119}},\
  \bibinfo {pages} {161101} (\bibinfo {year} {2017}{\natexlab{c}})}\BibitemShut
  {NoStop}%
\bibitem [{\citenamefont {{The LIGO Scientific Collaboration}}\ \emph
  {et~al.}(2019)\citenamefont {{The LIGO Scientific Collaboration}},
  \citenamefont {{The Virgo Collaboration}}, \citenamefont {{Abbott}},
  \citenamefont {{Abbott}}, \citenamefont {{Abbott}}, \citenamefont
  {{Acernese}}, \citenamefont {{Ackley}}, \citenamefont {{Adams}},
  \citenamefont {{Adams}}, \citenamefont {{Addesso}},\ and\ \citenamefont
  {et~al.}}]{LIGO-O2-Catalog}%
  \BibitemOpen
  \bibfield  {author} {\bibinfo {author} {\bibnamefont {{The LIGO Scientific
  Collaboration}}}, \bibinfo {author} {\bibnamefont {{The Virgo
  Collaboration}}}, \bibinfo {author} {\bibfnamefont {B.~P.}\ \bibnamefont
  {{Abbott}}}, \bibinfo {author} {\bibfnamefont {R.}~\bibnamefont {{Abbott}}},
  \bibinfo {author} {\bibfnamefont {T.~D.}\ \bibnamefont {{Abbott}}}, \bibinfo
  {author} {\bibfnamefont {F.}~\bibnamefont {{Acernese}}}, \bibinfo {author}
  {\bibfnamefont {K.}~\bibnamefont {{Ackley}}}, \bibinfo {author}
  {\bibfnamefont {C.}~\bibnamefont {{Adams}}}, \bibinfo {author} {\bibfnamefont
  {T.}~\bibnamefont {{Adams}}}, \bibinfo {author} {\bibfnamefont
  {P.}~\bibnamefont {{Addesso}}}, \ and\ \bibinfo {author} {\bibnamefont
  {et~al.}},\ }\href {\doibase 10.1103/PhysRevX.9.031040} {\bibfield  {journal}
  {\bibinfo  {journal} {\prx}\ }\textbf {\bibinfo {volume} {9}},\ \bibinfo
  {eid} {031040} (\bibinfo {year} {2019})}\BibitemShut {NoStop}%
\bibitem [{\citenamefont {{Pankow}}\ \emph
  {et~al.}(2015{\natexlab{a}})\citenamefont {{Pankow}}, \citenamefont
  {{Brady}}, \citenamefont {{Ochsner}},\ and\ \citenamefont
  {{O'Shaughnessy}}}]{gwastro-PE-AlternativeArchitectures}%
  \BibitemOpen
  \bibfield  {author} {\bibinfo {author} {\bibfnamefont {C.}~\bibnamefont
  {{Pankow}}}, \bibinfo {author} {\bibfnamefont {P.}~\bibnamefont {{Brady}}},
  \bibinfo {author} {\bibfnamefont {E.}~\bibnamefont {{Ochsner}}}, \ and\
  \bibinfo {author} {\bibfnamefont {R.}~\bibnamefont {{O'Shaughnessy}}},\
  }\href {\doibase 10.1103/PhysRevD.92.023002} {\bibfield  {journal} {\bibinfo
  {journal} {\prd}\ }\textbf {\bibinfo {volume} {92}},\ \bibinfo {eid} {023002}
  (\bibinfo {year} {2015}{\natexlab{a}})}\BibitemShut {NoStop}%
\bibitem [{\citenamefont {{Lange}}\ \emph {et~al.}(2018)\citenamefont
  {{Lange}}, \citenamefont {{O'Shaughnessy}},\ and\ \citenamefont
  {{Rizzo}}}]{gwastro-PENR-RIFT}%
  \BibitemOpen
  \bibfield  {author} {\bibinfo {author} {\bibfnamefont {J.}~\bibnamefont
  {{Lange}}}, \bibinfo {author} {\bibfnamefont {R.}~\bibnamefont
  {{O'Shaughnessy}}}, \ and\ \bibinfo {author} {\bibfnamefont {M.}~\bibnamefont
  {{Rizzo}}},\ }\href@noop {} {\bibfield  {journal} {\bibinfo  {journal}
  {Submitted to PRD; available at arxiv:1805.10457}\ } (\bibinfo {year}
  {2018})}\BibitemShut {NoStop}%
\bibitem [{\citenamefont {{Veitch}}\ \emph {et~al.}(2015)\citenamefont
  {{Veitch}}, \citenamefont {{Raymond}}, \citenamefont {{Farr}}, \citenamefont
  {{Farr}}, \citenamefont {{Graff}}, \citenamefont {{Vitale}}, \citenamefont
  {{Aylott}}, \citenamefont {{Blackburn}}, \citenamefont {{Christensen}},
  \citenamefont {{Coughlin}}, \citenamefont {{Pozzo}}, \citenamefont {{Feroz}},
  \citenamefont {{Gair}}, \citenamefont {{Haster}}, \citenamefont {{Kalogera}},
  \citenamefont {{Littenberg}}, \citenamefont {{Mandel}}, \citenamefont
  {{O'Shaughnessy}}, \citenamefont {{Pitkin}}, \citenamefont {{Rodriguez}},
  \citenamefont {{R\"over}}, \citenamefont {{Sidery}}, \citenamefont {{Smith}},
  \citenamefont {{Sluys}}, \citenamefont {{Vecchio}}, \citenamefont
  {{Vousden}},\ and\ \citenamefont {{Wade}}}]{gw-astro-PE-lalinference-v1}%
  \BibitemOpen
  \bibfield  {author} {\bibinfo {author} {\bibfnamefont {J.}~\bibnamefont
  {{Veitch}}}, \bibinfo {author} {\bibfnamefont {V.}~\bibnamefont {{Raymond}}},
  \bibinfo {author} {\bibfnamefont {B.}~\bibnamefont {{Farr}}}, \bibinfo
  {author} {\bibfnamefont {W.~M.}\ \bibnamefont {{Farr}}}, \bibinfo {author}
  {\bibfnamefont {P.}~\bibnamefont {{Graff}}}, \bibinfo {author} {\bibfnamefont
  {S.}~\bibnamefont {{Vitale}}}, \bibinfo {author} {\bibfnamefont
  {B.}~\bibnamefont {{Aylott}}}, \bibinfo {author} {\bibfnamefont
  {K.}~\bibnamefont {{Blackburn}}}, \bibinfo {author} {\bibfnamefont
  {N.}~\bibnamefont {{Christensen}}}, \bibinfo {author} {\bibfnamefont
  {M.}~\bibnamefont {{Coughlin}}}, \bibinfo {author} {\bibfnamefont {W.~D.}\
  \bibnamefont {{Pozzo}}}, \bibinfo {author} {\bibfnamefont {F.}~\bibnamefont
  {{Feroz}}}, \bibinfo {author} {\bibfnamefont {J.}~\bibnamefont {{Gair}}},
  \bibinfo {author} {\bibfnamefont {C.}~\bibnamefont {{Haster}}}, \bibinfo
  {author} {\bibfnamefont {V.}~\bibnamefont {{Kalogera}}}, \bibinfo {author}
  {\bibfnamefont {T.}~\bibnamefont {{Littenberg}}}, \bibinfo {author}
  {\bibfnamefont {I.}~\bibnamefont {{Mandel}}}, \bibinfo {author}
  {\bibfnamefont {R.}~\bibnamefont {{O'Shaughnessy}}}, \bibinfo {author}
  {\bibfnamefont {M.}~\bibnamefont {{Pitkin}}}, \bibinfo {author}
  {\bibfnamefont {C.}~\bibnamefont {{Rodriguez}}}, \bibinfo {author}
  {\bibfnamefont {C.}~\bibnamefont {{R\"over}}}, \bibinfo {author}
  {\bibfnamefont {T.}~\bibnamefont {{Sidery}}}, \bibinfo {author}
  {\bibfnamefont {R.}~\bibnamefont {{Smith}}}, \bibinfo {author} {\bibfnamefont
  {M.~V.~D.}\ \bibnamefont {{Sluys}}}, \bibinfo {author} {\bibfnamefont
  {A.}~\bibnamefont {{Vecchio}}}, \bibinfo {author} {\bibfnamefont
  {W.}~\bibnamefont {{Vousden}}}, \ and\ \bibinfo {author} {\bibfnamefont
  {L.}~\bibnamefont {{Wade}}},\ }\href {\doibase 10.1103/PhysRevD.91.042003}
  {\bibfield  {journal} {\bibinfo  {journal} {\prd}\ }\textbf {\bibinfo
  {volume} {91}},\ \bibinfo {pages} {042003} (\bibinfo {year}
  {2015})}\BibitemShut {NoStop}%
\bibitem [{\citenamefont {{Abbott}}\ \emph {et~al.}(2016)\citenamefont
  {{Abbott}}, \citenamefont {{Abbott}}, \citenamefont {{Abbott}}, \citenamefont
  {{Abernathy}}, \citenamefont {{Acernese}}, \citenamefont {{Ackley}},
  \citenamefont {{Adams}}, \citenamefont {{Adams}}, \citenamefont {{Addesso}},
  \citenamefont {{Adhikari}},\ and\ \citenamefont
  {et~al.}}]{2016LRR....19....1A}%
  \BibitemOpen
  \bibfield  {author} {\bibinfo {author} {\bibfnamefont {B.~P.}\ \bibnamefont
  {{Abbott}}}, \bibinfo {author} {\bibfnamefont {R.}~\bibnamefont {{Abbott}}},
  \bibinfo {author} {\bibfnamefont {T.~D.}\ \bibnamefont {{Abbott}}}, \bibinfo
  {author} {\bibfnamefont {M.~R.}\ \bibnamefont {{Abernathy}}}, \bibinfo
  {author} {\bibfnamefont {F.}~\bibnamefont {{Acernese}}}, \bibinfo {author}
  {\bibfnamefont {K.}~\bibnamefont {{Ackley}}}, \bibinfo {author}
  {\bibfnamefont {C.}~\bibnamefont {{Adams}}}, \bibinfo {author} {\bibfnamefont
  {T.}~\bibnamefont {{Adams}}}, \bibinfo {author} {\bibfnamefont
  {P.}~\bibnamefont {{Addesso}}}, \bibinfo {author} {\bibfnamefont {R.~X.}\
  \bibnamefont {{Adhikari}}}, \ and\ \bibinfo {author} {\bibnamefont
  {et~al.}},\ }\href {\doibase 10.1007/lrr-2016-1} {\bibfield  {journal}
  {\bibinfo  {journal} {Living Reviews in Relativity}\ }\textbf {\bibinfo
  {volume} {19}},\ \bibinfo {eid} {1} (\bibinfo {year} {2016})}\BibitemShut
  {NoStop}%
\bibitem [{\citenamefont {Pretorius}(2005)}]{Pretorius:2005gq}%
  \BibitemOpen
  \bibfield  {author} {\bibinfo {author} {\bibfnamefont {F.}~\bibnamefont
  {Pretorius}},\ }\href@noop {} {\bibfield  {journal} {\bibinfo  {journal}
  {Phys. Rev. Lett.}\ }\textbf {\bibinfo {volume} {95}},\ \bibinfo {pages}
  {121101} (\bibinfo {year} {2005})},\ \Eprint
  {http://arxiv.org/abs/gr-qc/0507014} {gr-qc/0507014} \BibitemShut {NoStop}%
\bibitem [{\citenamefont {Campanelli}\ \emph {et~al.}(2006)\citenamefont
  {Campanelli}, \citenamefont {Lousto}, \citenamefont {Marronetti},\ and\
  \citenamefont {Zlochower}}]{Campanelli:2005dd}%
  \BibitemOpen
  \bibfield  {author} {\bibinfo {author} {\bibfnamefont {M.}~\bibnamefont
  {Campanelli}}, \bibinfo {author} {\bibfnamefont {C.~O.}\ \bibnamefont
  {Lousto}}, \bibinfo {author} {\bibfnamefont {P.}~\bibnamefont {Marronetti}},
  \ and\ \bibinfo {author} {\bibfnamefont {Y.}~\bibnamefont {Zlochower}},\
  }\href@noop {} {\bibfield  {journal} {\bibinfo  {journal} {Phys. Rev. Lett.}\
  }\textbf {\bibinfo {volume} {96}},\ \bibinfo {pages} {111101} (\bibinfo
  {year} {2006})},\ \Eprint {http://arxiv.org/abs/gr-qc/0511048}
  {gr-qc/0511048} \BibitemShut {NoStop}%
\bibitem [{\citenamefont {Baker}\ \emph {et~al.}(2006)\citenamefont {Baker},
  \citenamefont {Centrella}, \citenamefont {Choi}, \citenamefont {Koppitz},\
  and\ \citenamefont {van Meter}}]{Baker:2005vv}%
  \BibitemOpen
  \bibfield  {author} {\bibinfo {author} {\bibfnamefont {J.~G.}\ \bibnamefont
  {Baker}}, \bibinfo {author} {\bibfnamefont {J.}~\bibnamefont {Centrella}},
  \bibinfo {author} {\bibfnamefont {D.-I.}\ \bibnamefont {Choi}}, \bibinfo
  {author} {\bibfnamefont {M.}~\bibnamefont {Koppitz}}, \ and\ \bibinfo
  {author} {\bibfnamefont {J.}~\bibnamefont {van Meter}},\ }\href@noop {}
  {\bibfield  {journal} {\bibinfo  {journal} {Phys. Rev. Lett.}\ }\textbf
  {\bibinfo {volume} {96}},\ \bibinfo {pages} {111102} (\bibinfo {year}
  {2006})},\ \Eprint {http://arxiv.org/abs/gr-qc/0511103} {gr-qc/0511103}
  \BibitemShut {NoStop}%
\bibitem [{\citenamefont {Abbott}\ \emph
  {et~al.}(2016{\natexlab{a}})\citenamefont {Abbott} \emph
  {et~al.}}]{TheLIGOScientific:2016wfe}%
  \BibitemOpen
  \bibfield  {author} {\bibinfo {author} {\bibfnamefont {B.~P.}\ \bibnamefont
  {Abbott}} \emph {et~al.} (\bibinfo {collaboration} {Virgo, LIGO
  Scientific}),\ }\href {\doibase 10.1103/PhysRevLett.116.241102} {\bibfield
  {journal} {\bibinfo  {journal} {Phys. Rev. Lett.}\ }\textbf {\bibinfo
  {volume} {116}},\ \bibinfo {pages} {241102} (\bibinfo {year}
  {2016}{\natexlab{a}})},\ \Eprint {http://arxiv.org/abs/1602.03840}
  {arXiv:1602.03840 [gr-qc]} \BibitemShut {NoStop}%
\bibitem [{\citenamefont {Abbott}\ \emph
  {et~al.}(2016{\natexlab{b}})\citenamefont {Abbott} \emph
  {et~al.}}]{Abbott:2016blz}%
  \BibitemOpen
  \bibfield  {author} {\bibinfo {author} {\bibfnamefont {B.}~\bibnamefont
  {Abbott}} \emph {et~al.} (\bibinfo {collaboration} {Virgo, LIGO
  Scientific}),\ }\href {\doibase 10.1103/PhysRevLett.116.061102} {\bibfield
  {journal} {\bibinfo  {journal} {Phys. Rev. Lett.}\ }\textbf {\bibinfo
  {volume} {116}},\ \bibinfo {pages} {061102} (\bibinfo {year}
  {2016}{\natexlab{b}})},\ \Eprint {http://arxiv.org/abs/1602.03837}
  {arXiv:1602.03837 [gr-qc]} \BibitemShut {NoStop}%
\bibitem [{\citenamefont {Abbott}\ \emph
  {et~al.}(2016{\natexlab{c}})\citenamefont {Abbott} \emph
  {et~al.}}]{Abbott:2016nmj}%
  \BibitemOpen
  \bibfield  {author} {\bibinfo {author} {\bibfnamefont {B.~P.}\ \bibnamefont
  {Abbott}} \emph {et~al.} (\bibinfo {collaboration} {Virgo, LIGO
  Scientific}),\ }\href {\doibase 10.1103/PhysRevLett.116.241103} {\bibfield
  {journal} {\bibinfo  {journal} {Phys. Rev. Lett.}\ }\textbf {\bibinfo
  {volume} {116}},\ \bibinfo {pages} {241103} (\bibinfo {year}
  {2016}{\natexlab{c}})},\ \Eprint {http://arxiv.org/abs/1606.04855}
  {arXiv:1606.04855 [gr-qc]} \BibitemShut {NoStop}%
\bibitem [{\citenamefont {Abbott}\ \emph
  {et~al.}(2016{\natexlab{d}})\citenamefont {Abbott} \emph
  {et~al.}}]{TheLIGOScientific:2016pea}%
  \BibitemOpen
  \bibfield  {author} {\bibinfo {author} {\bibfnamefont {B.~P.}\ \bibnamefont
  {Abbott}} \emph {et~al.} (\bibinfo {collaboration} {Virgo, LIGO
  Scientific}),\ }\href {\doibase 10.1103/PhysRevX.6.041015} {\bibfield
  {journal} {\bibinfo  {journal} {Phys. Rev.}\ }\textbf {\bibinfo {volume}
  {X6}},\ \bibinfo {pages} {041015} (\bibinfo {year} {2016}{\natexlab{d}})},\
  \Eprint {http://arxiv.org/abs/1606.04856} {arXiv:1606.04856 [gr-qc]}
  \BibitemShut {NoStop}%
\bibitem [{\citenamefont {Abbott}\ \emph
  {et~al.}(2017{\natexlab{a}})\citenamefont {Abbott} \emph
  {et~al.}}]{Abbott:2016wiq}%
  \BibitemOpen
  \bibfield  {author} {\bibinfo {author} {\bibfnamefont {B.~P.}\ \bibnamefont
  {Abbott}} \emph {et~al.} (\bibinfo {collaboration} {Virgo, LIGO
  Scientific}),\ }\href {\doibase 10.1088/1361-6382/aa6854} {\bibfield
  {journal} {\bibinfo  {journal} {Class. Quant. Grav.}\ }\textbf {\bibinfo
  {volume} {34}},\ \bibinfo {pages} {104002} (\bibinfo {year}
  {2017}{\natexlab{a}})},\ \Eprint {http://arxiv.org/abs/1611.07531}
  {arXiv:1611.07531 [gr-qc]} \BibitemShut {NoStop}%
\bibitem [{\citenamefont {Abbott}\ \emph
  {et~al.}(2016{\natexlab{e}})\citenamefont {Abbott} \emph
  {et~al.}}]{Abbott:2016apu}%
  \BibitemOpen
  \bibfield  {author} {\bibinfo {author} {\bibfnamefont {B.~P.}\ \bibnamefont
  {Abbott}} \emph {et~al.} (\bibinfo {collaboration} {Virgo, LIGO
  Scientific}),\ }\href {\doibase 10.1103/PhysRevD.94.064035} {\bibfield
  {journal} {\bibinfo  {journal} {Phys. Rev.}\ }\textbf {\bibinfo {volume}
  {D94}},\ \bibinfo {pages} {064035} (\bibinfo {year} {2016}{\natexlab{e}})},\
  \Eprint {http://arxiv.org/abs/1606.01262} {arXiv:1606.01262 [gr-qc]}
  \BibitemShut {NoStop}%
\bibitem [{\citenamefont {Abbott}\ \emph
  {et~al.}(2016{\natexlab{f}})\citenamefont {Abbott} \emph
  {et~al.}}]{TheLIGOScientific:2016uux}%
  \BibitemOpen
  \bibfield  {author} {\bibinfo {author} {\bibfnamefont {B.~P.}\ \bibnamefont
  {Abbott}} \emph {et~al.} (\bibinfo {collaboration} {Virgo, LIGO
  Scientific}),\ }\href {\doibase 10.1103/PhysRevD.94.069903,
  10.1103/PhysRevD.93.122004} {\bibfield  {journal} {\bibinfo  {journal} {Phys.
  Rev.}\ }\textbf {\bibinfo {volume} {D93}},\ \bibinfo {pages} {122004}
  (\bibinfo {year} {2016}{\natexlab{f}})},\ \bibinfo {note} {[Addendum: Phys.
  Rev.D94,no.6,069903(2016)]},\ \Eprint {http://arxiv.org/abs/1602.03843}
  {arXiv:1602.03843 [gr-qc]} \BibitemShut {NoStop}%
\bibitem [{\citenamefont {Lovelace}\ \emph {et~al.}(2016)\citenamefont
  {Lovelace} \emph {et~al.}}]{Lovelace:2016uwp}%
  \BibitemOpen
  \bibfield  {author} {\bibinfo {author} {\bibfnamefont {G.}~\bibnamefont
  {Lovelace}} \emph {et~al.},\ }\href {\doibase 10.1088/0264-9381/33/24/244002}
  {\bibfield  {journal} {\bibinfo  {journal} {Class. Quant. Grav.}\ }\textbf
  {\bibinfo {volume} {33}},\ \bibinfo {pages} {244002} (\bibinfo {year}
  {2016})},\ \Eprint {http://arxiv.org/abs/1607.05377} {arXiv:1607.05377
  [gr-qc]} \BibitemShut {NoStop}%
\bibitem [{\citenamefont {Aylott}\ \emph
  {et~al.}(2009{\natexlab{a}})\citenamefont {Aylott} \emph
  {et~al.}}]{Aylott:2009ya}%
  \BibitemOpen
  \bibfield  {author} {\bibinfo {author} {\bibfnamefont {B.}~\bibnamefont
  {Aylott}} \emph {et~al.},\ }\href {\doibase 10.1088/0264-9381/26/16/165008}
  {\bibfield  {journal} {\bibinfo  {journal} {Class. Quant. Grav.}\ }\textbf
  {\bibinfo {volume} {26}},\ \bibinfo {pages} {165008} (\bibinfo {year}
  {2009}{\natexlab{a}})},\ \Eprint {http://arxiv.org/abs/0901.4399}
  {arXiv:0901.4399 [gr-qc]} \BibitemShut {NoStop}%
\bibitem [{\citenamefont {Aylott}\ \emph
  {et~al.}(2009{\natexlab{b}})\citenamefont {Aylott} \emph
  {et~al.}}]{Aylott:2009tn}%
  \BibitemOpen
  \bibfield  {author} {\bibinfo {author} {\bibfnamefont {B.}~\bibnamefont
  {Aylott}} \emph {et~al.},\ }\href {\doibase 10.1088/0264-9381/26/11/114008}
  {\bibfield  {journal} {\bibinfo  {journal} {Class. Quant. Grav.}\ }\textbf
  {\bibinfo {volume} {26}},\ \bibinfo {pages} {114008} (\bibinfo {year}
  {2009}{\natexlab{b}})},\ \Eprint {http://arxiv.org/abs/0905.4227}
  {arXiv:0905.4227 [gr-qc]} \BibitemShut {NoStop}%
\bibitem [{\citenamefont {Ajith}\ \emph {et~al.}(2012)\citenamefont {Ajith}
  \emph {et~al.}}]{Ajith:2012az}%
  \BibitemOpen
  \bibfield  {author} {\bibinfo {author} {\bibfnamefont {P.}~\bibnamefont
  {Ajith}} \emph {et~al.},\ }\href {\doibase 10.1088/0264-9381/29/12/124001}
  {\bibfield  {journal} {\bibinfo  {journal} {Class. Quant. Grav.}\ }\textbf
  {\bibinfo {volume} {29}},\ \bibinfo {pages} {124001} (\bibinfo {year}
  {2012})},\ \Eprint {http://arxiv.org/abs/1201.5319} {arXiv:1201.5319 [gr-qc]}
  \BibitemShut {NoStop}%
\bibitem [{\citenamefont {Aasi}\ \emph {et~al.}(2014)\citenamefont {Aasi} \emph
  {et~al.}}]{Aasi:2014tra}%
  \BibitemOpen
  \bibfield  {author} {\bibinfo {author} {\bibfnamefont {J.}~\bibnamefont
  {Aasi}} \emph {et~al.} (\bibinfo {collaboration} {LIGO Scientific
  Collaboration, Virgo Collaboration, NINJA-2 Collaboration}),\ }\href
  {\doibase 10.1088/0264-9381/31/11/115004} {\bibfield  {journal} {\bibinfo
  {journal} {Class. Quant. Grav.}\ }\textbf {\bibinfo {volume} {31}},\ \bibinfo
  {pages} {115004} (\bibinfo {year} {2014})},\ \Eprint
  {http://arxiv.org/abs/1401.0939} {arXiv:1401.0939 [gr-qc]} \BibitemShut
  {NoStop}%
\bibitem [{\citenamefont {Hinder}\ \emph {et~al.}(2014)\citenamefont {Hinder},
  \citenamefont {Buonanno}, \citenamefont {Boyle}, \citenamefont {Etienne},
  \citenamefont {Healy}, \citenamefont {Johnson-McDaniel}, \citenamefont
  {Nagar}, \citenamefont {Nakano}, \citenamefont {Pan}, \citenamefont
  {Pfeiffer}, \citenamefont {P{\"u}rrer}, \citenamefont {Reisswig},
  \citenamefont {Scheel}, \citenamefont {Schnetter}, \citenamefont {Sperhake},
  \citenamefont {Szil{\'a}gyi}, \citenamefont {Tichy}, \citenamefont {Wardell},
  \citenamefont {Zengino\u{g}lu}, \citenamefont {Alic}, \citenamefont
  {Bernuzzi}, \citenamefont {Bode}, \citenamefont {Br{\"u}gmann}, \citenamefont
  {Buchman}, \citenamefont {Campanelli}, \citenamefont {Chu}, \citenamefont
  {Damour}, \citenamefont {Grigsby}, \citenamefont {Hannam}, \citenamefont
  {Haas}, \citenamefont {Hemberger}, \citenamefont {Husa}, \citenamefont
  {Kidder}, \citenamefont {Laguna}, \citenamefont {London}, \citenamefont
  {Lovelace}, \citenamefont {Lousto}, \citenamefont {Marronetti}, \citenamefont
  {Matzner}, \citenamefont {M{\"o}sta}, \citenamefont {Mrou{\'e}},
  \citenamefont {M{\"u}ller}, \citenamefont {Mundim}, \citenamefont {Nerozzi},
  \citenamefont {Paschalidis}, \citenamefont {Pollney}, \citenamefont
  {Reifenberger}, \citenamefont {Rezzolla}, \citenamefont {Shapiro},
  \citenamefont {Shoemaker}, \citenamefont {Taracchini}, \citenamefont
  {Taylor}, \citenamefont {Teukolsky}, \citenamefont {Thierfelder},
  \citenamefont {Witek},\ and\ \citenamefont {Zlochower}}]{Hinder:2013oqa}%
  \BibitemOpen
  \bibfield  {author} {\bibinfo {author} {\bibfnamefont {I.}~\bibnamefont
  {Hinder}}, \bibinfo {author} {\bibfnamefont {A.}~\bibnamefont {Buonanno}},
  \bibinfo {author} {\bibfnamefont {M.}~\bibnamefont {Boyle}}, \bibinfo
  {author} {\bibfnamefont {Z.~B.}\ \bibnamefont {Etienne}}, \bibinfo {author}
  {\bibfnamefont {J.}~\bibnamefont {Healy}}, \bibinfo {author} {\bibfnamefont
  {N.~K.}\ \bibnamefont {Johnson-McDaniel}}, \bibinfo {author} {\bibfnamefont
  {A.}~\bibnamefont {Nagar}}, \bibinfo {author} {\bibfnamefont
  {H.}~\bibnamefont {Nakano}}, \bibinfo {author} {\bibfnamefont
  {Y.}~\bibnamefont {Pan}}, \bibinfo {author} {\bibfnamefont {H.~P.}\
  \bibnamefont {Pfeiffer}}, \bibinfo {author} {\bibfnamefont {M.}~\bibnamefont
  {P{\"u}rrer}}, \bibinfo {author} {\bibfnamefont {C.}~\bibnamefont
  {Reisswig}}, \bibinfo {author} {\bibfnamefont {M.~A.}\ \bibnamefont
  {Scheel}}, \bibinfo {author} {\bibfnamefont {E.}~\bibnamefont {Schnetter}},
  \bibinfo {author} {\bibfnamefont {U.}~\bibnamefont {Sperhake}}, \bibinfo
  {author} {\bibfnamefont {B.}~\bibnamefont {Szil{\'a}gyi}}, \bibinfo {author}
  {\bibfnamefont {W.}~\bibnamefont {Tichy}}, \bibinfo {author} {\bibfnamefont
  {B.}~\bibnamefont {Wardell}}, \bibinfo {author} {\bibfnamefont
  {A.}~\bibnamefont {Zengino\u{g}lu}}, \bibinfo {author} {\bibfnamefont
  {D.}~\bibnamefont {Alic}}, \bibinfo {author} {\bibfnamefont {S.}~\bibnamefont
  {Bernuzzi}}, \bibinfo {author} {\bibfnamefont {T.}~\bibnamefont {Bode}},
  \bibinfo {author} {\bibfnamefont {B.}~\bibnamefont {Br{\"u}gmann}}, \bibinfo
  {author} {\bibfnamefont {L.~T.}\ \bibnamefont {Buchman}}, \bibinfo {author}
  {\bibfnamefont {M.}~\bibnamefont {Campanelli}}, \bibinfo {author}
  {\bibfnamefont {T.}~\bibnamefont {Chu}}, \bibinfo {author} {\bibfnamefont
  {T.}~\bibnamefont {Damour}}, \bibinfo {author} {\bibfnamefont {J.~D.}\
  \bibnamefont {Grigsby}}, \bibinfo {author} {\bibfnamefont {M.}~\bibnamefont
  {Hannam}}, \bibinfo {author} {\bibfnamefont {R.}~\bibnamefont {Haas}},
  \bibinfo {author} {\bibfnamefont {D.~A.}\ \bibnamefont {Hemberger}}, \bibinfo
  {author} {\bibfnamefont {S.}~\bibnamefont {Husa}}, \bibinfo {author}
  {\bibfnamefont {L.~E.}\ \bibnamefont {Kidder}}, \bibinfo {author}
  {\bibfnamefont {P.}~\bibnamefont {Laguna}}, \bibinfo {author} {\bibfnamefont
  {L.}~\bibnamefont {London}}, \bibinfo {author} {\bibfnamefont
  {G.}~\bibnamefont {Lovelace}}, \bibinfo {author} {\bibfnamefont {C.~O.}\
  \bibnamefont {Lousto}}, \bibinfo {author} {\bibfnamefont {P.}~\bibnamefont
  {Marronetti}}, \bibinfo {author} {\bibfnamefont {R.~A.}\ \bibnamefont
  {Matzner}}, \bibinfo {author} {\bibfnamefont {P.}~\bibnamefont {M{\"o}sta}},
  \bibinfo {author} {\bibfnamefont {A.}~\bibnamefont {Mrou{\'e}}}, \bibinfo
  {author} {\bibfnamefont {D.}~\bibnamefont {M{\"u}ller}}, \bibinfo {author}
  {\bibfnamefont {B.~C.}\ \bibnamefont {Mundim}}, \bibinfo {author}
  {\bibfnamefont {A.}~\bibnamefont {Nerozzi}}, \bibinfo {author} {\bibfnamefont
  {V.}~\bibnamefont {Paschalidis}}, \bibinfo {author} {\bibfnamefont
  {D.}~\bibnamefont {Pollney}}, \bibinfo {author} {\bibfnamefont
  {G.}~\bibnamefont {Reifenberger}}, \bibinfo {author} {\bibfnamefont
  {L.}~\bibnamefont {Rezzolla}}, \bibinfo {author} {\bibfnamefont {S.~L.}\
  \bibnamefont {Shapiro}}, \bibinfo {author} {\bibfnamefont {D.}~\bibnamefont
  {Shoemaker}}, \bibinfo {author} {\bibfnamefont {A.}~\bibnamefont
  {Taracchini}}, \bibinfo {author} {\bibfnamefont {N.~W.}\ \bibnamefont
  {Taylor}}, \bibinfo {author} {\bibfnamefont {S.~A.}\ \bibnamefont
  {Teukolsky}}, \bibinfo {author} {\bibfnamefont {M.}~\bibnamefont
  {Thierfelder}}, \bibinfo {author} {\bibfnamefont {H.}~\bibnamefont {Witek}},
  \ and\ \bibinfo {author} {\bibfnamefont {Y.}~\bibnamefont {Zlochower}},\
  }\href {\doibase 10.1088/0264-9381/31/2/025012} {\bibfield  {journal}
  {\bibinfo  {journal} {Class. Quant. Grav.}\ }\textbf {\bibinfo {volume}
  {31}},\ \bibinfo {pages} {025012} (\bibinfo {year} {2014})},\ \Eprint
  {http://arxiv.org/abs/1307.5307} {arXiv:1307.5307 [gr-qc]} \BibitemShut
  {NoStop}%
\bibitem [{\citenamefont {Mroue}\ \emph {et~al.}(2013)\citenamefont {Mroue},
  \citenamefont {Scheel}, \citenamefont {Szilagyi}, \citenamefont {Pfeiffer},
  \citenamefont {Boyle} \emph {et~al.}}]{Mroue:2013xna}%
  \BibitemOpen
  \bibfield  {author} {\bibinfo {author} {\bibfnamefont {A.~H.}\ \bibnamefont
  {Mroue}}, \bibinfo {author} {\bibfnamefont {M.~A.}\ \bibnamefont {Scheel}},
  \bibinfo {author} {\bibfnamefont {B.}~\bibnamefont {Szilagyi}}, \bibinfo
  {author} {\bibfnamefont {H.~P.}\ \bibnamefont {Pfeiffer}}, \bibinfo {author}
  {\bibfnamefont {M.}~\bibnamefont {Boyle}},  \emph {et~al.},\ }\href {\doibase
  10.1103/PhysRevLett.111.241104} {\bibfield  {journal} {\bibinfo  {journal}
  {Phys. Rev. Lett.}\ }\textbf {\bibinfo {volume} {111}},\ \bibinfo {pages}
  {241104} (\bibinfo {year} {2013})},\ \Eprint {http://arxiv.org/abs/1304.6077}
  {arXiv:1304.6077 [gr-qc]} \BibitemShut {NoStop}%
\bibitem [{\citenamefont {Blackman}\ \emph {et~al.}(2015)\citenamefont
  {Blackman}, \citenamefont {Field}, \citenamefont {Galley}, \citenamefont
  {Szil{\'a}gyi}, \citenamefont {Scheel}, \citenamefont {Tiglio},\ and\
  \citenamefont {Hemberger}}]{Blackman:2015pia}%
  \BibitemOpen
  \bibfield  {author} {\bibinfo {author} {\bibfnamefont {J.}~\bibnamefont
  {Blackman}}, \bibinfo {author} {\bibfnamefont {S.~E.}\ \bibnamefont {Field}},
  \bibinfo {author} {\bibfnamefont {C.~R.}\ \bibnamefont {Galley}}, \bibinfo
  {author} {\bibfnamefont {B.}~\bibnamefont {Szil{\'a}gyi}}, \bibinfo {author}
  {\bibfnamefont {M.~A.}\ \bibnamefont {Scheel}}, \bibinfo {author}
  {\bibfnamefont {M.}~\bibnamefont {Tiglio}}, \ and\ \bibinfo {author}
  {\bibfnamefont {D.~A.}\ \bibnamefont {Hemberger}},\ }\href {\doibase
  10.1103/PhysRevLett.115.121102} {\bibfield  {journal} {\bibinfo  {journal}
  {Phys. Rev. Lett.}\ }\textbf {\bibinfo {volume} {115}},\ \bibinfo {pages}
  {121102} (\bibinfo {year} {2015})},\ \Eprint
  {http://arxiv.org/abs/1502.07758} {arXiv:1502.07758 [gr-qc]} \BibitemShut
  {NoStop}%
\bibitem [{\citenamefont {Chu}\ \emph {et~al.}(2016)\citenamefont {Chu},
  \citenamefont {Fong}, \citenamefont {Kumar}, \citenamefont {Pfeiffer},
  \citenamefont {Boyle}, \citenamefont {Hemberger}, \citenamefont {Kidder},
  \citenamefont {Scheel},\ and\ \citenamefont {Szilagyi}}]{Chu:2015kft}%
  \BibitemOpen
  \bibfield  {author} {\bibinfo {author} {\bibfnamefont {T.}~\bibnamefont
  {Chu}}, \bibinfo {author} {\bibfnamefont {H.}~\bibnamefont {Fong}}, \bibinfo
  {author} {\bibfnamefont {P.}~\bibnamefont {Kumar}}, \bibinfo {author}
  {\bibfnamefont {H.~P.}\ \bibnamefont {Pfeiffer}}, \bibinfo {author}
  {\bibfnamefont {M.}~\bibnamefont {Boyle}}, \bibinfo {author} {\bibfnamefont
  {D.~A.}\ \bibnamefont {Hemberger}}, \bibinfo {author} {\bibfnamefont {L.~E.}\
  \bibnamefont {Kidder}}, \bibinfo {author} {\bibfnamefont {M.~A.}\
  \bibnamefont {Scheel}}, \ and\ \bibinfo {author} {\bibfnamefont
  {B.}~\bibnamefont {Szilagyi}},\ }\href {\doibase
  10.1088/0264-9381/33/16/165001} {\bibfield  {journal} {\bibinfo  {journal}
  {Class. Quant. Grav.}\ }\textbf {\bibinfo {volume} {33}},\ \bibinfo {pages}
  {165001} (\bibinfo {year} {2016})},\ \Eprint
  {http://arxiv.org/abs/1512.06800} {arXiv:1512.06800 [gr-qc]} \BibitemShut
  {NoStop}%
\bibitem [{\citenamefont {Boyle}\ \emph {et~al.}(2019)\citenamefont {Boyle}
  \emph {et~al.}}]{Boyle:2019kee}%
  \BibitemOpen
  \bibfield  {author} {\bibinfo {author} {\bibfnamefont {M.}~\bibnamefont
  {Boyle}} \emph {et~al.},\ }\href {\doibase 10.1088/1361-6382/ab34e2}
  {\bibfield  {journal} {\bibinfo  {journal} {Class. Quant. Grav.}\ }\textbf
  {\bibinfo {volume} {36}},\ \bibinfo {pages} {195006} (\bibinfo {year}
  {2019})},\ \Eprint {http://arxiv.org/abs/1904.04831} {arXiv:1904.04831
  [gr-qc]} \BibitemShut {NoStop}%
\bibitem [{\citenamefont {Jani}\ \emph {et~al.}(2016)\citenamefont {Jani},
  \citenamefont {Healy}, \citenamefont {Clark}, \citenamefont {London},
  \citenamefont {Laguna},\ and\ \citenamefont {Shoemaker}}]{Jani:2016wkt}%
  \BibitemOpen
  \bibfield  {author} {\bibinfo {author} {\bibfnamefont {K.}~\bibnamefont
  {Jani}}, \bibinfo {author} {\bibfnamefont {J.}~\bibnamefont {Healy}},
  \bibinfo {author} {\bibfnamefont {J.~A.}\ \bibnamefont {Clark}}, \bibinfo
  {author} {\bibfnamefont {L.}~\bibnamefont {London}}, \bibinfo {author}
  {\bibfnamefont {P.}~\bibnamefont {Laguna}}, \ and\ \bibinfo {author}
  {\bibfnamefont {D.}~\bibnamefont {Shoemaker}},\ }\href {\doibase
  10.1088/0264-9381/33/20/204001} {\bibfield  {journal} {\bibinfo  {journal}
  {Class. Quant. Grav.}\ }\textbf {\bibinfo {volume} {33}},\ \bibinfo {pages}
  {204001} (\bibinfo {year} {2016})},\ \Eprint
  {http://arxiv.org/abs/1605.03204} {arXiv:1605.03204 [gr-qc]} \BibitemShut
  {NoStop}%
\bibitem [{\citenamefont {Healy}\ \emph
  {et~al.}(2017{\natexlab{a}})\citenamefont {Healy}, \citenamefont {Lousto},
  \citenamefont {Zlochower},\ and\ \citenamefont {Campanelli}}]{Healy:2017psd}%
  \BibitemOpen
  \bibfield  {author} {\bibinfo {author} {\bibfnamefont {J.}~\bibnamefont
  {Healy}}, \bibinfo {author} {\bibfnamefont {C.~O.}\ \bibnamefont {Lousto}},
  \bibinfo {author} {\bibfnamefont {Y.}~\bibnamefont {Zlochower}}, \ and\
  \bibinfo {author} {\bibfnamefont {M.}~\bibnamefont {Campanelli}},\ }\href
  {\doibase 10.1088/1361-6382/aa91b1} {\bibfield  {journal} {\bibinfo
  {journal} {Class. Quant. Grav.}\ }\textbf {\bibinfo {volume} {34}},\ \bibinfo
  {pages} {224001} (\bibinfo {year} {2017}{\natexlab{a}})},\ \Eprint
  {http://arxiv.org/abs/1703.03423} {arXiv:1703.03423 [gr-qc]} \BibitemShut
  {NoStop}%
\bibitem [{\citenamefont {Healy}\ \emph {et~al.}(2019)\citenamefont {Healy},
  \citenamefont {Lousto}, \citenamefont {Lange}, \citenamefont {O'Shaughnessy},
  \citenamefont {Zlochower},\ and\ \citenamefont {Campanelli}}]{Healy:2019jyf}%
  \BibitemOpen
  \bibfield  {author} {\bibinfo {author} {\bibfnamefont {J.}~\bibnamefont
  {Healy}}, \bibinfo {author} {\bibfnamefont {C.~O.}\ \bibnamefont {Lousto}},
  \bibinfo {author} {\bibfnamefont {J.}~\bibnamefont {Lange}}, \bibinfo
  {author} {\bibfnamefont {R.}~\bibnamefont {O'Shaughnessy}}, \bibinfo {author}
  {\bibfnamefont {Y.}~\bibnamefont {Zlochower}}, \ and\ \bibinfo {author}
  {\bibfnamefont {M.}~\bibnamefont {Campanelli}},\ }\href {\doibase
  10.1103/PhysRevD.100.024021} {\bibfield  {journal} {\bibinfo  {journal}
  {Phys. Rev.}\ }\textbf {\bibinfo {volume} {D100}},\ \bibinfo {pages} {024021}
  (\bibinfo {year} {2019})},\ \Eprint {http://arxiv.org/abs/1901.02553}
  {arXiv:1901.02553 [gr-qc]} \BibitemShut {NoStop}%
\bibitem [{\citenamefont {Healy}\ and\ \citenamefont
  {Lousto}(2020)}]{Healy:2020vre}%
  \BibitemOpen
  \bibfield  {author} {\bibinfo {author} {\bibfnamefont {J.}~\bibnamefont
  {Healy}}\ and\ \bibinfo {author} {\bibfnamefont {C.~O.}\ \bibnamefont
  {Lousto}},\ }\href@noop {} {\  (\bibinfo {year} {2020})},\ \Eprint
  {http://arxiv.org/abs/2007.07910} {arXiv:2007.07910 [gr-qc]} \BibitemShut
  {NoStop}%
\bibitem [{\citenamefont {Abbott}\ \emph
  {et~al.}(2017{\natexlab{b}})\citenamefont {Abbott} \emph
  {et~al.}}]{Abbott:2017vtc}%
  \BibitemOpen
  \bibfield  {author} {\bibinfo {author} {\bibfnamefont {B.~P.}\ \bibnamefont
  {Abbott}} \emph {et~al.} (\bibinfo {collaboration} {VIRGO, LIGO
  Scientific}),\ }\href {\doibase 10.1103/PhysRevLett.118.221101} {\bibfield
  {journal} {\bibinfo  {journal} {Phys. Rev. Lett.}\ }\textbf {\bibinfo
  {volume} {118}},\ \bibinfo {pages} {221101} (\bibinfo {year}
  {2017}{\natexlab{b}})},\ \Eprint {http://arxiv.org/abs/1706.01812}
  {arXiv:1706.01812 [gr-qc]} \BibitemShut {NoStop}%
\bibitem [{\citenamefont {Healy}\ \emph
  {et~al.}(2018{\natexlab{a}})\citenamefont {Healy} \emph
  {et~al.}}]{Heal:2017abq}%
  \BibitemOpen
  \bibfield  {author} {\bibinfo {author} {\bibfnamefont {J.}~\bibnamefont
  {Healy}} \emph {et~al.},\ }\href {\doibase 10.1103/PhysRevD.97.064027}
  {\bibfield  {journal} {\bibinfo  {journal} {Phys. Rev.}\ }\textbf {\bibinfo
  {volume} {D97}},\ \bibinfo {pages} {064027} (\bibinfo {year}
  {2018}{\natexlab{a}})},\ \Eprint {http://arxiv.org/abs/1712.05836}
  {arXiv:1712.05836 [gr-qc]} \BibitemShut {NoStop}%
\bibitem [{\citenamefont {Abbott}\ \emph
  {et~al.}(2017{\natexlab{c}})\citenamefont {Abbott} \emph
  {et~al.}}]{Abbott:2017gyy}%
  \BibitemOpen
  \bibfield  {author} {\bibinfo {author} {\bibfnamefont {B.~P.}\ \bibnamefont
  {Abbott}} \emph {et~al.} (\bibinfo {collaboration} {Virgo, LIGO
  Scientific}),\ }\href {\doibase 10.3847/2041-8213/aa9f0c} {\bibfield
  {journal} {\bibinfo  {journal} {Astrophys. J.}\ }\textbf {\bibinfo {volume}
  {851}},\ \bibinfo {pages} {L35} (\bibinfo {year} {2017}{\natexlab{c}})},\
  \Eprint {http://arxiv.org/abs/1711.05578} {arXiv:1711.05578 [astro-ph.HE]}
  \BibitemShut {NoStop}%
\bibitem [{\citenamefont {{The LIGO Scientific Collaboration}}\ \emph
  {et~al.}(2020{\natexlab{a}})\citenamefont {{The LIGO Scientific
  Collaboration}}, \citenamefont {{the Virgo Collaboration}}, \citenamefont
  {{Abbott}}, \citenamefont {{Abbott}}, \citenamefont {{Abbott}}, \citenamefont
  {{Abraham}}, \citenamefont {{Acernese}}, \citenamefont {{Ackley}},
  \citenamefont {{Adams}}, \citenamefont {{Adya}},\ and\ \citenamefont
  {et~al.}}]{LIGO-O3-GW190521-discovery}%
  \BibitemOpen
  \bibfield  {author} {\bibinfo {author} {\bibnamefont {{The LIGO Scientific
  Collaboration}}}, \bibinfo {author} {\bibnamefont {{the Virgo
  Collaboration}}}, \bibinfo {author} {\bibfnamefont {B.~P.}\ \bibnamefont
  {{Abbott}}}, \bibinfo {author} {\bibfnamefont {R.}~\bibnamefont {{Abbott}}},
  \bibinfo {author} {\bibfnamefont {T.~D.}\ \bibnamefont {{Abbott}}}, \bibinfo
  {author} {\bibfnamefont {S.}~\bibnamefont {{Abraham}}}, \bibinfo {author}
  {\bibfnamefont {F.}~\bibnamefont {{Acernese}}}, \bibinfo {author}
  {\bibfnamefont {K.}~\bibnamefont {{Ackley}}}, \bibinfo {author}
  {\bibfnamefont {C.}~\bibnamefont {{Adams}}}, \bibinfo {author} {\bibfnamefont
  {V.~B.}\ \bibnamefont {{Adya}}}, \ and\ \bibinfo {author} {\bibnamefont
  {et~al.}},\ }\href {\doibase 10.1103/PhysRevLett.125.101102} {\bibfield
  {journal} {\bibinfo  {journal} {\prl}\ }\textbf {\bibinfo {volume} {125}},\
  \bibinfo {eid} {101102} (\bibinfo {year} {2020}{\natexlab{a}})}\BibitemShut
  {NoStop}%
\bibitem [{\citenamefont {{The LIGO Scientific Collaboration}}\ \emph
  {et~al.}(2020{\natexlab{b}})\citenamefont {{The LIGO Scientific
  Collaboration}}, \citenamefont {{the Virgo Collaboration}}, \citenamefont
  {{Abbott}}, \citenamefont {{Abbott}}, \citenamefont {{Abbott}}, \citenamefont
  {{Abraham}}, \citenamefont {{Acernese}}, \citenamefont {{Ackley}},
  \citenamefont {{Adams}}, \citenamefont {{Adya}},\ and\ \citenamefont
  {et~al.}}]{LIGO-O3-GW190521-implications}%
  \BibitemOpen
  \bibfield  {author} {\bibinfo {author} {\bibnamefont {{The LIGO Scientific
  Collaboration}}}, \bibinfo {author} {\bibnamefont {{the Virgo
  Collaboration}}}, \bibinfo {author} {\bibfnamefont {B.~P.}\ \bibnamefont
  {{Abbott}}}, \bibinfo {author} {\bibfnamefont {R.}~\bibnamefont {{Abbott}}},
  \bibinfo {author} {\bibfnamefont {T.~D.}\ \bibnamefont {{Abbott}}}, \bibinfo
  {author} {\bibfnamefont {S.}~\bibnamefont {{Abraham}}}, \bibinfo {author}
  {\bibfnamefont {F.}~\bibnamefont {{Acernese}}}, \bibinfo {author}
  {\bibfnamefont {K.}~\bibnamefont {{Ackley}}}, \bibinfo {author}
  {\bibfnamefont {C.}~\bibnamefont {{Adams}}}, \bibinfo {author} {\bibfnamefont
  {V.~B.}\ \bibnamefont {{Adya}}}, \ and\ \bibinfo {author} {\bibnamefont
  {et~al.}},\ }\href@noop {} {\bibfield  {journal} {\bibinfo  {journal} {arXiv
  e-prints}\ ,\ \bibinfo {eid} {arXiv:2009.01190}} (\bibinfo {year}
  {2020}{\natexlab{b}})},\ \Eprint {http://arxiv.org/abs/2009.01190}
  {arXiv:2009.01190 [astro-ph.HE]} \BibitemShut {NoStop}%
\bibitem [{\citenamefont {Lange}\ \emph {et~al.}(2017)\citenamefont {Lange}
  \emph {et~al.}}]{Lange:2017wki}%
  \BibitemOpen
  \bibfield  {author} {\bibinfo {author} {\bibfnamefont {J.}~\bibnamefont
  {Lange}} \emph {et~al.},\ }\href {\doibase 10.1103/PhysRevD.96.104041}
  {\bibfield  {journal} {\bibinfo  {journal} {Phys. Rev.}\ }\textbf {\bibinfo
  {volume} {D96}},\ \bibinfo {pages} {104041} (\bibinfo {year} {2017})},\
  \Eprint {http://arxiv.org/abs/1705.09833} {arXiv:1705.09833 [gr-qc]}
  \BibitemShut {NoStop}%
\bibitem [{\citenamefont {{Lange}}\ \emph {et~al.}(2017)\citenamefont
  {{Lange}}, \citenamefont {{O'Shaughnessy}}, \citenamefont {{Boyle}},
  \citenamefont {{Calder{\'o}n Bustillo}}, \citenamefont {{Campanelli}},
  \citenamefont {{Chu}}, \citenamefont {{Clark}}, \citenamefont {{Demos}},
  \citenamefont {{Fong}}, \citenamefont {{Healy}}, \citenamefont {{Hemberger}},
  \citenamefont {{Hinder}}, \citenamefont {{Jani}}, \citenamefont {{Khamesra}},
  \citenamefont {{Kidder}}, \citenamefont {{Kumar}}, \citenamefont {{Laguna}},
  \citenamefont {{Lousto}}, \citenamefont {{Lovelace}}, \citenamefont
  {{Ossokine}}, \citenamefont {{Pfeiffer}}, \citenamefont {{Scheel}},
  \citenamefont {{Shoemaker}}, \citenamefont {{Szilagyi}}, \citenamefont
  {{Teukolsky}},\ and\ \citenamefont {{Zlochower}}}]{2017PhRvD..96j4041L}%
  \BibitemOpen
  \bibfield  {author} {\bibinfo {author} {\bibfnamefont {J.}~\bibnamefont
  {{Lange}}}, \bibinfo {author} {\bibfnamefont {R.}~\bibnamefont
  {{O'Shaughnessy}}}, \bibinfo {author} {\bibfnamefont {M.}~\bibnamefont
  {{Boyle}}}, \bibinfo {author} {\bibfnamefont {J.}~\bibnamefont {{Calder{\'o}n
  Bustillo}}}, \bibinfo {author} {\bibfnamefont {M.}~\bibnamefont
  {{Campanelli}}}, \bibinfo {author} {\bibfnamefont {T.}~\bibnamefont {{Chu}}},
  \bibinfo {author} {\bibfnamefont {J.~A.}\ \bibnamefont {{Clark}}}, \bibinfo
  {author} {\bibfnamefont {N.}~\bibnamefont {{Demos}}}, \bibinfo {author}
  {\bibfnamefont {H.}~\bibnamefont {{Fong}}}, \bibinfo {author} {\bibfnamefont
  {J.}~\bibnamefont {{Healy}}}, \bibinfo {author} {\bibfnamefont {D.~A.}\
  \bibnamefont {{Hemberger}}}, \bibinfo {author} {\bibfnamefont
  {I.}~\bibnamefont {{Hinder}}}, \bibinfo {author} {\bibfnamefont
  {K.}~\bibnamefont {{Jani}}}, \bibinfo {author} {\bibfnamefont
  {B.}~\bibnamefont {{Khamesra}}}, \bibinfo {author} {\bibfnamefont {L.~E.}\
  \bibnamefont {{Kidder}}}, \bibinfo {author} {\bibfnamefont {P.}~\bibnamefont
  {{Kumar}}}, \bibinfo {author} {\bibfnamefont {P.}~\bibnamefont {{Laguna}}},
  \bibinfo {author} {\bibfnamefont {C.~O.}\ \bibnamefont {{Lousto}}}, \bibinfo
  {author} {\bibfnamefont {G.}~\bibnamefont {{Lovelace}}}, \bibinfo {author}
  {\bibfnamefont {S.}~\bibnamefont {{Ossokine}}}, \bibinfo {author}
  {\bibfnamefont {H.}~\bibnamefont {{Pfeiffer}}}, \bibinfo {author}
  {\bibfnamefont {M.~A.}\ \bibnamefont {{Scheel}}}, \bibinfo {author}
  {\bibfnamefont {D.~M.}\ \bibnamefont {{Shoemaker}}}, \bibinfo {author}
  {\bibfnamefont {B.}~\bibnamefont {{Szilagyi}}}, \bibinfo {author}
  {\bibfnamefont {S.}~\bibnamefont {{Teukolsky}}}, \ and\ \bibinfo {author}
  {\bibfnamefont {Y.}~\bibnamefont {{Zlochower}}},\ }\href {\doibase
  10.1103/PhysRevD.96.104041} {\bibfield  {journal} {\bibinfo  {journal}
  {\prd}\ }\textbf {\bibinfo {volume} {96}},\ \bibinfo {eid} {104041} (\bibinfo
  {year} {2017})},\ \Eprint {http://arxiv.org/abs/1705.09833} {arXiv:1705.09833
  [gr-qc]} \BibitemShut {NoStop}%
\bibitem [{\citenamefont
  {{Lange}}(2016)}]{gwastro-mergers-nr-LangeMastersThesis}%
  \BibitemOpen
  \bibfield  {author} {\bibinfo {author} {\bibfnamefont {J.}~\bibnamefont
  {{Lange}}},\ }\href@noop {} {\enquote {\bibinfo {title} {{ Reconstructing
  gravitational wave source parameters via direct comparisons to numerical
  relativity}},}\ } (\bibinfo {year} {2016}),\ \bibinfo {note} {master's thesis
  submitted to the Rochester Institute of Technology, available as
  LIGO-P1600281 at https://dcc.ligo.org/LIGO-P1600281}\BibitemShut {NoStop}%
\bibitem [{\citenamefont {Lange}\ \emph {et~al.}(2018)\citenamefont {Lange},
  \citenamefont {O'Shaughnessy},\ and\ \citenamefont {Rizzo}}]{Lange:2018pyp}%
  \BibitemOpen
  \bibfield  {author} {\bibinfo {author} {\bibfnamefont {J.}~\bibnamefont
  {Lange}}, \bibinfo {author} {\bibfnamefont {R.}~\bibnamefont
  {O'Shaughnessy}}, \ and\ \bibinfo {author} {\bibfnamefont {M.}~\bibnamefont
  {Rizzo}},\ }\href@noop {} {\  (\bibinfo {year} {2018})},\ \Eprint
  {http://arxiv.org/abs/1805.10457} {arXiv:1805.10457 [gr-qc]} \BibitemShut
  {NoStop}%
\bibitem [{\citenamefont {Abbott}\ \emph {et~al.}(2019)\citenamefont {Abbott}
  \emph {et~al.}}]{LIGOScientific:2018mvr}%
  \BibitemOpen
  \bibfield  {author} {\bibinfo {author} {\bibfnamefont {B.~P.}\ \bibnamefont
  {Abbott}} \emph {et~al.} (\bibinfo {collaboration} {LIGO Scientific,
  Virgo}),\ }\href {\doibase 10.1103/PhysRevX.9.031040} {\bibfield  {journal}
  {\bibinfo  {journal} {Phys. Rev.}\ }\textbf {\bibinfo {volume} {X9}},\
  \bibinfo {pages} {031040} (\bibinfo {year} {2019})},\ \Eprint
  {http://arxiv.org/abs/1811.12907} {arXiv:1811.12907 [astro-ph.HE]}
  \BibitemShut {NoStop}%
\bibitem [{\citenamefont {Vitale}\ \emph {et~al.}(2017)\citenamefont {Vitale},
  \citenamefont {Gerosa}, \citenamefont {Haster}, \citenamefont
  {Chatziioannou},\ and\ \citenamefont {Zimmerman}}]{Vitale:2017cfs}%
  \BibitemOpen
  \bibfield  {author} {\bibinfo {author} {\bibfnamefont {S.}~\bibnamefont
  {Vitale}}, \bibinfo {author} {\bibfnamefont {D.}~\bibnamefont {Gerosa}},
  \bibinfo {author} {\bibfnamefont {C.-J.}\ \bibnamefont {Haster}}, \bibinfo
  {author} {\bibfnamefont {K.}~\bibnamefont {Chatziioannou}}, \ and\ \bibinfo
  {author} {\bibfnamefont {A.}~\bibnamefont {Zimmerman}},\ }\href {\doibase
  10.1103/PhysRevLett.119.251103} {\bibfield  {journal} {\bibinfo  {journal}
  {Phys. Rev. Lett.}\ }\textbf {\bibinfo {volume} {119}},\ \bibinfo {pages}
  {251103} (\bibinfo {year} {2017})},\ \Eprint
  {http://arxiv.org/abs/1707.04637} {arXiv:1707.04637 [gr-qc]} \BibitemShut
  {NoStop}%
\bibitem [{\citenamefont {Lousto}\ and\ \citenamefont
  {Zlochower}(2013)}]{Lousto:2012gt}%
  \BibitemOpen
  \bibfield  {author} {\bibinfo {author} {\bibfnamefont {C.~O.}\ \bibnamefont
  {Lousto}}\ and\ \bibinfo {author} {\bibfnamefont {Y.}~\bibnamefont
  {Zlochower}},\ }\href {\doibase 10.1103/PhysRevD.87.084027} {\bibfield
  {journal} {\bibinfo  {journal} {Phys. Rev.}\ }\textbf {\bibinfo {volume}
  {D87}},\ \bibinfo {pages} {084027} (\bibinfo {year} {2013})},\ \Eprint
  {http://arxiv.org/abs/1211.7099} {arXiv:1211.7099 [gr-qc]} \BibitemShut
  {NoStop}%
\bibitem [{\citenamefont {Zlochower}\ and\ \citenamefont
  {Lousto}(2015)}]{Zlochower:2015wga}%
  \BibitemOpen
  \bibfield  {author} {\bibinfo {author} {\bibfnamefont {Y.}~\bibnamefont
  {Zlochower}}\ and\ \bibinfo {author} {\bibfnamefont {C.~O.}\ \bibnamefont
  {Lousto}},\ }\href {\doibase 10.1103/PhysRevD.92.024022} {\bibfield
  {journal} {\bibinfo  {journal} {Phys. Rev.}\ }\textbf {\bibinfo {volume}
  {D92}},\ \bibinfo {pages} {024022} (\bibinfo {year} {2015})},\ \Eprint
  {http://arxiv.org/abs/1503.07536} {arXiv:1503.07536 [gr-qc]} \BibitemShut
  {NoStop}%
\bibitem [{\citenamefont {{Pankow}}\ \emph
  {et~al.}(2015{\natexlab{b}})\citenamefont {{Pankow}}, \citenamefont
  {{Brady}}, \citenamefont {{Ochsner}},\ and\ \citenamefont
  {{O'Shaughnessy}}}]{2015PhRvD..92b3002P}%
  \BibitemOpen
  \bibfield  {author} {\bibinfo {author} {\bibfnamefont {C.}~\bibnamefont
  {{Pankow}}}, \bibinfo {author} {\bibfnamefont {P.}~\bibnamefont {{Brady}}},
  \bibinfo {author} {\bibfnamefont {E.}~\bibnamefont {{Ochsner}}}, \ and\
  \bibinfo {author} {\bibfnamefont {R.}~\bibnamefont {{O'Shaughnessy}}},\
  }\href {\doibase 10.1103/PhysRevD.92.023002} {\bibfield  {journal} {\bibinfo
  {journal} {\prd}\ }\textbf {\bibinfo {volume} {92}},\ \bibinfo {eid} {023002}
  (\bibinfo {year} {2015}{\natexlab{b}})},\ \Eprint
  {http://arxiv.org/abs/1502.04370} {arXiv:1502.04370 [gr-qc]} \BibitemShut
  {NoStop}%
\bibitem [{\citenamefont {{O'Shaughnessy}}\ \emph {et~al.}(2017)\citenamefont
  {{O'Shaughnessy}}, \citenamefont {{Blackman}},\ and\ \citenamefont
  {{Field}}}]{2017CQGra..34n4002O}%
  \BibitemOpen
  \bibfield  {author} {\bibinfo {author} {\bibfnamefont {R.}~\bibnamefont
  {{O'Shaughnessy}}}, \bibinfo {author} {\bibfnamefont {J.}~\bibnamefont
  {{Blackman}}}, \ and\ \bibinfo {author} {\bibfnamefont {S.~E.}\ \bibnamefont
  {{Field}}},\ }\href {\doibase 10.1088/1361-6382/aa7649} {\bibfield  {journal}
  {\bibinfo  {journal} {Classical and Quantum Gravity}\ }\textbf {\bibinfo
  {volume} {34}},\ \bibinfo {eid} {144002} (\bibinfo {year} {2017})},\ \Eprint
  {http://arxiv.org/abs/1701.01137} {arXiv:1701.01137 [gr-qc]} \BibitemShut
  {NoStop}%
\bibitem [{\citenamefont {{Lange}}\ \emph {et~al.}(2018)\citenamefont
  {{Lange}}, \citenamefont {{O'Shaughnessy}},\ and\ \citenamefont
  {{Rizzo}}}]{2018arXiv180510457L}%
  \BibitemOpen
  \bibfield  {author} {\bibinfo {author} {\bibfnamefont {J.}~\bibnamefont
  {{Lange}}}, \bibinfo {author} {\bibfnamefont {R.}~\bibnamefont
  {{O'Shaughnessy}}}, \ and\ \bibinfo {author} {\bibfnamefont {M.}~\bibnamefont
  {{Rizzo}}},\ }\href@noop {} {\bibfield  {journal} {\bibinfo  {journal} {ArXiv
  e-prints}\ } (\bibinfo {year} {2018})},\ \Eprint
  {http://arxiv.org/abs/1805.10457} {arXiv:1805.10457 [gr-qc]} \BibitemShut
  {NoStop}%
\bibitem [{\citenamefont {Healy}\ and\ \citenamefont
  {Lousto}(2018)}]{Healy:2018swt}%
  \BibitemOpen
  \bibfield  {author} {\bibinfo {author} {\bibfnamefont {J.}~\bibnamefont
  {Healy}}\ and\ \bibinfo {author} {\bibfnamefont {C.~O.}\ \bibnamefont
  {Lousto}},\ }\href {\doibase 10.1103/PhysRevD.97.084002} {\bibfield
  {journal} {\bibinfo  {journal} {Phys. Rev.}\ }\textbf {\bibinfo {volume}
  {D97}},\ \bibinfo {pages} {084002} (\bibinfo {year} {2018})},\ \Eprint
  {http://arxiv.org/abs/1801.08162} {arXiv:1801.08162 [gr-qc]} \BibitemShut
  {NoStop}%
\bibitem [{\citenamefont {Ajith}\ \emph {et~al.}(2011)\citenamefont {Ajith}
  \emph {et~al.}}]{Ajith:2009bn}%
  \BibitemOpen
  \bibfield  {author} {\bibinfo {author} {\bibfnamefont {P.}~\bibnamefont
  {Ajith}} \emph {et~al.},\ }\href {\doibase 10.1103/PhysRevLett.106.241101}
  {\bibfield  {journal} {\bibinfo  {journal} {Phys. Rev. Lett.}\ }\textbf
  {\bibinfo {volume} {106}},\ \bibinfo {pages} {241101} (\bibinfo {year}
  {2011})},\ \Eprint {http://arxiv.org/abs/0909.2867} {arXiv:0909.2867 [gr-qc]}
  \BibitemShut {NoStop}%
\bibitem [{\citenamefont {Healy}\ \emph
  {et~al.}(2018{\natexlab{b}})\citenamefont {Healy} \emph
  {et~al.}}]{Healy:2017abq}%
  \BibitemOpen
  \bibfield  {author} {\bibinfo {author} {\bibfnamefont {J.}~\bibnamefont
  {Healy}} \emph {et~al.},\ }\href {\doibase 10.1103/PhysRevD.97.064027}
  {\bibfield  {journal} {\bibinfo  {journal} {Phys. Rev.}\ }\textbf {\bibinfo
  {volume} {D97}},\ \bibinfo {pages} {064027} (\bibinfo {year}
  {2018}{\natexlab{b}})},\ \Eprint {http://arxiv.org/abs/1712.05836}
  {arXiv:1712.05836 [gr-qc]} \BibitemShut {NoStop}%
\bibitem [{\citenamefont {Ade}\ \emph {et~al.}(2016)\citenamefont {Ade} \emph
  {et~al.}}]{Ade:2015xua}%
  \BibitemOpen
  \bibfield  {author} {\bibinfo {author} {\bibfnamefont {P.~A.~R.}\
  \bibnamefont {Ade}} \emph {et~al.} (\bibinfo {collaboration} {Planck}),\
  }\href {\doibase 10.1051/0004-6361/201525830} {\bibfield  {journal} {\bibinfo
   {journal} {Astron. Astrophys.}\ }\textbf {\bibinfo {volume} {594}},\
  \bibinfo {pages} {A13} (\bibinfo {year} {2016})},\ \Eprint
  {http://arxiv.org/abs/1502.01589} {arXiv:1502.01589 [astro-ph.CO]}
  \BibitemShut {NoStop}%
\bibitem [{\citenamefont {Mateos}\ \emph {et~al.}(2017)\citenamefont {Mateos},
  \citenamefont {Riveaud},\ and\ \citenamefont {Lamberti}}]{Mateos:2017gvm}%
  \BibitemOpen
  \bibfield  {author} {\bibinfo {author} {\bibfnamefont {D.}~\bibnamefont
  {Mateos}}, \bibinfo {author} {\bibfnamefont {L.}~\bibnamefont {Riveaud}}, \
  and\ \bibinfo {author} {\bibfnamefont {P.}~\bibnamefont {Lamberti}},\ }\href
  {\doibase 10.1063/1.4999613} {\bibfield  {journal} {\bibinfo  {journal}
  {Chaos}\ }\textbf {\bibinfo {volume} {27}},\ \bibinfo {pages} {083118}
  (\bibinfo {year} {2017})},\ \Eprint {http://arxiv.org/abs/1702.08276}
  {arXiv:1702.08276 [physics.data-an]} \BibitemShut {NoStop}%
\bibitem [{\citenamefont {{Chatziioannou}}\ \emph {et~al.}(2019)\citenamefont
  {{Chatziioannou}}, \citenamefont {{Cotesta}}, \citenamefont {{Ghonge}},
  \citenamefont {{Lange}}, \citenamefont {{Ng}}, \citenamefont {{Calderon
  Bustillo}}, \citenamefont {{Clark}}, \citenamefont {{Haster}}, \citenamefont
  {{Khan}}, \citenamefont {{Puerrer}}, \citenamefont {{Raymond}}, \citenamefont
  {{Vitale}}, \citenamefont {{Afshari}}, \citenamefont {{Babak}}, \citenamefont
  {{Barkett}}, \citenamefont {{Blackman}}, \citenamefont {{Bohe}},
  \citenamefont {{Boyle}}, \citenamefont {{Buonanno}}, \citenamefont
  {{Campanelli}}, \citenamefont {{Carullo}}, \citenamefont {{Chu}},
  \citenamefont {{Flynn}}, \citenamefont {{Fong}}, \citenamefont {{Garcia}},
  \citenamefont {{Giesler}}, \citenamefont {{Haney}}, \citenamefont {{Hannam}},
  \citenamefont {{Harry}}, \citenamefont {{Healy}}, \citenamefont
  {{Hemberger}}, \citenamefont {{Hinder}}, \citenamefont {{Jani}},
  \citenamefont {{Khamersa}}, \citenamefont {{Kidder}}, \citenamefont
  {{Kumar}}, \citenamefont {{Laguna}}, \citenamefont {{Lousto}}, \citenamefont
  {{Lovelace}}, \citenamefont {{Littenberg}}, \citenamefont {{London}},
  \citenamefont {{Millhouse}}, \citenamefont {{Nuttall}}, \citenamefont
  {{Ohme}}, \citenamefont {{O'Shaughnessy}}, \citenamefont {{Ossokine}},
  \citenamefont {{Pannarale}}, \citenamefont {{Schmidt}}, \citenamefont
  {{Pfeiffer}}, \citenamefont {{Scheel}}, \citenamefont {{Shao}}, \citenamefont
  {{Shoemaker}}, \citenamefont {{Szilagyi}}, \citenamefont {{Taracchini}},
  \citenamefont {{Teukolsky}},\ and\ \citenamefont
  {{Zlochower}}}]{gwastro-170729HM-Katerina}%
  \BibitemOpen
  \bibfield  {author} {\bibinfo {author} {\bibfnamefont {K.}~\bibnamefont
  {{Chatziioannou}}}, \bibinfo {author} {\bibfnamefont {R.}~\bibnamefont
  {{Cotesta}}}, \bibinfo {author} {\bibfnamefont {S.}~\bibnamefont {{Ghonge}}},
  \bibinfo {author} {\bibfnamefont {J.}~\bibnamefont {{Lange}}}, \bibinfo
  {author} {\bibfnamefont {K.~K.-Y.}\ \bibnamefont {{Ng}}}, \bibinfo {author}
  {\bibfnamefont {J.}~\bibnamefont {{Calderon Bustillo}}}, \bibinfo {author}
  {\bibfnamefont {J.}~\bibnamefont {{Clark}}}, \bibinfo {author} {\bibfnamefont
  {C.-J.}\ \bibnamefont {{Haster}}}, \bibinfo {author} {\bibfnamefont
  {S.}~\bibnamefont {{Khan}}}, \bibinfo {author} {\bibfnamefont
  {M.}~\bibnamefont {{Puerrer}}}, \bibinfo {author} {\bibfnamefont
  {V.}~\bibnamefont {{Raymond}}}, \bibinfo {author} {\bibfnamefont
  {S.}~\bibnamefont {{Vitale}}}, \bibinfo {author} {\bibfnamefont
  {N.}~\bibnamefont {{Afshari}}}, \bibinfo {author} {\bibfnamefont
  {S.}~\bibnamefont {{Babak}}}, \bibinfo {author} {\bibfnamefont
  {K.}~\bibnamefont {{Barkett}}}, \bibinfo {author} {\bibfnamefont
  {J.}~\bibnamefont {{Blackman}}}, \bibinfo {author} {\bibfnamefont
  {A.}~\bibnamefont {{Bohe}}}, \bibinfo {author} {\bibfnamefont
  {M.}~\bibnamefont {{Boyle}}}, \bibinfo {author} {\bibfnamefont
  {A.}~\bibnamefont {{Buonanno}}}, \bibinfo {author} {\bibfnamefont
  {M.}~\bibnamefont {{Campanelli}}}, \bibinfo {author} {\bibfnamefont
  {G.}~\bibnamefont {{Carullo}}}, \bibinfo {author} {\bibfnamefont
  {T.}~\bibnamefont {{Chu}}}, \bibinfo {author} {\bibfnamefont
  {E.}~\bibnamefont {{Flynn}}}, \bibinfo {author} {\bibfnamefont
  {H.}~\bibnamefont {{Fong}}}, \bibinfo {author} {\bibfnamefont
  {A.}~\bibnamefont {{Garcia}}}, \bibinfo {author} {\bibfnamefont
  {M.}~\bibnamefont {{Giesler}}}, \bibinfo {author} {\bibfnamefont
  {M.}~\bibnamefont {{Haney}}}, \bibinfo {author} {\bibfnamefont
  {M.}~\bibnamefont {{Hannam}}}, \bibinfo {author} {\bibfnamefont
  {I.}~\bibnamefont {{Harry}}}, \bibinfo {author} {\bibfnamefont
  {J.}~\bibnamefont {{Healy}}}, \bibinfo {author} {\bibfnamefont
  {D.}~\bibnamefont {{Hemberger}}}, \bibinfo {author} {\bibfnamefont
  {I.}~\bibnamefont {{Hinder}}}, \bibinfo {author} {\bibfnamefont
  {K.}~\bibnamefont {{Jani}}}, \bibinfo {author} {\bibfnamefont
  {B.}~\bibnamefont {{Khamersa}}}, \bibinfo {author} {\bibfnamefont {L.~E.}\
  \bibnamefont {{Kidder}}}, \bibinfo {author} {\bibfnamefont {P.}~\bibnamefont
  {{Kumar}}}, \bibinfo {author} {\bibfnamefont {P.}~\bibnamefont {{Laguna}}},
  \bibinfo {author} {\bibfnamefont {C.~O.}\ \bibnamefont {{Lousto}}}, \bibinfo
  {author} {\bibfnamefont {G.}~\bibnamefont {{Lovelace}}}, \bibinfo {author}
  {\bibfnamefont {T.~B.}\ \bibnamefont {{Littenberg}}}, \bibinfo {author}
  {\bibfnamefont {L.}~\bibnamefont {{London}}}, \bibinfo {author}
  {\bibfnamefont {M.}~\bibnamefont {{Millhouse}}}, \bibinfo {author}
  {\bibfnamefont {L.~K.}\ \bibnamefont {{Nuttall}}}, \bibinfo {author}
  {\bibfnamefont {F.}~\bibnamefont {{Ohme}}}, \bibinfo {author} {\bibfnamefont
  {R.}~\bibnamefont {{O'Shaughnessy}}}, \bibinfo {author} {\bibfnamefont
  {S.}~\bibnamefont {{Ossokine}}}, \bibinfo {author} {\bibfnamefont
  {F.}~\bibnamefont {{Pannarale}}}, \bibinfo {author} {\bibfnamefont
  {P.}~\bibnamefont {{Schmidt}}}, \bibinfo {author} {\bibfnamefont {H.~P.}\
  \bibnamefont {{Pfeiffer}}}, \bibinfo {author} {\bibfnamefont {M.~A.}\
  \bibnamefont {{Scheel}}}, \bibinfo {author} {\bibfnamefont {L.}~\bibnamefont
  {{Shao}}}, \bibinfo {author} {\bibfnamefont {D.}~\bibnamefont {{Shoemaker}}},
  \bibinfo {author} {\bibfnamefont {B.}~\bibnamefont {{Szilagyi}}}, \bibinfo
  {author} {\bibfnamefont {A.}~\bibnamefont {{Taracchini}}}, \bibinfo {author}
  {\bibfnamefont {S.~A.}\ \bibnamefont {{Teukolsky}}}, \ and\ \bibinfo {author}
  {\bibfnamefont {Y.}~\bibnamefont {{Zlochower}}},\ }\href {\doibase
  https://doi.org/10.1103/PhysRevD.100.104015} {\bibfield  {journal} {\bibinfo
  {journal} {\prd}\ }\textbf {\bibinfo {volume} {100}},\ \bibinfo {pages}
  {104015} (\bibinfo {year} {2019})},\ \Eprint
  {http://arxiv.org/abs/1903.06742} {1903.06742 [gr-qc]} \BibitemShut {NoStop}%
\bibitem [{\citenamefont {{O'Shaughnessy}}\ and\ \citenamefont
  {Lange}(2019)}]{gwastro-170729HM-TechDoc}%
  \BibitemOpen
  \bibfield  {author} {\bibinfo {author} {\bibfnamefont {R.}~\bibnamefont
  {{O'Shaughnessy}}}\ and\ \bibinfo {author} {\bibfnamefont {J.}~\bibnamefont
  {Lange}},\ }\href@noop {} {\bibfield  {journal} {\bibinfo  {journal}
  {Available as LIGO-T1900096}\ } (\bibinfo {year} {2019})}\BibitemShut
  {NoStop}%
\bibitem [{\citenamefont {Healy}\ \emph {et~al.}(2014)\citenamefont {Healy},
  \citenamefont {Lousto},\ and\ \citenamefont {Zlochower}}]{Healy:2014yta}%
  \BibitemOpen
  \bibfield  {author} {\bibinfo {author} {\bibfnamefont {J.}~\bibnamefont
  {Healy}}, \bibinfo {author} {\bibfnamefont {C.~O.}\ \bibnamefont {Lousto}}, \
  and\ \bibinfo {author} {\bibfnamefont {Y.}~\bibnamefont {Zlochower}},\ }\href
  {\doibase 10.1103/PhysRevD.90.104004} {\bibfield  {journal} {\bibinfo
  {journal} {Phys. Rev.}\ }\textbf {\bibinfo {volume} {D90}},\ \bibinfo {pages}
  {104004} (\bibinfo {year} {2014})},\ \Eprint {http://arxiv.org/abs/1406.7295}
  {arXiv:1406.7295 [gr-qc]} \BibitemShut {NoStop}%
\bibitem [{\citenamefont {Healy}\ and\ \citenamefont
  {Lousto}(2017)}]{Healy:2016lce}%
  \BibitemOpen
  \bibfield  {author} {\bibinfo {author} {\bibfnamefont {J.}~\bibnamefont
  {Healy}}\ and\ \bibinfo {author} {\bibfnamefont {C.~O.}\ \bibnamefont
  {Lousto}},\ }\href {\doibase 10.1103/PhysRevD.95.024037} {\bibfield
  {journal} {\bibinfo  {journal} {Phys. Rev.}\ }\textbf {\bibinfo {volume}
  {D95}},\ \bibinfo {pages} {024037} (\bibinfo {year} {2017})},\ \Eprint
  {http://arxiv.org/abs/1610.09713} {arXiv:1610.09713 [gr-qc]} \BibitemShut
  {NoStop}%
\bibitem [{\citenamefont {{Abbott}}\ \emph {et~al.}(2016)\citenamefont
  {{Abbott}}, \citenamefont {{Abbott}}, \citenamefont {{Abbott}}, \citenamefont
  {{Abernathy}}, \citenamefont {{Acernese}}, \citenamefont {{Ackley}},
  \citenamefont {{Adams}}, \citenamefont {{Adams}}, \citenamefont {{Addesso}},
  \citenamefont {{Adhikari}},\ and\ \citenamefont
  {et~al.}}]{2016PhRvL.116x1103A}%
  \BibitemOpen
  \bibfield  {author} {\bibinfo {author} {\bibfnamefont {B.~P.}\ \bibnamefont
  {{Abbott}}}, \bibinfo {author} {\bibfnamefont {R.}~\bibnamefont {{Abbott}}},
  \bibinfo {author} {\bibfnamefont {T.~D.}\ \bibnamefont {{Abbott}}}, \bibinfo
  {author} {\bibfnamefont {M.~R.}\ \bibnamefont {{Abernathy}}}, \bibinfo
  {author} {\bibfnamefont {F.}~\bibnamefont {{Acernese}}}, \bibinfo {author}
  {\bibfnamefont {K.}~\bibnamefont {{Ackley}}}, \bibinfo {author}
  {\bibfnamefont {C.}~\bibnamefont {{Adams}}}, \bibinfo {author} {\bibfnamefont
  {T.}~\bibnamefont {{Adams}}}, \bibinfo {author} {\bibfnamefont
  {P.}~\bibnamefont {{Addesso}}}, \bibinfo {author} {\bibfnamefont {R.~X.}\
  \bibnamefont {{Adhikari}}}, \ and\ \bibinfo {author} {\bibnamefont
  {et~al.}},\ }\href {\doibase 10.1103/PhysRevLett.116.241103} {\bibfield
  {journal} {\bibinfo  {journal} {Physical Review Letters}\ }\textbf {\bibinfo
  {volume} {116}},\ \bibinfo {eid} {241103} (\bibinfo {year} {2016})},\ \Eprint
  {http://arxiv.org/abs/1606.04855} {arXiv:1606.04855 [gr-qc]} \BibitemShut
  {NoStop}%
\bibitem [{\citenamefont {Kumar}\ \emph {et~al.}(2019)\citenamefont {Kumar},
  \citenamefont {Blackman}, \citenamefont {Field}, \citenamefont {Scheel},
  \citenamefont {Galley}, \citenamefont {Boyle}, \citenamefont {Kidder},
  \citenamefont {Pfeiffer}, \citenamefont {Szilagyi},\ and\ \citenamefont
  {Teukolsky}}]{Kumar:2018hml}%
  \BibitemOpen
  \bibfield  {author} {\bibinfo {author} {\bibfnamefont {P.}~\bibnamefont
  {Kumar}}, \bibinfo {author} {\bibfnamefont {J.}~\bibnamefont {Blackman}},
  \bibinfo {author} {\bibfnamefont {S.~E.}\ \bibnamefont {Field}}, \bibinfo
  {author} {\bibfnamefont {M.}~\bibnamefont {Scheel}}, \bibinfo {author}
  {\bibfnamefont {C.~R.}\ \bibnamefont {Galley}}, \bibinfo {author}
  {\bibfnamefont {M.}~\bibnamefont {Boyle}}, \bibinfo {author} {\bibfnamefont
  {L.~E.}\ \bibnamefont {Kidder}}, \bibinfo {author} {\bibfnamefont {H.~P.}\
  \bibnamefont {Pfeiffer}}, \bibinfo {author} {\bibfnamefont {B.}~\bibnamefont
  {Szilagyi}}, \ and\ \bibinfo {author} {\bibfnamefont {S.~A.}\ \bibnamefont
  {Teukolsky}},\ }\href {\doibase 10.1103/PhysRevD.99.124005} {\bibfield
  {journal} {\bibinfo  {journal} {Phys. Rev. D}\ }\textbf {\bibinfo {volume}
  {99}},\ \bibinfo {pages} {124005} (\bibinfo {year} {2019})},\ \Eprint
  {http://arxiv.org/abs/1808.08004} {arXiv:1808.08004 [gr-qc]} \BibitemShut
  {NoStop}%
\bibitem [{\citenamefont {Babak}\ \emph {et~al.}(2017)\citenamefont {Babak},
  \citenamefont {Taracchini},\ and\ \citenamefont {Buonanno}}]{Babak:2016tgq}%
  \BibitemOpen
  \bibfield  {author} {\bibinfo {author} {\bibfnamefont {S.}~\bibnamefont
  {Babak}}, \bibinfo {author} {\bibfnamefont {A.}~\bibnamefont {Taracchini}}, \
  and\ \bibinfo {author} {\bibfnamefont {A.}~\bibnamefont {Buonanno}},\ }\href
  {\doibase 10.1103/PhysRevD.95.024010} {\bibfield  {journal} {\bibinfo
  {journal} {Phys. Rev.}\ }\textbf {\bibinfo {volume} {D95}},\ \bibinfo {pages}
  {024010} (\bibinfo {year} {2017})},\ \Eprint
  {http://arxiv.org/abs/1607.05661} {arXiv:1607.05661 [gr-qc]} \BibitemShut
  {NoStop}%
\bibitem [{\citenamefont {Hannam}\ \emph {et~al.}(2014)\citenamefont {Hannam},
  \citenamefont {Schmidt}, \citenamefont {Boh{\'e}}, \citenamefont {Haegel},
  \citenamefont {Husa}, \citenamefont {Ohme}, \citenamefont {Pratten},\ and\
  \citenamefont {P{\"u}rrer}}]{Hannam:2013oca}%
  \BibitemOpen
  \bibfield  {author} {\bibinfo {author} {\bibfnamefont {M.}~\bibnamefont
  {Hannam}}, \bibinfo {author} {\bibfnamefont {P.}~\bibnamefont {Schmidt}},
  \bibinfo {author} {\bibfnamefont {A.}~\bibnamefont {Boh{\'e}}}, \bibinfo
  {author} {\bibfnamefont {L.}~\bibnamefont {Haegel}}, \bibinfo {author}
  {\bibfnamefont {S.}~\bibnamefont {Husa}}, \bibinfo {author} {\bibfnamefont
  {F.}~\bibnamefont {Ohme}}, \bibinfo {author} {\bibfnamefont {G.}~\bibnamefont
  {Pratten}}, \ and\ \bibinfo {author} {\bibfnamefont {M.}~\bibnamefont
  {P{\"u}rrer}},\ }\href {\doibase 10.1103/PhysRevLett.113.151101} {\bibfield
  {journal} {\bibinfo  {journal} {Phys. Rev. Lett.}\ }\textbf {\bibinfo
  {volume} {113}},\ \bibinfo {pages} {151101} (\bibinfo {year} {2014})},\
  \Eprint {http://arxiv.org/abs/1308.3271} {arXiv:1308.3271 [gr-qc]}
  \BibitemShut {NoStop}%
\bibitem [{\citenamefont {Sadiq}\ \emph {et~al.}(2020)\citenamefont {Sadiq},
  \citenamefont {Zlochower}, \citenamefont {O'Shaughnessy},\ and\ \citenamefont
  {Lange}}]{Sadiq:2020hti}%
  \BibitemOpen
  \bibfield  {author} {\bibinfo {author} {\bibfnamefont {J.}~\bibnamefont
  {Sadiq}}, \bibinfo {author} {\bibfnamefont {Y.}~\bibnamefont {Zlochower}},
  \bibinfo {author} {\bibfnamefont {R.}~\bibnamefont {O'Shaughnessy}}, \ and\
  \bibinfo {author} {\bibfnamefont {J.}~\bibnamefont {Lange}},\ }\href
  {\doibase 10.1103/PhysRevD.102.024012} {\bibfield  {journal} {\bibinfo
  {journal} {Phys. Rev. D}\ }\textbf {\bibinfo {volume} {102}},\ \bibinfo
  {pages} {024012} (\bibinfo {year} {2020})},\ \Eprint
  {http://arxiv.org/abs/2001.07109} {arXiv:2001.07109 [gr-qc]} \BibitemShut
  {NoStop}%
\bibitem [{\citenamefont {Healy}\ \emph
  {et~al.}(2017{\natexlab{b}})\citenamefont {Healy}, \citenamefont {Lousto},\
  and\ \citenamefont {Zlochower}}]{Healy:2017mvh}%
  \BibitemOpen
  \bibfield  {author} {\bibinfo {author} {\bibfnamefont {J.}~\bibnamefont
  {Healy}}, \bibinfo {author} {\bibfnamefont {C.~O.}\ \bibnamefont {Lousto}}, \
  and\ \bibinfo {author} {\bibfnamefont {Y.}~\bibnamefont {Zlochower}},\ }\href
  {\doibase 10.1103/PhysRevD.96.024031} {\bibfield  {journal} {\bibinfo
  {journal} {Phys. Rev.}\ }\textbf {\bibinfo {volume} {D96}},\ \bibinfo {pages}
  {024031} (\bibinfo {year} {2017}{\natexlab{b}})},\ \Eprint
  {http://arxiv.org/abs/1705.07034} {arXiv:1705.07034 [gr-qc]} \BibitemShut
  {NoStop}%
\end{thebibliography}%


\end{document}